\documentclass[reprint,twocolumn,showpacs,showkeys,preprintnumbers,amsmath,amssymb,floatfix,aps,prl,longbibliography]{revtex4-1}

\usepackage{graphicx}% Include figure files
\usepackage{bm}% bold math

\usepackage{physics}
\usepackage{placeins}
\usepackage{graphics}
\usepackage{color}
\usepackage{hyperref}
\usepackage{multirow}
\usepackage{blindtext}
\usepackage{pdfpages}
\makeatletter
\AtBeginDocument{\let\LS@rot\@undefined}
\makeatother 

\begin{document}

%\preprint{APS/123-QED}

\title{Engineering proximity exchange by twisting: Reversal of ferromagnetic and emergence of antiferromagnetic Dirac bands in graphene/Cr$_2$Ge$_2$Te$_6$}

\author{Klaus Zollner}
\email{klaus.zollner@physik.uni-regensburg.de}
\affiliation{Institute for Theoretical Physics, University of Regensburg, 93040 Regensburg, Germany}
\author{Jaroslav Fabian}
\affiliation{Institute for Theoretical Physics, University of Regensburg, 93040 Regensburg, Germany}

\begin{abstract}
We investigate the twist-angle and gate dependence of the proximity exchange coupling in twisted graphene on monolayer Cr$_2$Ge$_2$Te$_6$ from first principles. The proximitized Dirac band dispersions of graphene are fitted to a model Hamiltonian, yielding effective sublattice-resolved proximity-induced exchange parameters ($\lambda_{\textrm{ex}}^\textrm{A}$ and $\lambda_{\textrm{ex}}^\textrm{B}$) for a series of twist angles between 0$^{\circ}$ and 30$^{\circ}$. For aligned layers (0$^{\circ}$ twist angle), the exchange coupling of graphene is the same on both sublattices,  \mbox{$\lambda_{\textrm{ex}}^\textrm{A} \approx \lambda_{\textrm{ex}}^\textrm{B} \approx 4$~meV}, while the coupling is reversed
at 30$^{\circ}$ (with \mbox{$\lambda_{\textrm{ex}}^\textrm{A} \approx \lambda_{\textrm{ex}}^\textrm{B} \approx -4$~meV}). Remarkably, at 19.1$^{\circ}$ the induced
exchange coupling becomes antiferromagnetic:  $\lambda_{\textrm{ex}}^\textrm{A} < 0, \lambda_{\textrm{ex}}^\textrm{B} > 0$. 
Further tuning is provided by a transverse electric field and the interlayer distance. The predicted proximity magnetization reversal and emergence of an antiferromagnetic Dirac dispersion make twisted graphene/Cr$_2$Ge$_2$Te$_6$ bilayers a versatile platform for realizing topological phases and for spintronics applications. 
\end{abstract}

\maketitle

%------------------------------------------------------------
%\section{Introduction}
%------------------------------------------------------------

Van der Waals (vdW) heterostructures composed of twisted monolayers 
\cite{Carr2017:PRB,Hennighausen2021:ES,Ribeiro2018:SC,Carr2020:NRM}
promise great tunability of electronic, optical, and magnetic properties. The most prominent example is magic-angle twisted bilayer graphene, exhibiting magnetism and superconductivity due to strong correlations \cite{Cao2018:Nat,Cao2018a:Nat,Arora2020:arxiv,Stepanov2020:Nat,Lu2019:Nat,Sharpe2019:SC, Saito2021:Nat,Serlin2020:S,Nimbalkar2020:NML,Bultinck2020:PRL,Repellin2020:PRL,Choi2019:NP,Lisi2021:NP,Balents2020:NP,Wolf2019:PRL}. Other platforms for correlated physics are offered by trilayer graphene \cite{Zhu2020:PRL,Park2021:Nat,Chen2019:NP,Chen2020:Nat,Chen2019:Nat,Zhou2021:arxiv,Chou2021:arxiv,Phong2021:arxiv,Zhou2021:arxiv2,Qin2021:arxiv} and twisted transition metal dichalcogenides (TMDCs) \cite{Tang2020:Nat}.

However, twistronics is yet to demonstrate its potential for proximity effects \cite{Sierra2021:NN}, enabling phenomena such as superconductivity \cite{Moriya2020:PRB,Han2021:APL}, magnetism \cite{Zollner2016:PRB,Zollner2018:NJP,Zollner2019a:PRB,Zollner2020:PRB,Hallal2017:2DM, Zhang2015:PRB,Zhang2018:PRB,Yang2013:PRL,Dyrdal2017:2DM,Song2017:JPD, Haugen2008:PRB, Zhang2015:SR, Su2017:PRB,Gibertini2019:NN,Klein2018:SC,Cardoso2018:PRL,Henriques2020:PRB,Singh2017:PRL,Swartz2012:ACS}, and strong spin-orbit coupling (SOC) \cite{Gmitra2015:PRB,Gmitra2016:PRB,Zollner2019b:PRB,Zollner2021:PSSB,Khokhriakov2020:NC,Karpiak2019:arxiv,Zihlmann2018:PRB,Song2018:NL,Khokhriakov2018:SA,Garcia2018:CSR,Safeer2019:NL,Herlin2020:APL,Khoo2017:NL,Wang2015:NC,Omar2018:PRB,Omar2017:PRB,Fulop2021:arxiv} in materials --- most notably graphene --- lacking them. Magnetism in graphene can be induced by proximity exchange coupling with 
a ferro- or antiferromagnet. Of particular interest are magnetic insulators
(semiconductors) such as Cr$_2$Ge$_2$Te$_6$ \cite{Zollner2019:PRR,Zhang2015:PRB,Karpiak2019:arxiv} (CGT) or CrI$_3$ \cite{Zhang2018:PRB,Farooq2019:NPJ,Cardoso2018:PRL,Seyler2018:NL},
which can modulate the band structure of graphene (or another nonmagnetic material) without significant charge transfer and without contributing additional transport channels. Proximity exchange effects in graphene can be observed by quantum anomalous Hall effect \cite{Wang2015:PRL}, magnetoresistance~\cite{Mendes2015:PRL}, or nonlocal spin transport experiments~\cite{Leutenantsmeyer2016:2DM}. Joined with 
strong SOC in ex-so-tic heterostructures \cite{Zollner2020:PRL} proximity exchange can also induce spin-orbit torque \cite{Zollner2019:PRR,Dolui2020:NL,Alghamdi2019:NL,Wang2019:SA}.

We already know that proximity exchange coupling in graphene can be tuned by gate  \cite{Zollner2016:PRB,Lazic2016:PRB}. Can we also tune it by twisting? A recent study
shows the sensitivity of the spin polarization, magnetic anisotropy, and Dzyaloshinskii-Moriya interaction to the twist angle in graphene/2H-VSeTe heterostructures \cite{Yan2021:PE}. Similarly, tight-binding studies
predict that the strength of proximity SOC in graphene/TMDC heterostructures \cite{David2019:arxiv,Li2019:PRB} can be tuned by the twist angle.
It is then natural to expect that the strength of the proximity exchange could change depending on the twist angle. 

We show here that not only the magnitude, but also the orientation and even the character (ferro- or antiferromagnetic) of the proximity exchange can depend on the twist angle. 
Employing first-principles calculations we study the twist-angle dependence of the proximity exchange coupling in large graphene/CGT supercells.
From the proximitized Dirac band dispersions of graphene, which we fit to a model Hamiltonian, we extract sublattice-resolved exchange parameters, $\lambda_{\textrm{ex}}^\textrm{A}$ and $\lambda_{\textrm{ex}}^\textrm{B}$, for a series of twist angles between 0$^{\circ}$ and 30$^{\circ}$. We find that one can tune the ferromagnetic (uniform) exchange couplings ($\lambda_{\textrm{ex}}^\textrm{A} \approx \lambda_{\textrm{ex}}^\textrm{B}$) from about $4$ to $-4$~meV by twisting the layers. This reversal of the induced spin polarization by the twist angle is surprising when considering that the CGT magnetization orientation is unchanged.

Even more surprising is the emergence of antiferromagnetic (staggered) proximity exchange coupling at 19.1$^{\circ}$, where $\lambda_{\textrm{ex}}^\textrm{A} < 0$ and  $\lambda_{\textrm{ex}}^\textrm{B} > 0$. At 
this twist angle there is a delicate balance in the
orbital hybridization of the spin up and spin down CGT bands with the carbon $p_z$ orbitals, which makes the
exchange coupling highly sensitive to the atomic registry. By laterally shifting the two layers, ferromagnetic couplings can be realized as well. Graphene/CGT stacks thus form a versatile platform for engineering proximity exchange coupling in graphene.

   %-----------------------------------------------------------------------
    \begin{figure}[htb]
     \includegraphics[width=.95\columnwidth]{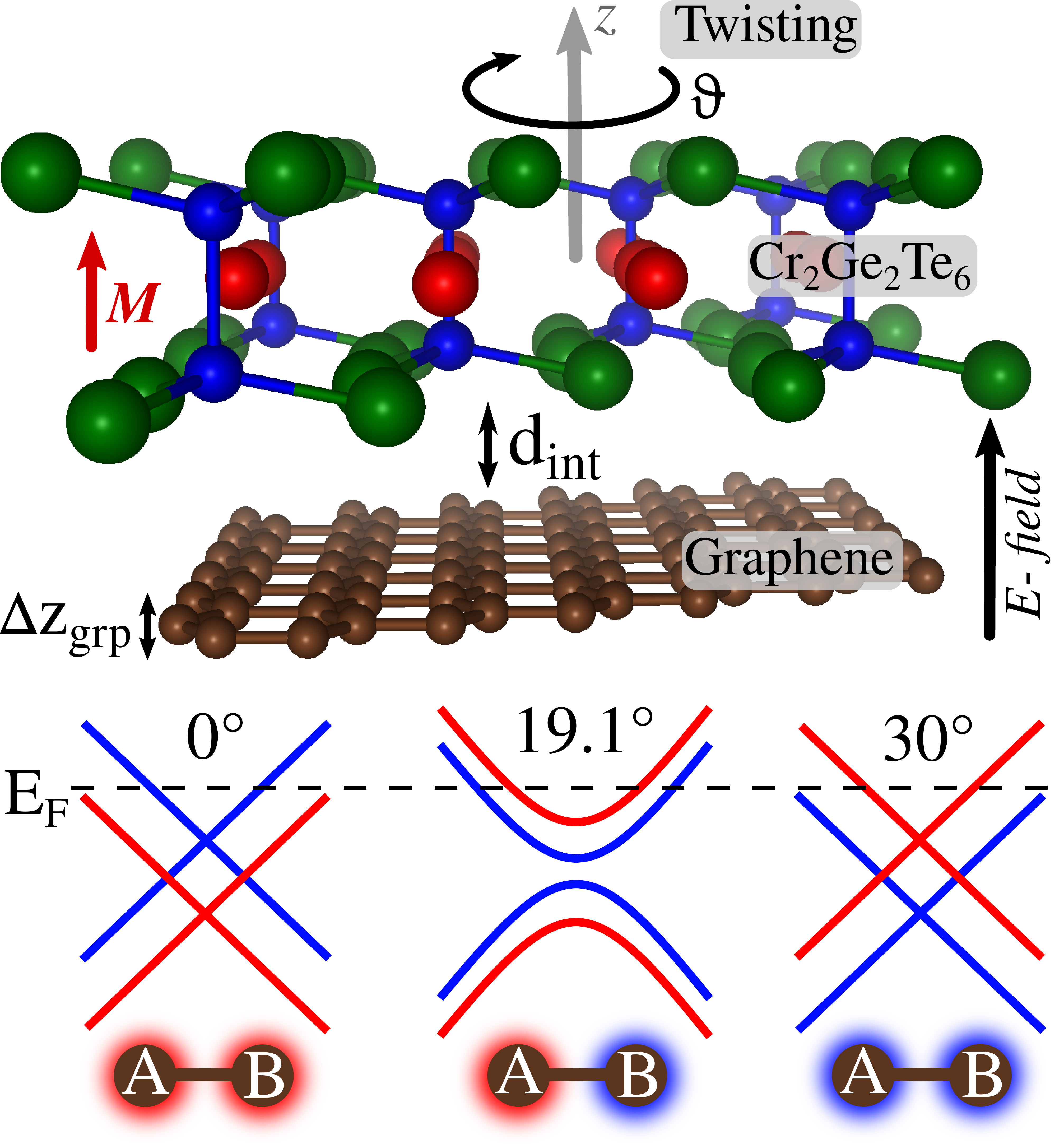}
     \caption{3D view of CGT on graphene, where we define the interlayer distance, $\mathrm{d}_{\mathrm{int}}$, and the rippling of the graphene layer, $\Delta z_{\textrm{grp}}$. We twist CGT by an angle $\vartheta$ around the $z$ axis, with respect to graphene.
     The proximitized Dirac dispersions are sketched for three most relevant twist angles. Red bands are polarized spin up (defined by the CGT magnetization $\bm{M}$ along $z$ direction), while blue bands are polarized spin down.
     The spin polarizations on the graphene lattice, resulting from these Dirac bands at the given Fermi level (dashed line), are also sketched. 
     }\label{Fig1}
    \end{figure}
%------------------------------------------------------------------------

Finally, we also study the influence of strain, interlayer distance, and (transverse) electric field on the doping level, band offsets, and proximity exchange parameters, for different twist angles. We point out the crucial role of both momentum backfolding and interlayer orbital hybridization when tracing the microscopic mechanism for the observed proximity exchange tunability. One important message that our results convey is that the knowledge of the twist angle is crucial when reporting  experiments on magnetic proximity effects: not only the orientation of the induced spin polarization, but also the apparent magnetic ordering (ferro- or antiferromagnetic) need not correspond to the substrate magnetic layer.

 \paragraph{Crystal structures.}  We consider vdW heterostructures of graphene and CGT, with a series of twist angles, ranging from 0$^{\circ}$ to 30$^{\circ}$ in steps of roughly 3$^{\circ}$, between the two monolayers, see Fig.~\ref{Fig1}.
   In order to form commensurate supercells for periodic density functional theory (DFT) calculations, we strain the monolayers in the twisted heterostructures.
    In Table~S1 we summarize the main structural information for the twist angles we consider, see Supplemental Material (SM) \footnotemark[1].
    After relaxation of the heterostructures, we find an average interlayer distance, $d_{\textrm{int}}\approx 3.55$~\AA, and a graphene rippling, $\Delta z_{\textrm{grp}}< 1$~pm, nearly independent of the twist angle.

     \paragraph{Effective low-energy Hamiltonian.} From our first-principles calculations we extract the low-energy band structure of the proximitized graphene. The systems we consider have broken time-reversal
    symmetry with $C_3$ structural symmetry. The following Hamiltonian,
    derived from symmetry~\cite{Kochan2017:PRB,Phong2017:2DM, Zollner2016:PRB}, is able to describe 
    the graphene bands in the vicinity of the Dirac points when proximity exchange is present
    \begin{flalign}
    \label{Eq:Hamiltonian}
    &\mathcal{H} = \mathcal{H}_{0}+\mathcal{H}_{\Delta}+\mathcal{H}_{\textrm{ex}}+\mathrm{E_D},\\
   &\mathcal{H}_{0} = \hbar v_{\textrm{F}}(\tau k_x \sigma_x - k_y \sigma_y)\otimes s_0, \\
    &\mathcal{H}_{\Delta} =\Delta \sigma_z \otimes s_0,\\
    &\mathcal{H}_{\textrm{ex}} =  (-\lambda_{\textrm{ex}}^\textrm{A} \sigma_{+}	+\lambda_{\textrm{ex}}^\textrm{B} \sigma_{-}) \otimes s_z.
    \end{flalign}
    Here $v_{\textrm{F}}$ is the Fermi velocity and the in-plane wave vector 
components $k_x$ and $k_y$ are measured from $\pm$K, 
corresponding to the valley index $\tau = \pm 1$.
The Pauli spin matrices are $s_i$, 
acting on spin space ($\uparrow, \downarrow$), and $\sigma_i$ are pseudospin 
matrices, acting on sublattice space (C$_\textrm{A}$, C$_\textrm{B}$), 
with $i = \{ 0,x,y,z \}$ and $\sigma_{\pm} = \frac{1}{2}(\sigma_z \pm \sigma_0)$. 
The staggered potential gap is $\Delta$ and
    the sublattice-resolved proximity-induced exchange parameters are
    $\lambda_{\textrm{ex}}^\textrm{A}$ and
    $\lambda_{\textrm{ex}}^\textrm{B}$. The four basis states are
    $|\Psi_{\textrm{A}}, \uparrow\rangle$,  $|\Psi_{\textrm{A}},
    \downarrow\rangle$,  $|\Psi_{\textrm{B}}, \uparrow\rangle$,  and
    $|\Psi_{\textrm{B}}, \downarrow\rangle$.  The model
    Hamiltonian is valid close to the Fermi level at zero energy.
    Charge transfer between the monolayers in the DFT calculation is captured by the Dirac point energy, $\mathrm{E_D}$, which adjusts the Dirac point with respect to the Fermi level.

\begin{figure*}[htb]
     \includegraphics[width=.95\textwidth]{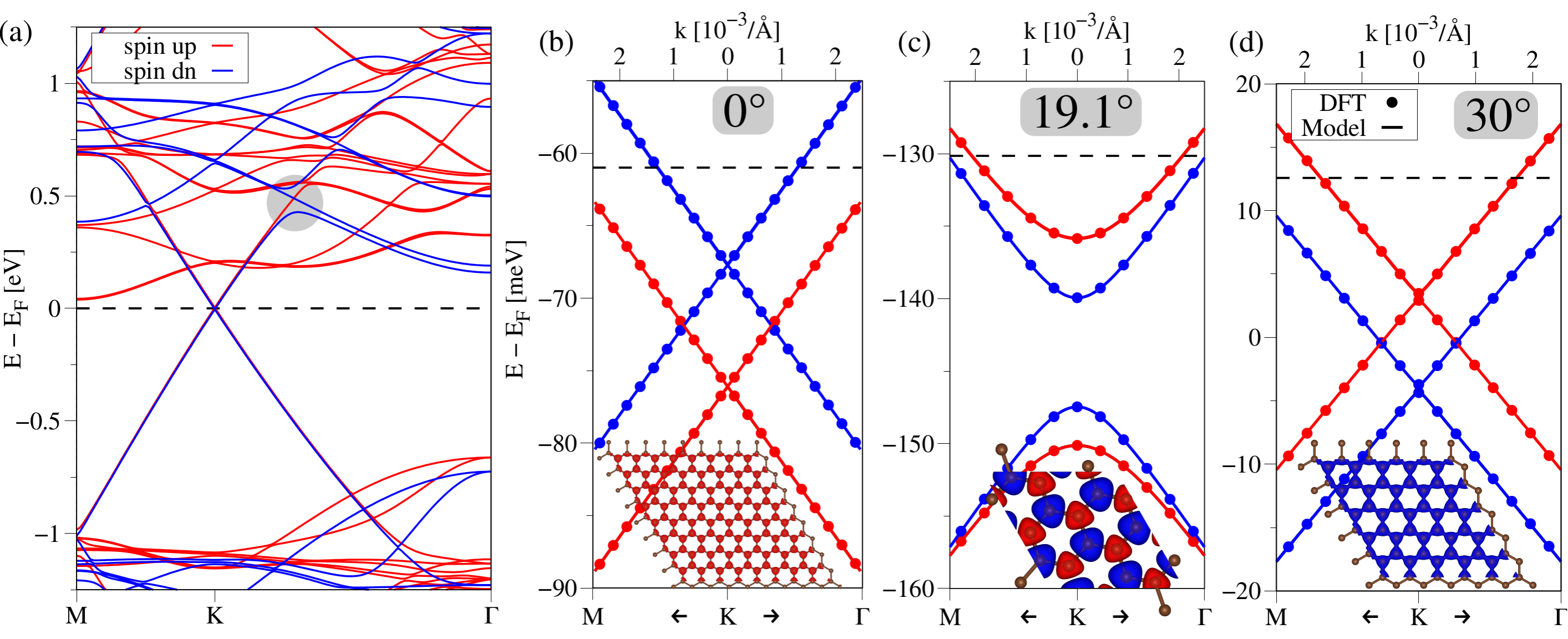}
     \caption{(a) DFT-calculated band structure of the graphene/CGT heterostructure along the high-symmetry path M-K-$\Gamma$ for a twist angle of 30$^{\circ}$. Red (blue) solid  lines correspond to spin up (spin down). Grey disk indicates anticrossing of Dirac and CGT bands. 
     (b) Zoom to the DFT-calculated (symbols) low energy Dirac bands near the K point with a fit to the model Hamiltonian (solid lines) for a twist angle of 0$^{\circ}$. The inset shows the calculated spin polarization on graphene, considering Dirac states in the energy window of about $\pm 2.5$~meV around the indicated Fermi level (dashed line). (c) and (d) are the same as (b) but for 19.1$^{\circ}$ and 30$^{\circ}$.
     }\label{Fig2}
    \end{figure*}

\paragraph{Proximity induced exchange in twisted structures.}
In Fig.~\ref{Fig2}(a), we show the global band structure for the graphene/CGT heterostructure for a twist angle of 30$^{\circ}$; the results for other angles and effects of interlayer charge transfer are summarized in the SM \footnotemark[1].
In agreement with recent calculations \cite{Zollner2019:PRR,Zhang2015:PRB,Karpiak2019:arxiv}, we find the Dirac cone located at the Fermi level and close to the conduction band edge of the CGT.

In Figs.~\ref{Fig2}(b)-(d) we present zooms to the Dirac bands, which exhibit 
proximity exchange splitting, along with the calculated spin polarizations on graphene. For the aligned heterostructure (0$^{\circ}$) the exchange splitting is ferromagnetic, with uniform spin polarization
on A and B sublattices. The fitted exchange parameters are 
$\lambda_{\textrm{ex}}^\textrm{A} \approx \lambda_{\textrm{ex}}^\textrm{B} \approx 4.2$~meV. The Dirac bands look 
similar for the 30$^{\circ}$ twist angle, but the spin polarization on graphene is \textit{reversed}, with the parameter values of $\lambda_{\textrm{ex}}^\textrm{A} \approx \lambda_{\textrm{ex}}^\textrm{B} \approx -3.6$~meV.
This is rather surprising considering that the ferromagnet in both cases has the same magnetization orientation.

%------------------------------------------------------------------------
%------------------------------------------------------------------------

However, the most remarkable case is the 19.1$^{\circ}$ twist angle, shown in 
Fig.~\ref{Fig2}(c). The Dirac band structure does not resemble a ferromagnetic graphene at all. Instead, the spin splittings of the bands are 
compatible with \textit{antiferromagnetic} exchange. Indeed, a fit to the low-energy Hamiltonian, Eq.~\eqref{Eq:Hamiltonian}, yields staggered exchange couplings,
$\lambda_{\textrm{ex}}^\textrm{A} \approx -2.1$~meV and $\lambda_{\textrm{ex}}^\textrm{B} \approx 1.3$~meV. In other words, 
graphene proximitized by a ferromagnetic substrate
can behave as an antiferromagnet, with alternating spin polarization on A and B sublattices.

\begin{figure}[htb]
     \includegraphics[width=.95\columnwidth]{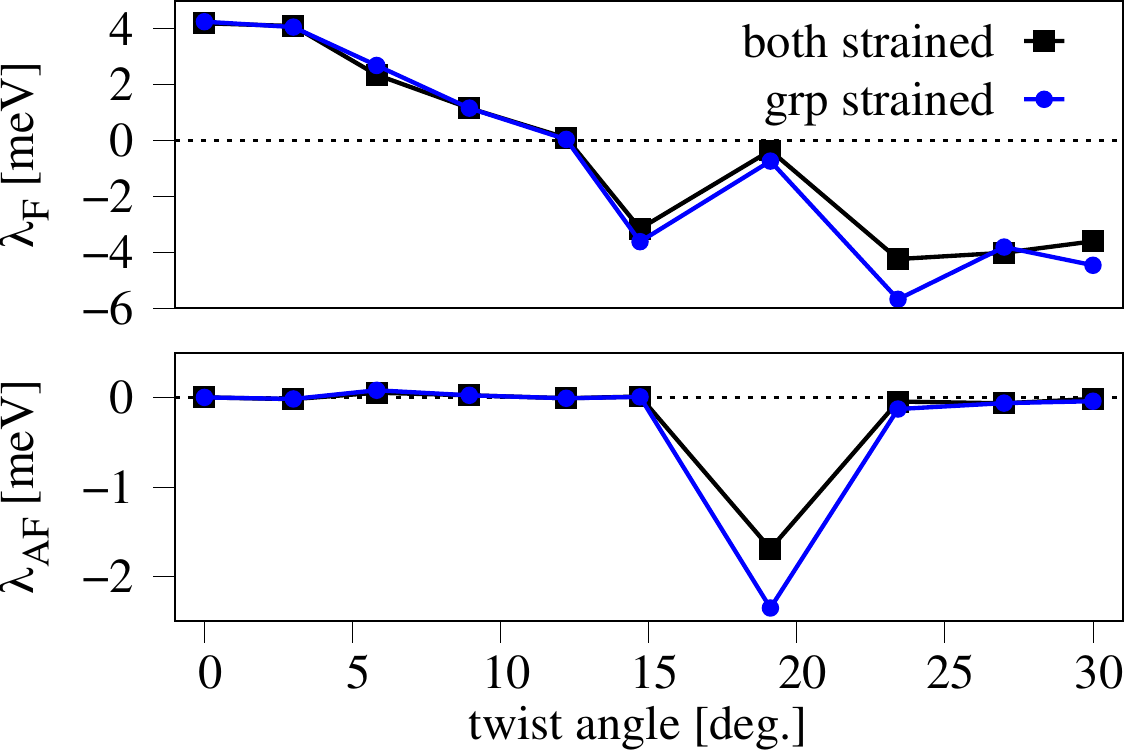} 
     \caption{Calculated twist-angle dependence of the ferromagnetic (top), $\lambda_{\textrm{F}}$, and antiferromagnetic (bottom), $\lambda_{\textrm{AF}}$, proximity exchange coupling of the graphene/CGT bilayers. We summarize the results for heterostructures with both layers strained
     and with only graphene strained.
     }\label{Fig3}
    \end{figure}

To get the full picture of proximity exchange we plot 
in Fig.~\ref{Fig3} the twist-angle dependence of ferromagnetic, \mbox{$\lambda_{\textrm{F}} = (\lambda_{\textrm{ex}}^\textrm{A}+\lambda_{\textrm{ex}}^\textrm{B}$)/2}, and antiferromagnetic, \mbox{$\lambda_{\textrm{AF}} = (\lambda_{\textrm{ex}}^\textrm{A}-\lambda_{\textrm{ex}}^\textrm{B}$)/2}, couplings (listed in Tab.~S2 and Tab.~S4); the magnetization of 
CGT is kept in the same direction for all studied angles.  
We find a rather continuous tunability of the ferromagnetic exchange 
from $4$ to $-4$~meV, when twisting from 0$^{\circ}$ to 30$^{\circ}$. 
Antiferromagnetic exchange emerges only at 19.1$^{\circ}$. Figure~\ref{Fig3}
also shows data for structures, where only graphene is strained (CGT is kept unstrained), to demonstrate the robustness of these findings against strain; we note that
strain controls mainly the band offsets and related charge transfer, see SM \footnotemark[1].

%------------------------------------------------------
%--------------------

\paragraph{Tunability by electric field.}
We now consider the graphene/CGT stacks with different twist angles 
and apply a transverse electric field between $\pm 1.5$~V/nm.
The positive direction of the field is indicated in Fig. \ref{Fig1}. We wish to answer the question: Can one tune the 
proximity-induced exchange coupling by gating? 

\begin{figure}[htb]
     \includegraphics[width=.95\columnwidth]{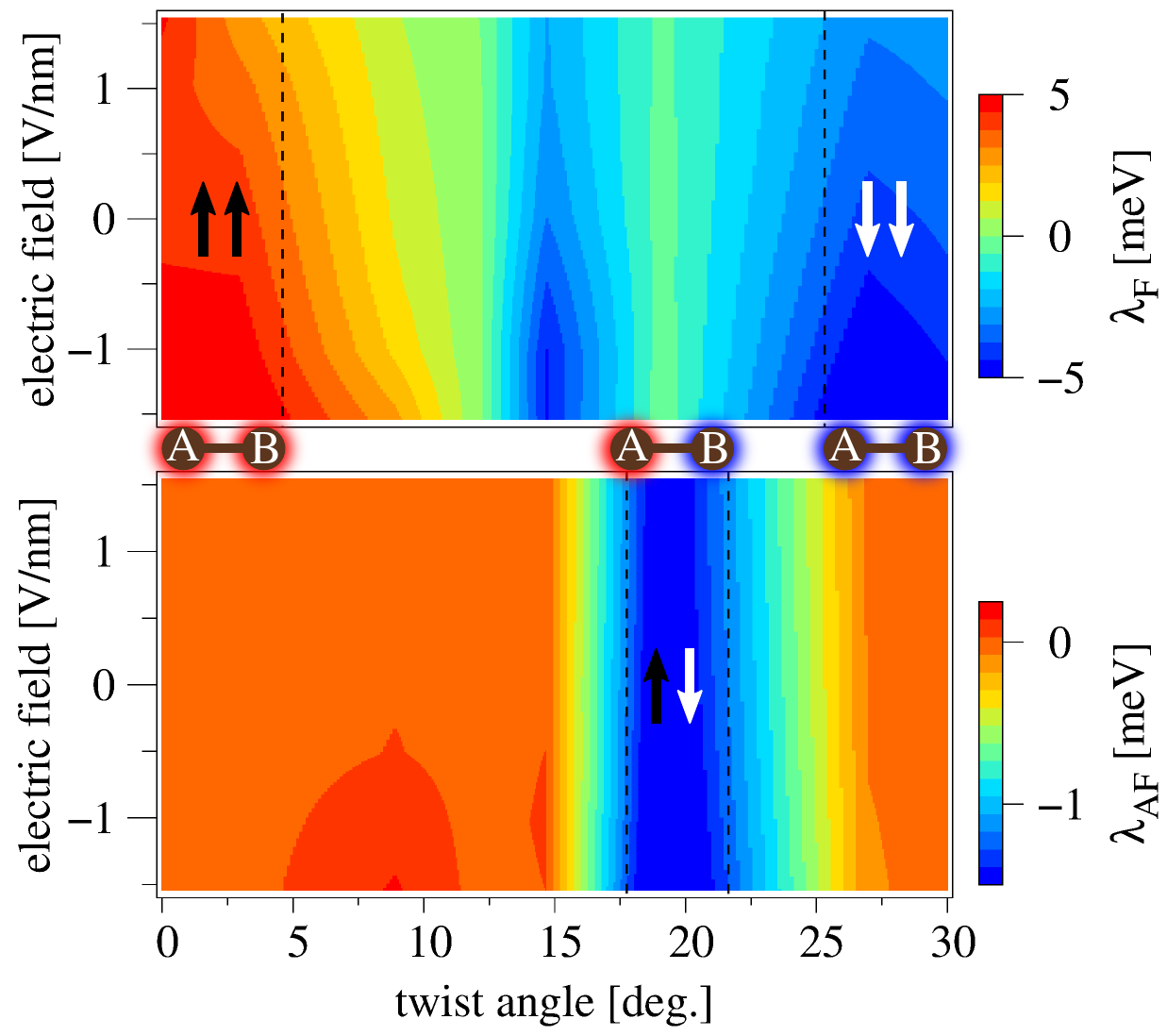}
     \caption{Calculated electric-field and twist-angle dependence of the ferromagnetic (top), $\lambda_{\textrm{F}}$, and antiferromagnetic (bottom), $\lambda_{\textrm{AF}}$, proximity-induced exchange coupling (interpolated from Table~S5). Vertical dashed lines indicate regions of strong ferromagnetic/antiferromagnetic exchange. The spin polarizations on the graphene lattice are sketched (see also Fig.~\ref{Fig1}).
     }\label{Fig4}
    \end{figure}

In Fig.~\ref{Fig4} we summarize the calculated electric-field and twist-angle dependence of the proximity-induced ferromagnetic and antiferromagnetic exchange coupling, as listed in Table~S5. 
While the qualitative picture of spin-polarization reversal at 
30$^{\circ}$ and appearance of antiferromagnetic polarization
at 19.1$^{\circ}$ remains unchanged, the applied electric field
can tune the proximity magnetization rather significantly for some
twist angles. A striking example is the 12.2$^{\circ}$ twist angle: 
Even though the proximity exchange parameters are small, they 
can be tuned from positive to negative by the gate field. 
The antiferromagnetic proximity exchange at 19.1$^{\circ}$ stays,
but is weakly tunable by the field. 
Overall, we find that both gating and twisting are two efficient knobs to tailor the signs and magnitudes of the proximity-induced exchange couplings in graphene/CGT bilayers. We expect similar tunabilities (although at different twist angles) 
for other graphene/ferromagnetic-insulator heterostructures.

\paragraph{Sensitivity to interlayer distance and atomic registry.} 
We find that the interlayer distance strongly influences the proximity exchange, see Table~S6. Tuning $d_{\textrm{int}}$ by $\pm0.1$~\AA, the exchange parameters can be tuned by about $\mp$30\%. Such tunability has recently been measured for the proximity SOC in graphene/WSe$_2$ heterostructures \cite{Fulop2021:arxiv,Fulop2021:arxiv2}.
The atomic registry does not play a role for proximity exchange couplings for $0^{\circ}$ and $30^{\circ}$ twist angles.  
However, as we show in the SM \footnotemark[1], shifting the layers relative to each other, while keeping the twist angle at $19.1^{\circ}$, one can get both staggered and uniform exchange couplings. At this angle
the heterostructure supercell is relatively small (24  atoms), making the 
proximity exchange coupling particularly sensitive to the atomic registry. Further encapsulation of graphene within two CGT layers provides additional
boost and tailoring of proximity exchange, see SM \footnotemark[1].

%------------------------------------------------------
%-------------------------------------------------

\paragraph{Mechanism of twist-angle dependence of proximity exchange.}
The twist-angle dependence of proximity SOC in graphene/TMDC heterostructures has been explained by downfolding the tight-binding model of coupled bilayers \cite{Li2019:PRB,David2019:arxiv}.
The main mechanism there is the tunability of the interlayer interaction connecting the graphene K point with TMDC Bloch states at different $k$ points for different twist angles. Comparison with 
recent large scale DFT calculations \cite{Naimer2021:arXiv,Pezo2021:arXiv}
points to the importance of both spectral variations of the TMDC band structure in the Brillouin zone, but also of the interlayer orbital hybridization.

\begin{figure}[htb]
     \includegraphics[width=.95\columnwidth]{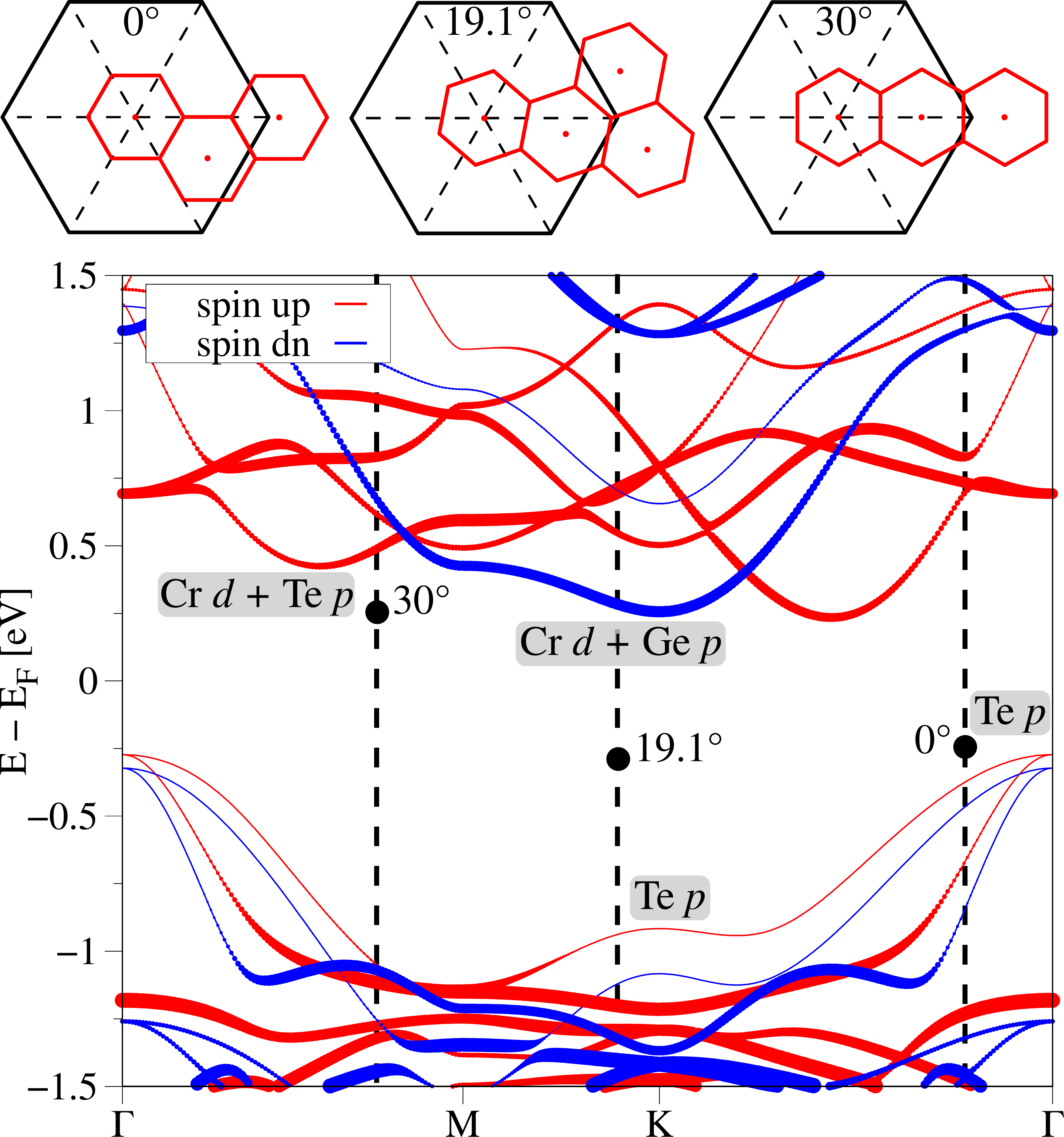} 
     \caption{Top: Backfolding of the graphene Dirac point at K to $k$ points of CGT for different twist angles. The black (red) hexagons represent the graphene (CGT) Brillouin zones. Bottom: The DFT-calculated band structure of monolayer CGT, where the vertical dashed lines indicate the $k$-points, to which the Dirac states couple to, according to the backfolding. The black dots are the locations of the Dirac point for the different twist angles from Table~S4, when CGT is unstrained. 
     We also indicate the main orbital contribution of the bands close to the black dots. The line thicknesses are weighted by the sum of projections onto $z$-extended orbitals (Te $p_z$, Ge $p_z$, and Cr $d_{z^2}$+$d_{xz+yz}$). 
     }\label{Fig:Backfolding}
    \end{figure}

Can we deduce the rather striking reversal of the spin polarization of the Dirac electrons by considering the spectral variations only? The calculated spin-resolved electronic band structure of monolayer CGT, with backfolded graphene K points, is shown in  Fig.~\ref{Fig:Backfolding}.
Second-order perturbation theory predicts level repulsion, so  considering energy bands only, Fig.~\ref{Fig:Backfolding} indicates for 0$^\circ$ that spin up Dirac bands are pushed above spin down bands, and vice versa for 30$^\circ$. This is opposite to what is predicted in Fig. \ref{Fig2}(b) and (d). 

The CGT bands in Fig.~\ref{Fig:Backfolding} are weighted by their $z$-like orbitals content. Those are most likely to overlap with the lobes of graphene's $p_z$ orbitals. There does not appear any discernable pattern here that would predict the DFT calculated behavior in Fig.~\ref{Fig2}. But what Fig.~\ref{Fig:Backfolding} does reveal is that one would need to consider many bands---and both the energies and overlaps with Dirac band $p_z$ orbitals---around the 
CGT gap at the corresponding backfolded K point, to be able to reproduce the DFT results. For example, for 0$^\circ$ the nearest valence bands are formed by Te $p_x+p_y$ orbitals whose overlap with graphene $p_z$ is weak.
We elaborate more on this point in the SM (see Fig.~S19) \footnotemark[1]. 

%------------------------------------------------------
%--------------------

The relevance of high-energy bands for the spin polarization at the Dirac point is revealed by the heterostructure dispersion of, for example, the 30$^\circ$ structure in Fig.~\ref{Fig2}(a). One finds a pronounced anticrossing (grey disk) signalling a particularly strong coupling of spin down carbon $p_z$ orbitals and the lowest CGT spin down conduction band states (formed by Ge $p_z$ and Cr $d$ orbitals). This coupling,  which is nicely seen in the density plots in Fig.~S12 \footnotemark[1], justifying the effective model Hamiltonian,  lowers the spin down more than spin up Dirac states, in agreement with Fig.~\ref{Fig2}(d). Even though the anticrossing is at about 500~meV above the Dirac point, the corresponding high-lying CGT band provides a sizable spin splitting of the Dirac band structure due to the strong coupling. Similar observations hold for 0$^{\circ}$ and 19.1$^{\circ}$ cases, see SM \footnotemark[1]. 

Since the heterostructure unit cells comprise many carbon atoms, it is not obvious that the calculated spin-split Dirac bands, which arise due to couplings to high-lying CGT bands, map to a local proximity magnetization pattern in the graphene layer. The DFT calculated local spin polarizations of the conduction-band electrons in proximitized graphene are plotted as insets in Figs.~\ref{Fig2}(b)-(d). There is a perfect correspondence between the spin-split bands and the local spin polarization pattern---with the emerging pseudospin-resolved polarization--- justifying our effective Hamiltonian, Eq.~\eqref{Eq:Hamiltonian}. 

%------------------------------------------------------------
%\section{Summary}
%------------------------------------------------------------

\paragraph{Conclusions.} Employing DFT on large supercells we show that one can engineer the proximity exchange of Dirac electrons in graphene/CGT stacks by twisting, which should be useful for spin transport experiments --- spin Hanle effect, spin relaxation anisotropy, spin torque --- as well as for realizing topological states \cite{Hogl2020:PRL,Zhang2018:PRB,Vila2021:arxiv} requiring both SOC and (ferro- or antiferromagnetic) exchange in graphene. Our results also stress the importance of documenting the twist angle when employing magnetic vdW heterostructures in experiments.

%------------------------------------------------------------
\acknowledgments
%------------------------------------------------------------

This work was funded by the Deutsche Forschungsgemeinschaft (DFG, German Research Foundation) SFB 1277 (Project No. 314695032), SPP 2244 (Project No. 443416183), and the European Union Horizon 2020 Research and Innovation Program under contract number 881603 (Graphene Flagship).
    
    \footnotetext[1]{See Supplemental Material [url] where we show a more extended summary of
     results, which includes
      Refs.~\cite{ASE,Lazic2015:CPC,Koda2016:JPCC,Carr2020:NRM,Baskin1955:PR,Carteaux1995:JPCM,
    Hohenberg1964:PRB,Giannozzi2009:JPCM,Gong2017:Nat,Kresse1999:PRB,Perdew1996:PRL,Zhang2015:PRB,
     Zollner2019:PRR,Karpiak2019:arxiv,Grimme2006:JCC,Grimme2010:JCP,Barone2009:JCC,Bengtsson1999:PRB,
     Li2014:JMCC,Chen2015:PRB,Liu2021:N,Leutenantsmeyer2016:2DM,Hogl2020:PRL,
     Ghiasi2021:NN,Rosenberger2020:ACS,Weston2020:NN,Fulop2021:arxiv,Fulop2021:arxiv2,Tribhuwan2016:S,
     Li2019:PRB,David2019:arxiv,Zollner2021:arxiv,Zollner2016:PRB} }

\bibliography{references}

\begin{thebibliography}{113}%
\makeatletter
\providecommand \@ifxundefined [1]{%
 \@ifx{#1\undefined}
}%
\providecommand \@ifnum [1]{%
 \ifnum #1\expandafter \@firstoftwo
 \else \expandafter \@secondoftwo
 \fi
}%
\providecommand \@ifx [1]{%
 \ifx #1\expandafter \@firstoftwo
 \else \expandafter \@secondoftwo
 \fi
}%
\providecommand \natexlab [1]{#1}%
\providecommand \enquote  [1]{``#1''}%
\providecommand \bibnamefont  [1]{#1}%
\providecommand \bibfnamefont [1]{#1}%
\providecommand \citenamefont [1]{#1}%
\providecommand \href@noop [0]{\@secondoftwo}%
\providecommand \href [0]{\begingroup \@sanitize@url \@href}%
\providecommand \@href[1]{\@@startlink{#1}\@@href}%
\providecommand \@@href[1]{\endgroup#1\@@endlink}%
\providecommand \@sanitize@url [0]{\catcode `\\12\catcode `\$12\catcode
  `\&12\catcode `\#12\catcode `\^12\catcode `\_12\catcode `\%12\relax}%
\providecommand \@@startlink[1]{}%
\providecommand \@@endlink[0]{}%
\providecommand \url  [0]{\begingroup\@sanitize@url \@url }%
\providecommand \@url [1]{\endgroup\@href {#1}{\urlprefix }}%
\providecommand \urlprefix  [0]{URL }%
\providecommand \Eprint [0]{\href }%
\providecommand \doibase [0]{http://dx.doi.org/}%
\providecommand \selectlanguage [0]{\@gobble}%
\providecommand \bibinfo  [0]{\@secondoftwo}%
\providecommand \bibfield  [0]{\@secondoftwo}%
\providecommand \translation [1]{[#1]}%
\providecommand \BibitemOpen [0]{}%
\providecommand \bibitemStop [0]{}%
\providecommand \bibitemNoStop [0]{.\EOS\space}%
\providecommand \EOS [0]{\spacefactor3000\relax}%
\providecommand \BibitemShut  [1]{\csname bibitem#1\endcsname}%
\let\auto@bib@innerbib\@empty
%</preamble>
\bibitem [{\citenamefont {Carr}\ \emph {et~al.}(2017)\citenamefont {Carr},
  \citenamefont {Massatt}, \citenamefont {Fang}, \citenamefont {Cazeaux},
  \citenamefont {Luskin},\ and\ \citenamefont {Kaxiras}}]{Carr2017:PRB}%
  \BibitemOpen
  \bibfield  {author} {\bibinfo {author} {\bibfnamefont {Stephen}\ \bibnamefont
  {Carr}}, \bibinfo {author} {\bibfnamefont {Daniel}\ \bibnamefont {Massatt}},
  \bibinfo {author} {\bibfnamefont {Shiang}\ \bibnamefont {Fang}}, \bibinfo
  {author} {\bibfnamefont {Paul}\ \bibnamefont {Cazeaux}}, \bibinfo {author}
  {\bibfnamefont {Mitchell}\ \bibnamefont {Luskin}}, \ and\ \bibinfo {author}
  {\bibfnamefont {Efthimios}\ \bibnamefont {Kaxiras}},\ }\bibfield  {title}
  {\enquote {\bibinfo {title} {Twistronics: Manipulating the electronic
  properties of two-dimensional layered structures through their twist
  angle},}\ }\href {\doibase 10.1103/PhysRevB.95.075420} {\bibfield  {journal}
  {\bibinfo  {journal} {Phys. Rev. B}\ }\textbf {\bibinfo {volume} {95}},\
  \bibinfo {pages} {075420} (\bibinfo {year} {2017})}\BibitemShut {NoStop}%
\bibitem [{\citenamefont {Hennighausen}\ and\ \citenamefont
  {Kar}(2021)}]{Hennighausen2021:ES}%
  \BibitemOpen
  \bibfield  {author} {\bibinfo {author} {\bibfnamefont {Zachariah}\
  \bibnamefont {Hennighausen}}\ and\ \bibinfo {author} {\bibfnamefont
  {Swastik}\ \bibnamefont {Kar}},\ }\bibfield  {title} {\enquote {\bibinfo
  {title} {Twistronics: a turning point in 2d quantum materials},}\ }\href
  {\doibase 10.1088/2516-1075/abd957} {\bibfield  {journal} {\bibinfo
  {journal} {Electronic Structure}\ }\textbf {\bibinfo {volume} {3}},\ \bibinfo
  {pages} {014004} (\bibinfo {year} {2021})}\BibitemShut {NoStop}%
\bibitem [{\citenamefont {Ribeiro-Palau}\ \emph {et~al.}(2018)\citenamefont
  {Ribeiro-Palau}, \citenamefont {Zhang}, \citenamefont {Watanabe},
  \citenamefont {Taniguchi}, \citenamefont {Hone},\ and\ \citenamefont
  {Dean}}]{Ribeiro2018:SC}%
  \BibitemOpen
  \bibfield  {author} {\bibinfo {author} {\bibfnamefont {Rebeca}\ \bibnamefont
  {Ribeiro-Palau}}, \bibinfo {author} {\bibfnamefont {Changjian}\ \bibnamefont
  {Zhang}}, \bibinfo {author} {\bibfnamefont {Kenji}\ \bibnamefont {Watanabe}},
  \bibinfo {author} {\bibfnamefont {Takashi}\ \bibnamefont {Taniguchi}},
  \bibinfo {author} {\bibfnamefont {James}\ \bibnamefont {Hone}}, \ and\
  \bibinfo {author} {\bibfnamefont {Cory~R.}\ \bibnamefont {Dean}},\ }\bibfield
   {title} {\enquote {\bibinfo {title} {Twistable electronics with dynamically
  rotatable heterostructures},}\ }\href {\doibase 10.1126/science.aat6981}
  {\bibfield  {journal} {\bibinfo  {journal} {Science}\ }\textbf {\bibinfo
  {volume} {361}},\ \bibinfo {pages} {690--693} (\bibinfo {year}
  {2018})}\BibitemShut {NoStop}%
\bibitem [{\citenamefont {Carr}\ \emph {et~al.}(2020)\citenamefont {Carr},
  \citenamefont {Fang},\ and\ \citenamefont {Kaxiras}}]{Carr2020:NRM}%
  \BibitemOpen
  \bibfield  {author} {\bibinfo {author} {\bibfnamefont {Stephen}\ \bibnamefont
  {Carr}}, \bibinfo {author} {\bibfnamefont {Shiang}\ \bibnamefont {Fang}}, \
  and\ \bibinfo {author} {\bibfnamefont {Efthimios}\ \bibnamefont {Kaxiras}},\
  }\bibfield  {title} {\enquote {\bibinfo {title} {Electronic-structure methods
  for twisted moir{\'e} layers},}\ }\href {\doibase 10.1038/s41578-020-0214-0}
  {\bibfield  {journal} {\bibinfo  {journal} {Nature Reviews Materials}\
  }\textbf {\bibinfo {volume} {5}},\ \bibinfo {pages} {748--763} (\bibinfo
  {year} {2020})}\BibitemShut {NoStop}%
\bibitem [{\citenamefont {Cao}\ \emph {et~al.}(2018{\natexlab{a}})\citenamefont
  {Cao}, \citenamefont {Fatemi}, \citenamefont {Fang}, \citenamefont
  {Watanabe}, \citenamefont {Taniguchi}, \citenamefont {Kaxiras},\ and\
  \citenamefont {Jarillo-Herrero}}]{Cao2018:Nat}%
  \BibitemOpen
  \bibfield  {author} {\bibinfo {author} {\bibfnamefont {Yuan}\ \bibnamefont
  {Cao}}, \bibinfo {author} {\bibfnamefont {Valla}\ \bibnamefont {Fatemi}},
  \bibinfo {author} {\bibfnamefont {Shiang}\ \bibnamefont {Fang}}, \bibinfo
  {author} {\bibfnamefont {Kenji}\ \bibnamefont {Watanabe}}, \bibinfo {author}
  {\bibfnamefont {Takashi}\ \bibnamefont {Taniguchi}}, \bibinfo {author}
  {\bibfnamefont {Efthimios}\ \bibnamefont {Kaxiras}}, \ and\ \bibinfo {author}
  {\bibfnamefont {Pablo}\ \bibnamefont {Jarillo-Herrero}},\ }\bibfield  {title}
  {\enquote {\bibinfo {title} {{Unconventional superconductivity in magic-angle
  graphene superlattices}},}\ }\href {\doibase 10.1038/nature26160} {\bibfield
  {journal} {\bibinfo  {journal} {Nature}\ }\textbf {\bibinfo {volume} {556}},\
  \bibinfo {pages} {43} (\bibinfo {year} {2018}{\natexlab{a}})}\BibitemShut
  {NoStop}%
\bibitem [{\citenamefont {Cao}\ \emph {et~al.}(2018{\natexlab{b}})\citenamefont
  {Cao}, \citenamefont {Fatemi}, \citenamefont {Demir}, \citenamefont {Fang},
  \citenamefont {Tomarken}, \citenamefont {Luo}, \citenamefont
  {Sanchez-Yamagishi}, \citenamefont {Watanabe}, \citenamefont {Taniguchi},
  \citenamefont {Kaxiras}, \citenamefont {Ashoori},\ and\ \citenamefont
  {Jarillo-Herrero}}]{Cao2018a:Nat}%
  \BibitemOpen
  \bibfield  {author} {\bibinfo {author} {\bibfnamefont {Yuan}\ \bibnamefont
  {Cao}}, \bibinfo {author} {\bibfnamefont {Valla}\ \bibnamefont {Fatemi}},
  \bibinfo {author} {\bibfnamefont {Ahmet}\ \bibnamefont {Demir}}, \bibinfo
  {author} {\bibfnamefont {Shiang}\ \bibnamefont {Fang}}, \bibinfo {author}
  {\bibfnamefont {Spencer~L.}\ \bibnamefont {Tomarken}}, \bibinfo {author}
  {\bibfnamefont {Jason~Y.}\ \bibnamefont {Luo}}, \bibinfo {author}
  {\bibfnamefont {Javier~D.}\ \bibnamefont {Sanchez-Yamagishi}}, \bibinfo
  {author} {\bibfnamefont {Kenji}\ \bibnamefont {Watanabe}}, \bibinfo {author}
  {\bibfnamefont {Takashi}\ \bibnamefont {Taniguchi}}, \bibinfo {author}
  {\bibfnamefont {Efthimios}\ \bibnamefont {Kaxiras}}, \bibinfo {author}
  {\bibfnamefont {Ray~C.}\ \bibnamefont {Ashoori}}, \ and\ \bibinfo {author}
  {\bibfnamefont {Pablo}\ \bibnamefont {Jarillo-Herrero}},\ }\bibfield  {title}
  {\enquote {\bibinfo {title} {{Correlated insulator behaviour at half-filling
  in magic-angle graphene superlattices}},}\ }\href {\doibase
  10.1038/nature26154} {\bibfield  {journal} {\bibinfo  {journal} {Nature}\
  }\textbf {\bibinfo {volume} {556}},\ \bibinfo {pages} {80} (\bibinfo {year}
  {2018}{\natexlab{b}})}\BibitemShut {NoStop}%
\bibitem [{\citenamefont {Arora}\ \emph {et~al.}(2020)\citenamefont {Arora},
  \citenamefont {Polski}, \citenamefont {Zhang}, \citenamefont {Thomson},
  \citenamefont {Choi}, \citenamefont {Kim}, \citenamefont {Lin}, \citenamefont
  {Wilson}, \citenamefont {Xu}, \citenamefont {Chu} \emph
  {et~al.}}]{Arora2020:arxiv}%
  \BibitemOpen
  \bibfield  {author} {\bibinfo {author} {\bibfnamefont {Harpreet~Singh}\
  \bibnamefont {Arora}}, \bibinfo {author} {\bibfnamefont {Robert}\
  \bibnamefont {Polski}}, \bibinfo {author} {\bibfnamefont {Yiran}\
  \bibnamefont {Zhang}}, \bibinfo {author} {\bibfnamefont {Alex}\ \bibnamefont
  {Thomson}}, \bibinfo {author} {\bibfnamefont {Youngjoon}\ \bibnamefont
  {Choi}}, \bibinfo {author} {\bibfnamefont {Hyunjin}\ \bibnamefont {Kim}},
  \bibinfo {author} {\bibfnamefont {Zhong}\ \bibnamefont {Lin}}, \bibinfo
  {author} {\bibfnamefont {Ilham~Zaky}\ \bibnamefont {Wilson}}, \bibinfo
  {author} {\bibfnamefont {Xiaodong}\ \bibnamefont {Xu}}, \bibinfo {author}
  {\bibfnamefont {Jiun-Haw}\ \bibnamefont {Chu}},  \emph {et~al.},\ }\bibfield
  {title} {\enquote {\bibinfo {title} {Superconductivity in metallic twisted
  bilayer graphene stabilized by wse 2},}\ }\href {\doibase
  https://doi.org/10.1038/s41586-020-2473-8} {\bibfield  {journal} {\bibinfo
  {journal} {Nature}\ }\textbf {\bibinfo {volume} {583}},\ \bibinfo {pages}
  {379--384} (\bibinfo {year} {2020})}\BibitemShut {NoStop}%
\bibitem [{\citenamefont {Stepanov}\ \emph {et~al.}(2020)\citenamefont
  {Stepanov}, \citenamefont {Das}, \citenamefont {Lu}, \citenamefont
  {Fahimniya}, \citenamefont {Watanabe}, \citenamefont {Taniguchi},
  \citenamefont {Koppens}, \citenamefont {Lischner}, \citenamefont {Levitov},\
  and\ \citenamefont {Efetov}}]{Stepanov2020:Nat}%
  \BibitemOpen
  \bibfield  {author} {\bibinfo {author} {\bibfnamefont {Petr}\ \bibnamefont
  {Stepanov}}, \bibinfo {author} {\bibfnamefont {Ipsita}\ \bibnamefont {Das}},
  \bibinfo {author} {\bibfnamefont {Xiaobo}\ \bibnamefont {Lu}}, \bibinfo
  {author} {\bibfnamefont {Ali}\ \bibnamefont {Fahimniya}}, \bibinfo {author}
  {\bibfnamefont {Kenji}\ \bibnamefont {Watanabe}}, \bibinfo {author}
  {\bibfnamefont {Takashi}\ \bibnamefont {Taniguchi}}, \bibinfo {author}
  {\bibfnamefont {Frank~HL}\ \bibnamefont {Koppens}}, \bibinfo {author}
  {\bibfnamefont {Johannes}\ \bibnamefont {Lischner}}, \bibinfo {author}
  {\bibfnamefont {Leonid}\ \bibnamefont {Levitov}}, \ and\ \bibinfo {author}
  {\bibfnamefont {Dmitri~K}\ \bibnamefont {Efetov}},\ }\bibfield  {title}
  {\enquote {\bibinfo {title} {Untying the insulating and superconducting
  orders in magic-angle graphene},}\ }\href {\doibase
  https://doi.org/10.1038/s41586-020-2459-6} {\bibfield  {journal} {\bibinfo
  {journal} {Nature}\ }\textbf {\bibinfo {volume} {583}},\ \bibinfo {pages}
  {375--378} (\bibinfo {year} {2020})}\BibitemShut {NoStop}%
\bibitem [{\citenamefont {Lu}\ \emph {et~al.}(2019)\citenamefont {Lu},
  \citenamefont {Stepanov}, \citenamefont {Yang}, \citenamefont {Xie},
  \citenamefont {Aamir}, \citenamefont {Das}, \citenamefont {Urgell},
  \citenamefont {Watanabe}, \citenamefont {Taniguchi}, \citenamefont {Zhang}
  \emph {et~al.}}]{Lu2019:Nat}%
  \BibitemOpen
  \bibfield  {author} {\bibinfo {author} {\bibfnamefont {Xiaobo}\ \bibnamefont
  {Lu}}, \bibinfo {author} {\bibfnamefont {Petr}\ \bibnamefont {Stepanov}},
  \bibinfo {author} {\bibfnamefont {Wei}\ \bibnamefont {Yang}}, \bibinfo
  {author} {\bibfnamefont {Ming}\ \bibnamefont {Xie}}, \bibinfo {author}
  {\bibfnamefont {Mohammed~Ali}\ \bibnamefont {Aamir}}, \bibinfo {author}
  {\bibfnamefont {Ipsita}\ \bibnamefont {Das}}, \bibinfo {author}
  {\bibfnamefont {Carles}\ \bibnamefont {Urgell}}, \bibinfo {author}
  {\bibfnamefont {Kenji}\ \bibnamefont {Watanabe}}, \bibinfo {author}
  {\bibfnamefont {Takashi}\ \bibnamefont {Taniguchi}}, \bibinfo {author}
  {\bibfnamefont {Guangyu}\ \bibnamefont {Zhang}},  \emph {et~al.},\ }\bibfield
   {title} {\enquote {\bibinfo {title} {Superconductors, orbital magnets and
  correlated states in magic-angle bilayer graphene},}\ }\href {\doibase
  https://doi.org/10.1038/s41586-019-1695-0} {\bibfield  {journal} {\bibinfo
  {journal} {Nature}\ }\textbf {\bibinfo {volume} {574}},\ \bibinfo {pages}
  {653--657} (\bibinfo {year} {2019})}\BibitemShut {NoStop}%
\bibitem [{\citenamefont {Sharpe}\ \emph {et~al.}(2019)\citenamefont {Sharpe},
  \citenamefont {Fox}, \citenamefont {Barnard}, \citenamefont {Finney},
  \citenamefont {Watanabe}, \citenamefont {Taniguchi}, \citenamefont
  {Kastner},\ and\ \citenamefont {Goldhaber-Gordon}}]{Sharpe2019:SC}%
  \BibitemOpen
  \bibfield  {author} {\bibinfo {author} {\bibfnamefont {Aaron~L.}\
  \bibnamefont {Sharpe}}, \bibinfo {author} {\bibfnamefont {Eli~J.}\
  \bibnamefont {Fox}}, \bibinfo {author} {\bibfnamefont {Arthur~W.}\
  \bibnamefont {Barnard}}, \bibinfo {author} {\bibfnamefont {Joe}\ \bibnamefont
  {Finney}}, \bibinfo {author} {\bibfnamefont {Kenji}\ \bibnamefont
  {Watanabe}}, \bibinfo {author} {\bibfnamefont {Takashi}\ \bibnamefont
  {Taniguchi}}, \bibinfo {author} {\bibfnamefont {M.~A.}\ \bibnamefont
  {Kastner}}, \ and\ \bibinfo {author} {\bibfnamefont {David}\ \bibnamefont
  {Goldhaber-Gordon}},\ }\bibfield  {title} {\enquote {\bibinfo {title}
  {Emergent ferromagnetism near three-quarters filling in twisted bilayer
  graphene},}\ }\href {\doibase 10.1126/science.aaw3780} {\bibfield  {journal}
  {\bibinfo  {journal} {Science}\ }\textbf {\bibinfo {volume} {365}},\ \bibinfo
  {pages} {605--608} (\bibinfo {year} {2019})}\BibitemShut {NoStop}%
\bibitem [{\citenamefont {Saito}\ \emph {et~al.}(2021)\citenamefont {Saito},
  \citenamefont {Yang}, \citenamefont {Ge}, \citenamefont {Liu}, \citenamefont
  {Taniguchi}, \citenamefont {Watanabe}, \citenamefont {Li}, \citenamefont
  {Berg},\ and\ \citenamefont {Young}}]{Saito2021:Nat}%
  \BibitemOpen
  \bibfield  {author} {\bibinfo {author} {\bibfnamefont {Yu}~\bibnamefont
  {Saito}}, \bibinfo {author} {\bibfnamefont {Fangyuan}\ \bibnamefont {Yang}},
  \bibinfo {author} {\bibfnamefont {Jingyuan}\ \bibnamefont {Ge}}, \bibinfo
  {author} {\bibfnamefont {Xiaoxue}\ \bibnamefont {Liu}}, \bibinfo {author}
  {\bibfnamefont {Takashi}\ \bibnamefont {Taniguchi}}, \bibinfo {author}
  {\bibfnamefont {Kenji}\ \bibnamefont {Watanabe}}, \bibinfo {author}
  {\bibfnamefont {JIA}\ \bibnamefont {Li}}, \bibinfo {author} {\bibfnamefont
  {Erez}\ \bibnamefont {Berg}}, \ and\ \bibinfo {author} {\bibfnamefont
  {Andrea~F}\ \bibnamefont {Young}},\ }\bibfield  {title} {\enquote {\bibinfo
  {title} {Isospin pomeranchuk effect in twisted bilayer graphene},}\ }\href
  {\doibase https://doi.org/10.1038/s41586-021-03409-2} {\bibfield  {journal}
  {\bibinfo  {journal} {Nature}\ }\textbf {\bibinfo {volume} {592}},\ \bibinfo
  {pages} {220--224} (\bibinfo {year} {2021})}\BibitemShut {NoStop}%
\bibitem [{\citenamefont {Serlin}\ \emph {et~al.}(2020)\citenamefont {Serlin},
  \citenamefont {Tschirhart}, \citenamefont {Polshyn}, \citenamefont {Zhang},
  \citenamefont {Zhu}, \citenamefont {Watanabe}, \citenamefont {Taniguchi},
  \citenamefont {Balents},\ and\ \citenamefont {Young}}]{Serlin2020:S}%
  \BibitemOpen
  \bibfield  {author} {\bibinfo {author} {\bibfnamefont {M.}~\bibnamefont
  {Serlin}}, \bibinfo {author} {\bibfnamefont {C.~L.}\ \bibnamefont
  {Tschirhart}}, \bibinfo {author} {\bibfnamefont {H.}~\bibnamefont {Polshyn}},
  \bibinfo {author} {\bibfnamefont {Y.}~\bibnamefont {Zhang}}, \bibinfo
  {author} {\bibfnamefont {J.}~\bibnamefont {Zhu}}, \bibinfo {author}
  {\bibfnamefont {K.}~\bibnamefont {Watanabe}}, \bibinfo {author}
  {\bibfnamefont {T.}~\bibnamefont {Taniguchi}}, \bibinfo {author}
  {\bibfnamefont {L.}~\bibnamefont {Balents}}, \ and\ \bibinfo {author}
  {\bibfnamefont {A.~F.}\ \bibnamefont {Young}},\ }\bibfield  {title} {\enquote
  {\bibinfo {title} {Intrinsic quantized anomalous hall effect in a moire
  heterostructure},}\ }\href {\doibase 10.1126/science.aay5533} {\bibfield
  {journal} {\bibinfo  {journal} {Science}\ }\textbf {\bibinfo {volume}
  {367}},\ \bibinfo {pages} {900--903} (\bibinfo {year} {2020})}\BibitemShut
  {NoStop}%
\bibitem [{\citenamefont {Nimbalkar}\ and\ \citenamefont
  {Kim}(2020)}]{Nimbalkar2020:NML}%
  \BibitemOpen
  \bibfield  {author} {\bibinfo {author} {\bibfnamefont {Amol}\ \bibnamefont
  {Nimbalkar}}\ and\ \bibinfo {author} {\bibfnamefont {Hyunmin}\ \bibnamefont
  {Kim}},\ }\bibfield  {title} {\enquote {\bibinfo {title} {Opportunities and
  challenges in twisted bilayer graphene: a review},}\ }\href {\doibase
  https://doi.org/10.1007/s40820-020-00464-8} {\bibfield  {journal} {\bibinfo
  {journal} {Nano-Micro Letters}\ }\textbf {\bibinfo {volume} {12}},\ \bibinfo
  {pages} {126} (\bibinfo {year} {2020})}\BibitemShut {NoStop}%
\bibitem [{\citenamefont {Bultinck}\ \emph {et~al.}(2020)\citenamefont
  {Bultinck}, \citenamefont {Chatterjee},\ and\ \citenamefont
  {Zaletel}}]{Bultinck2020:PRL}%
  \BibitemOpen
  \bibfield  {author} {\bibinfo {author} {\bibfnamefont {Nick}\ \bibnamefont
  {Bultinck}}, \bibinfo {author} {\bibfnamefont {Shubhayu}\ \bibnamefont
  {Chatterjee}}, \ and\ \bibinfo {author} {\bibfnamefont {Michael~P.}\
  \bibnamefont {Zaletel}},\ }\bibfield  {title} {\enquote {\bibinfo {title}
  {Mechanism for anomalous hall ferromagnetism in twisted bilayer graphene},}\
  }\href {\doibase 10.1103/PhysRevLett.124.166601} {\bibfield  {journal}
  {\bibinfo  {journal} {Phys. Rev. Lett.}\ }\textbf {\bibinfo {volume} {124}},\
  \bibinfo {pages} {166601} (\bibinfo {year} {2020})}\BibitemShut {NoStop}%
\bibitem [{\citenamefont {Repellin}\ \emph {et~al.}(2020)\citenamefont
  {Repellin}, \citenamefont {Dong}, \citenamefont {Zhang},\ and\ \citenamefont
  {Senthil}}]{Repellin2020:PRL}%
  \BibitemOpen
  \bibfield  {author} {\bibinfo {author} {\bibfnamefont {C\'ecile}\
  \bibnamefont {Repellin}}, \bibinfo {author} {\bibfnamefont {Zhihuan}\
  \bibnamefont {Dong}}, \bibinfo {author} {\bibfnamefont {Ya-Hui}\ \bibnamefont
  {Zhang}}, \ and\ \bibinfo {author} {\bibfnamefont {T.}~\bibnamefont
  {Senthil}},\ }\bibfield  {title} {\enquote {\bibinfo {title} {Ferromagnetism
  in narrow bands of moir\'e superlattices},}\ }\href {\doibase
  10.1103/PhysRevLett.124.187601} {\bibfield  {journal} {\bibinfo  {journal}
  {Phys. Rev. Lett.}\ }\textbf {\bibinfo {volume} {124}},\ \bibinfo {pages}
  {187601} (\bibinfo {year} {2020})}\BibitemShut {NoStop}%
\bibitem [{\citenamefont {Choi}\ \emph {et~al.}(2019)\citenamefont {Choi},
  \citenamefont {Kemmer}, \citenamefont {Peng}, \citenamefont {Thomson},
  \citenamefont {Arora}, \citenamefont {Polski}, \citenamefont {Zhang},
  \citenamefont {Ren}, \citenamefont {Alicea}, \citenamefont {Refael} \emph
  {et~al.}}]{Choi2019:NP}%
  \BibitemOpen
  \bibfield  {author} {\bibinfo {author} {\bibfnamefont {Youngjoon}\
  \bibnamefont {Choi}}, \bibinfo {author} {\bibfnamefont {Jeannette}\
  \bibnamefont {Kemmer}}, \bibinfo {author} {\bibfnamefont {Yang}\ \bibnamefont
  {Peng}}, \bibinfo {author} {\bibfnamefont {Alex}\ \bibnamefont {Thomson}},
  \bibinfo {author} {\bibfnamefont {Harpreet}\ \bibnamefont {Arora}}, \bibinfo
  {author} {\bibfnamefont {Robert}\ \bibnamefont {Polski}}, \bibinfo {author}
  {\bibfnamefont {Yiran}\ \bibnamefont {Zhang}}, \bibinfo {author}
  {\bibfnamefont {Hechen}\ \bibnamefont {Ren}}, \bibinfo {author}
  {\bibfnamefont {Jason}\ \bibnamefont {Alicea}}, \bibinfo {author}
  {\bibfnamefont {Gil}\ \bibnamefont {Refael}},  \emph {et~al.},\ }\bibfield
  {title} {\enquote {\bibinfo {title} {Electronic correlations in twisted
  bilayer graphene near the magic angle},}\ }\href {\doibase
  https://doi.org/10.1038/s41567-019-0606-5} {\bibfield  {journal} {\bibinfo
  {journal} {Nature Physics}\ }\textbf {\bibinfo {volume} {15}},\ \bibinfo
  {pages} {1174--1180} (\bibinfo {year} {2019})}\BibitemShut {NoStop}%
\bibitem [{\citenamefont {Lisi}\ \emph {et~al.}(2021)\citenamefont {Lisi},
  \citenamefont {Lu}, \citenamefont {Benschop}, \citenamefont {de~Jong},
  \citenamefont {Stepanov}, \citenamefont {Duran}, \citenamefont {Margot},
  \citenamefont {Cucchi}, \citenamefont {Cappelli}, \citenamefont {Hunter}
  \emph {et~al.}}]{Lisi2021:NP}%
  \BibitemOpen
  \bibfield  {author} {\bibinfo {author} {\bibfnamefont {Simone}\ \bibnamefont
  {Lisi}}, \bibinfo {author} {\bibfnamefont {Xiaobo}\ \bibnamefont {Lu}},
  \bibinfo {author} {\bibfnamefont {Tjerk}\ \bibnamefont {Benschop}}, \bibinfo
  {author} {\bibfnamefont {Tobias~A}\ \bibnamefont {de~Jong}}, \bibinfo
  {author} {\bibfnamefont {Petr}\ \bibnamefont {Stepanov}}, \bibinfo {author}
  {\bibfnamefont {Jose~R}\ \bibnamefont {Duran}}, \bibinfo {author}
  {\bibfnamefont {Florian}\ \bibnamefont {Margot}}, \bibinfo {author}
  {\bibfnamefont {Ir{\`e}ne}\ \bibnamefont {Cucchi}}, \bibinfo {author}
  {\bibfnamefont {Edoardo}\ \bibnamefont {Cappelli}}, \bibinfo {author}
  {\bibfnamefont {Andrew}\ \bibnamefont {Hunter}},  \emph {et~al.},\ }\bibfield
   {title} {\enquote {\bibinfo {title} {Observation of flat bands in twisted
  bilayer graphene},}\ }\href {\doibase
  https://doi.org/10.1038/s41567-020-01041-x} {\bibfield  {journal} {\bibinfo
  {journal} {Nature Physics}\ }\textbf {\bibinfo {volume} {17}},\ \bibinfo
  {pages} {189--193} (\bibinfo {year} {2021})}\BibitemShut {NoStop}%
\bibitem [{\citenamefont {Balents}\ \emph {et~al.}(2020)\citenamefont
  {Balents}, \citenamefont {Dean}, \citenamefont {Efetov},\ and\ \citenamefont
  {Young}}]{Balents2020:NP}%
  \BibitemOpen
  \bibfield  {author} {\bibinfo {author} {\bibfnamefont {Leon}\ \bibnamefont
  {Balents}}, \bibinfo {author} {\bibfnamefont {Cory~R}\ \bibnamefont {Dean}},
  \bibinfo {author} {\bibfnamefont {Dmitri~K}\ \bibnamefont {Efetov}}, \ and\
  \bibinfo {author} {\bibfnamefont {Andrea~F}\ \bibnamefont {Young}},\
  }\bibfield  {title} {\enquote {\bibinfo {title} {Superconductivity and strong
  correlations in moir{\'e} flat bands},}\ }\href {\doibase
  https://doi.org/10.1038/s41567-020-0906-9} {\bibfield  {journal} {\bibinfo
  {journal} {Nature Physics}\ }\textbf {\bibinfo {volume} {16}},\ \bibinfo
  {pages} {725--733} (\bibinfo {year} {2020})}\BibitemShut {NoStop}%
\bibitem [{\citenamefont {Wolf}\ \emph {et~al.}(2019)\citenamefont {Wolf},
  \citenamefont {Lado}, \citenamefont {Blatter},\ and\ \citenamefont
  {Zilberberg}}]{Wolf2019:PRL}%
  \BibitemOpen
  \bibfield  {author} {\bibinfo {author} {\bibfnamefont {T.~M.~R.}\
  \bibnamefont {Wolf}}, \bibinfo {author} {\bibfnamefont {J.~L.}\ \bibnamefont
  {Lado}}, \bibinfo {author} {\bibfnamefont {G.}~\bibnamefont {Blatter}}, \
  and\ \bibinfo {author} {\bibfnamefont {O.}~\bibnamefont {Zilberberg}},\
  }\bibfield  {title} {\enquote {\bibinfo {title} {Electrically tunable flat
  bands and magnetism in twisted bilayer graphene},}\ }\href {\doibase
  10.1103/PhysRevLett.123.096802} {\bibfield  {journal} {\bibinfo  {journal}
  {Phys. Rev. Lett.}\ }\textbf {\bibinfo {volume} {123}},\ \bibinfo {pages}
  {096802} (\bibinfo {year} {2019})}\BibitemShut {NoStop}%
\bibitem [{\citenamefont {Zhu}\ \emph {et~al.}(2020)\citenamefont {Zhu},
  \citenamefont {Carr}, \citenamefont {Massatt}, \citenamefont {Luskin},\ and\
  \citenamefont {Kaxiras}}]{Zhu2020:PRL}%
  \BibitemOpen
  \bibfield  {author} {\bibinfo {author} {\bibfnamefont {Ziyan}\ \bibnamefont
  {Zhu}}, \bibinfo {author} {\bibfnamefont {Stephen}\ \bibnamefont {Carr}},
  \bibinfo {author} {\bibfnamefont {Daniel}\ \bibnamefont {Massatt}}, \bibinfo
  {author} {\bibfnamefont {Mitchell}\ \bibnamefont {Luskin}}, \ and\ \bibinfo
  {author} {\bibfnamefont {Efthimios}\ \bibnamefont {Kaxiras}},\ }\bibfield
  {title} {\enquote {\bibinfo {title} {Twisted trilayer graphene: A precisely
  tunable platform for correlated electrons},}\ }\href {\doibase
  10.1103/PhysRevLett.125.116404} {\bibfield  {journal} {\bibinfo  {journal}
  {Phys. Rev. Lett.}\ }\textbf {\bibinfo {volume} {125}},\ \bibinfo {pages}
  {116404} (\bibinfo {year} {2020})}\BibitemShut {NoStop}%
\bibitem [{\citenamefont {Park}\ \emph {et~al.}(2021)\citenamefont {Park},
  \citenamefont {Cao}, \citenamefont {Watanabe}, \citenamefont {Taniguchi},\
  and\ \citenamefont {Jarillo-Herrero}}]{Park2021:Nat}%
  \BibitemOpen
  \bibfield  {author} {\bibinfo {author} {\bibfnamefont {Jeong~Min}\
  \bibnamefont {Park}}, \bibinfo {author} {\bibfnamefont {Yuan}\ \bibnamefont
  {Cao}}, \bibinfo {author} {\bibfnamefont {Kenji}\ \bibnamefont {Watanabe}},
  \bibinfo {author} {\bibfnamefont {Takashi}\ \bibnamefont {Taniguchi}}, \ and\
  \bibinfo {author} {\bibfnamefont {Pablo}\ \bibnamefont {Jarillo-Herrero}},\
  }\bibfield  {title} {\enquote {\bibinfo {title} {Tunable strongly coupled
  superconductivity in magic-angle twisted trilayer graphene},}\ }\href
  {\doibase https://doi.org/10.1038/s41586-021-03192-0} {\bibfield  {journal}
  {\bibinfo  {journal} {Nature}\ }\textbf {\bibinfo {volume} {590}},\ \bibinfo
  {pages} {249--255} (\bibinfo {year} {2021})}\BibitemShut {NoStop}%
\bibitem [{\citenamefont {Chen}\ \emph
  {et~al.}(2019{\natexlab{a}})\citenamefont {Chen}, \citenamefont {Jiang},
  \citenamefont {Wu}, \citenamefont {Lyu}, \citenamefont {Li}, \citenamefont
  {Chittari}, \citenamefont {Watanabe}, \citenamefont {Taniguchi},
  \citenamefont {Shi}, \citenamefont {Jung} \emph {et~al.}}]{Chen2019:NP}%
  \BibitemOpen
  \bibfield  {author} {\bibinfo {author} {\bibfnamefont {Guorui}\ \bibnamefont
  {Chen}}, \bibinfo {author} {\bibfnamefont {Lili}\ \bibnamefont {Jiang}},
  \bibinfo {author} {\bibfnamefont {Shuang}\ \bibnamefont {Wu}}, \bibinfo
  {author} {\bibfnamefont {Bosai}\ \bibnamefont {Lyu}}, \bibinfo {author}
  {\bibfnamefont {Hongyuan}\ \bibnamefont {Li}}, \bibinfo {author}
  {\bibfnamefont {Bheema~Lingam}\ \bibnamefont {Chittari}}, \bibinfo {author}
  {\bibfnamefont {Kenji}\ \bibnamefont {Watanabe}}, \bibinfo {author}
  {\bibfnamefont {Takashi}\ \bibnamefont {Taniguchi}}, \bibinfo {author}
  {\bibfnamefont {Zhiwen}\ \bibnamefont {Shi}}, \bibinfo {author}
  {\bibfnamefont {Jeil}\ \bibnamefont {Jung}},  \emph {et~al.},\ }\bibfield
  {title} {\enquote {\bibinfo {title} {Evidence of a gate-tunable mott
  insulator in a trilayer graphene moir{\'e} superlattice},}\ }\href {\doibase
  https://doi.org/10.1038/s41567-018-0387-2} {\bibfield  {journal} {\bibinfo
  {journal} {Nature Physics}\ }\textbf {\bibinfo {volume} {15}},\ \bibinfo
  {pages} {237--241} (\bibinfo {year} {2019}{\natexlab{a}})}\BibitemShut
  {NoStop}%
\bibitem [{\citenamefont {Chen}\ \emph {et~al.}(2020)\citenamefont {Chen},
  \citenamefont {Sharpe}, \citenamefont {Fox}, \citenamefont {Zhang},
  \citenamefont {Wang}, \citenamefont {Jiang}, \citenamefont {Lyu},
  \citenamefont {Li}, \citenamefont {Watanabe}, \citenamefont {Taniguchi} \emph
  {et~al.}}]{Chen2020:Nat}%
  \BibitemOpen
  \bibfield  {author} {\bibinfo {author} {\bibfnamefont {Guorui}\ \bibnamefont
  {Chen}}, \bibinfo {author} {\bibfnamefont {Aaron~L}\ \bibnamefont {Sharpe}},
  \bibinfo {author} {\bibfnamefont {Eli~J}\ \bibnamefont {Fox}}, \bibinfo
  {author} {\bibfnamefont {Ya-Hui}\ \bibnamefont {Zhang}}, \bibinfo {author}
  {\bibfnamefont {Shaoxin}\ \bibnamefont {Wang}}, \bibinfo {author}
  {\bibfnamefont {Lili}\ \bibnamefont {Jiang}}, \bibinfo {author}
  {\bibfnamefont {Bosai}\ \bibnamefont {Lyu}}, \bibinfo {author} {\bibfnamefont
  {Hongyuan}\ \bibnamefont {Li}}, \bibinfo {author} {\bibfnamefont {Kenji}\
  \bibnamefont {Watanabe}}, \bibinfo {author} {\bibfnamefont {Takashi}\
  \bibnamefont {Taniguchi}},  \emph {et~al.},\ }\bibfield  {title} {\enquote
  {\bibinfo {title} {Tunable correlated chern insulator and ferromagnetism in a
  moir{\'e} superlattice},}\ }\href {\doibase
  https://doi.org/10.1038/s41586-020-2049-7} {\bibfield  {journal} {\bibinfo
  {journal} {Nature}\ }\textbf {\bibinfo {volume} {579}},\ \bibinfo {pages}
  {56--61} (\bibinfo {year} {2020})}\BibitemShut {NoStop}%
\bibitem [{\citenamefont {Chen}\ \emph
  {et~al.}(2019{\natexlab{b}})\citenamefont {Chen}, \citenamefont {Sharpe},
  \citenamefont {Gallagher}, \citenamefont {Rosen}, \citenamefont {Fox},
  \citenamefont {Jiang}, \citenamefont {Lyu}, \citenamefont {Li}, \citenamefont
  {Watanabe}, \citenamefont {Taniguchi} \emph {et~al.}}]{Chen2019:Nat}%
  \BibitemOpen
  \bibfield  {author} {\bibinfo {author} {\bibfnamefont {Guorui}\ \bibnamefont
  {Chen}}, \bibinfo {author} {\bibfnamefont {Aaron~L}\ \bibnamefont {Sharpe}},
  \bibinfo {author} {\bibfnamefont {Patrick}\ \bibnamefont {Gallagher}},
  \bibinfo {author} {\bibfnamefont {Ilan~T}\ \bibnamefont {Rosen}}, \bibinfo
  {author} {\bibfnamefont {Eli~J}\ \bibnamefont {Fox}}, \bibinfo {author}
  {\bibfnamefont {Lili}\ \bibnamefont {Jiang}}, \bibinfo {author}
  {\bibfnamefont {Bosai}\ \bibnamefont {Lyu}}, \bibinfo {author} {\bibfnamefont
  {Hongyuan}\ \bibnamefont {Li}}, \bibinfo {author} {\bibfnamefont {Kenji}\
  \bibnamefont {Watanabe}}, \bibinfo {author} {\bibfnamefont {Takashi}\
  \bibnamefont {Taniguchi}},  \emph {et~al.},\ }\bibfield  {title} {\enquote
  {\bibinfo {title} {Signatures of tunable superconductivity in a trilayer
  graphene moir{\'e} superlattice},}\ }\href {\doibase
  https://doi.org/10.1038/s41586-019-1393-y} {\bibfield  {journal} {\bibinfo
  {journal} {Nature}\ }\textbf {\bibinfo {volume} {572}},\ \bibinfo {pages}
  {215--219} (\bibinfo {year} {2019}{\natexlab{b}})}\BibitemShut {NoStop}%
\bibitem [{\citenamefont {Zhou}\ \emph
  {et~al.}(2021{\natexlab{a}})\citenamefont {Zhou}, \citenamefont {Xie},
  \citenamefont {Ghazaryan}, \citenamefont {Holder}, \citenamefont {Ehrets},
  \citenamefont {Spanton}, \citenamefont {Taniguchi}, \citenamefont {Watanabe},
  \citenamefont {Berg}, \citenamefont {Serbyn},\ and\ \citenamefont
  {et~al.}}]{Zhou2021:arxiv}%
  \BibitemOpen
  \bibfield  {author} {\bibinfo {author} {\bibfnamefont {Haoxin}\ \bibnamefont
  {Zhou}}, \bibinfo {author} {\bibfnamefont {Tian}\ \bibnamefont {Xie}},
  \bibinfo {author} {\bibfnamefont {Areg}\ \bibnamefont {Ghazaryan}}, \bibinfo
  {author} {\bibfnamefont {Tobias}\ \bibnamefont {Holder}}, \bibinfo {author}
  {\bibfnamefont {James~R.}\ \bibnamefont {Ehrets}}, \bibinfo {author}
  {\bibfnamefont {Eric~M.}\ \bibnamefont {Spanton}}, \bibinfo {author}
  {\bibfnamefont {Takashi}\ \bibnamefont {Taniguchi}}, \bibinfo {author}
  {\bibfnamefont {Kenji}\ \bibnamefont {Watanabe}}, \bibinfo {author}
  {\bibfnamefont {Erez}\ \bibnamefont {Berg}}, \bibinfo {author} {\bibfnamefont
  {Maksym}\ \bibnamefont {Serbyn}}, \ and\ \bibinfo {author} {\bibnamefont
  {et~al.}},\ }\bibfield  {title} {\enquote {\bibinfo {title} {Half- and
  quarter-metals in rhombohedral trilayer graphene},}\ }\href {\doibase
  10.1038/s41586-021-03938-w} {\bibfield  {journal} {\bibinfo  {journal}
  {Nature}\ }\textbf {\bibinfo {volume} {598}},\ \bibinfo {pages} {429–433}
  (\bibinfo {year} {2021}{\natexlab{a}})}\BibitemShut {NoStop}%
\bibitem [{\citenamefont {Chou}\ \emph {et~al.}(2021)\citenamefont {Chou},
  \citenamefont {Wu}, \citenamefont {Sau},\ and\ \citenamefont
  {Das~Sarma}}]{Chou2021:arxiv}%
  \BibitemOpen
  \bibfield  {author} {\bibinfo {author} {\bibfnamefont {Yang-Zhi}\
  \bibnamefont {Chou}}, \bibinfo {author} {\bibfnamefont {Fengcheng}\
  \bibnamefont {Wu}}, \bibinfo {author} {\bibfnamefont {Jay~D.}\ \bibnamefont
  {Sau}}, \ and\ \bibinfo {author} {\bibfnamefont {Sankar}\ \bibnamefont
  {Das~Sarma}},\ }\bibfield  {title} {\enquote {\bibinfo {title}
  {Acoustic-phonon-mediated superconductivity in rhombohedral trilayer
  graphene},}\ }\href {\doibase 10.1103/PhysRevLett.127.187001} {\bibfield
  {journal} {\bibinfo  {journal} {Phys. Rev. Lett.}\ }\textbf {\bibinfo
  {volume} {127}},\ \bibinfo {pages} {187001} (\bibinfo {year}
  {2021})}\BibitemShut {NoStop}%
\bibitem [{\citenamefont {Phong}\ \emph {et~al.}(2021)\citenamefont {Phong},
  \citenamefont {Pantale\'on}, \citenamefont {Cea},\ and\ \citenamefont
  {Guinea}}]{Phong2021:arxiv}%
  \BibitemOpen
  \bibfield  {author} {\bibinfo {author} {\bibfnamefont {V\~o~Tien}\
  \bibnamefont {Phong}}, \bibinfo {author} {\bibfnamefont {Pierre~A.}\
  \bibnamefont {Pantale\'on}}, \bibinfo {author} {\bibfnamefont {Tommaso}\
  \bibnamefont {Cea}}, \ and\ \bibinfo {author} {\bibfnamefont {Francisco}\
  \bibnamefont {Guinea}},\ }\bibfield  {title} {\enquote {\bibinfo {title}
  {Band structure and superconductivity in twisted trilayer graphene},}\ }\href
  {\doibase 10.1103/PhysRevB.104.L121116} {\bibfield  {journal} {\bibinfo
  {journal} {Phys. Rev. B}\ }\textbf {\bibinfo {volume} {104}},\ \bibinfo
  {pages} {L121116} (\bibinfo {year} {2021})}\BibitemShut {NoStop}%
\bibitem [{\citenamefont {Zhou}\ \emph
  {et~al.}(2021{\natexlab{b}})\citenamefont {Zhou}, \citenamefont {Xie},
  \citenamefont {Taniguchi}, \citenamefont {Watanabe},\ and\ \citenamefont
  {Young}}]{Zhou2021:arxiv2}%
  \BibitemOpen
  \bibfield  {author} {\bibinfo {author} {\bibfnamefont {Haoxin}\ \bibnamefont
  {Zhou}}, \bibinfo {author} {\bibfnamefont {Tian}\ \bibnamefont {Xie}},
  \bibinfo {author} {\bibfnamefont {Takashi}\ \bibnamefont {Taniguchi}},
  \bibinfo {author} {\bibfnamefont {Kenji}\ \bibnamefont {Watanabe}}, \ and\
  \bibinfo {author} {\bibfnamefont {Andrea~F.}\ \bibnamefont {Young}},\
  }\bibfield  {title} {\enquote {\bibinfo {title} {Superconductivity in
  rhombohedral trilayer graphene},}\ }\href {\doibase
  10.1038/s41586-021-03926-0} {\bibfield  {journal} {\bibinfo  {journal}
  {Nature}\ }\textbf {\bibinfo {volume} {598}},\ \bibinfo {pages} {434–438}
  (\bibinfo {year} {2021}{\natexlab{b}})}\BibitemShut {NoStop}%
\bibitem [{\citenamefont {Qin}\ and\ \citenamefont
  {MacDonald}(2021)}]{Qin2021:arxiv}%
  \BibitemOpen
  \bibfield  {author} {\bibinfo {author} {\bibfnamefont {Wei}\ \bibnamefont
  {Qin}}\ and\ \bibinfo {author} {\bibfnamefont {Allan~H.}\ \bibnamefont
  {MacDonald}},\ }\bibfield  {title} {\enquote {\bibinfo {title} {In-plane
  critical magnetic fields in magic-angle twisted trilayer graphene},}\ }\href
  {\doibase 10.1103/PhysRevLett.127.097001} {\bibfield  {journal} {\bibinfo
  {journal} {Phys. Rev. Lett.}\ }\textbf {\bibinfo {volume} {127}},\ \bibinfo
  {pages} {097001} (\bibinfo {year} {2021})}\BibitemShut {NoStop}%
\bibitem [{\citenamefont {Tang}\ \emph {et~al.}(2020)\citenamefont {Tang},
  \citenamefont {Li}, \citenamefont {Li}, \citenamefont {Xu}, \citenamefont
  {Liu}, \citenamefont {Barmak}, \citenamefont {Watanabe}, \citenamefont
  {Taniguchi}, \citenamefont {MacDonald}, \citenamefont {Shan} \emph
  {et~al.}}]{Tang2020:Nat}%
  \BibitemOpen
  \bibfield  {author} {\bibinfo {author} {\bibfnamefont {Yanhao}\ \bibnamefont
  {Tang}}, \bibinfo {author} {\bibfnamefont {Lizhong}\ \bibnamefont {Li}},
  \bibinfo {author} {\bibfnamefont {Tingxin}\ \bibnamefont {Li}}, \bibinfo
  {author} {\bibfnamefont {Yang}\ \bibnamefont {Xu}}, \bibinfo {author}
  {\bibfnamefont {Song}\ \bibnamefont {Liu}}, \bibinfo {author} {\bibfnamefont
  {Katayun}\ \bibnamefont {Barmak}}, \bibinfo {author} {\bibfnamefont {Kenji}\
  \bibnamefont {Watanabe}}, \bibinfo {author} {\bibfnamefont {Takashi}\
  \bibnamefont {Taniguchi}}, \bibinfo {author} {\bibfnamefont {Allan~H}\
  \bibnamefont {MacDonald}}, \bibinfo {author} {\bibfnamefont {Jie}\
  \bibnamefont {Shan}},  \emph {et~al.},\ }\bibfield  {title} {\enquote
  {\bibinfo {title} {Simulation of hubbard model physics in wse 2/ws 2
  moir{\'e} superlattices},}\ }\href {\doibase
  https://doi.org/10.1038/s41586-020-2085-3} {\bibfield  {journal} {\bibinfo
  {journal} {Nature}\ }\textbf {\bibinfo {volume} {579}},\ \bibinfo {pages}
  {353--358} (\bibinfo {year} {2020})}\BibitemShut {NoStop}%
\bibitem [{\citenamefont {Sierra}\ \emph {et~al.}(2021)\citenamefont {Sierra},
  \citenamefont {Fabian}, \citenamefont {Kawakami}, \citenamefont {Roche},\
  and\ \citenamefont {Valenzuela}}]{Sierra2021:NN}%
  \BibitemOpen
  \bibfield  {author} {\bibinfo {author} {\bibfnamefont {Juan~F}\ \bibnamefont
  {Sierra}}, \bibinfo {author} {\bibfnamefont {Jaroslav}\ \bibnamefont
  {Fabian}}, \bibinfo {author} {\bibfnamefont {Roland~K}\ \bibnamefont
  {Kawakami}}, \bibinfo {author} {\bibfnamefont {Stephan}\ \bibnamefont
  {Roche}}, \ and\ \bibinfo {author} {\bibfnamefont {Sergio~O}\ \bibnamefont
  {Valenzuela}},\ }\bibfield  {title} {\enquote {\bibinfo {title} {Van der
  waals heterostructures for spintronics and opto-spintronics},}\ }\href
  {\doibase https://doi.org/10.1038/s41565-021-00936-x} {\bibfield  {journal}
  {\bibinfo  {journal} {Nature Nanotechnology}\ }\textbf {\bibinfo {volume}
  {16}},\ \bibinfo {pages} {856} (\bibinfo {year} {2021})}\BibitemShut
  {NoStop}%
\bibitem [{\citenamefont {Moriya}\ \emph {et~al.}(2020)\citenamefont {Moriya},
  \citenamefont {Yabuki},\ and\ \citenamefont {Machida}}]{Moriya2020:PRB}%
  \BibitemOpen
  \bibfield  {author} {\bibinfo {author} {\bibfnamefont {Rai}\ \bibnamefont
  {Moriya}}, \bibinfo {author} {\bibfnamefont {Naoto}\ \bibnamefont {Yabuki}},
  \ and\ \bibinfo {author} {\bibfnamefont {Tomoki}\ \bibnamefont {Machida}},\
  }\bibfield  {title} {\enquote {\bibinfo {title} {Superconducting proximity
  effect in a $\mathrm{Nb}{\mathrm{se}}_{2}/\text{graphene}$ van der waals
  junction},}\ }\href {\doibase 10.1103/PhysRevB.101.054503} {\bibfield
  {journal} {\bibinfo  {journal} {Phys. Rev. B}\ }\textbf {\bibinfo {volume}
  {101}},\ \bibinfo {pages} {054503} (\bibinfo {year} {2020})}\BibitemShut
  {NoStop}%
\bibitem [{\citenamefont {Han}\ \emph {et~al.}(2021)\citenamefont {Han},
  \citenamefont {Ling}, \citenamefont {Liu}, \citenamefont {Li}, \citenamefont
  {Zhang},\ and\ \citenamefont {Wang}}]{Han2021:APL}%
  \BibitemOpen
  \bibfield  {author} {\bibinfo {author} {\bibfnamefont {Hui}\ \bibnamefont
  {Han}}, \bibinfo {author} {\bibfnamefont {Jie}\ \bibnamefont {Ling}},
  \bibinfo {author} {\bibfnamefont {Wenhui}\ \bibnamefont {Liu}}, \bibinfo
  {author} {\bibfnamefont {Hui}\ \bibnamefont {Li}}, \bibinfo {author}
  {\bibfnamefont {Changjin}\ \bibnamefont {Zhang}}, \ and\ \bibinfo {author}
  {\bibfnamefont {Jiannong}\ \bibnamefont {Wang}},\ }\bibfield  {title}
  {\enquote {\bibinfo {title} {Superconducting proximity effect in a van der
  waals 2h-tas2/nbse2 heterostructure},}\ }\href {\doibase 10.1063/5.0051968}
  {\bibfield  {journal} {\bibinfo  {journal} {Applied Physics Letters}\
  }\textbf {\bibinfo {volume} {118}},\ \bibinfo {pages} {253101} (\bibinfo
  {year} {2021})}\BibitemShut {NoStop}%
\bibitem [{\citenamefont {Zollner}\ \emph {et~al.}(2016)\citenamefont
  {Zollner}, \citenamefont {Gmitra}, \citenamefont {Frank},\ and\ \citenamefont
  {Fabian}}]{Zollner2016:PRB}%
  \BibitemOpen
  \bibfield  {author} {\bibinfo {author} {\bibfnamefont {Klaus}\ \bibnamefont
  {Zollner}}, \bibinfo {author} {\bibfnamefont {Martin}\ \bibnamefont
  {Gmitra}}, \bibinfo {author} {\bibfnamefont {Tobias}\ \bibnamefont {Frank}},
  \ and\ \bibinfo {author} {\bibfnamefont {Jaroslav}\ \bibnamefont {Fabian}},\
  }\bibfield  {title} {\enquote {\bibinfo {title} {Theory of proximity-induced
  exchange coupling in graphene on hbn/(co, ni)},}\ }\href {\doibase
  10.1103/PhysRevB.94.155441} {\bibfield  {journal} {\bibinfo  {journal} {Phys.
  Rev. B}\ }\textbf {\bibinfo {volume} {94}},\ \bibinfo {pages} {155441}
  (\bibinfo {year} {2016})}\BibitemShut {NoStop}%
\bibitem [{\citenamefont {Zollner}\ \emph {et~al.}(2018)\citenamefont
  {Zollner}, \citenamefont {Gmitra},\ and\ \citenamefont
  {Fabian}}]{Zollner2018:NJP}%
  \BibitemOpen
  \bibfield  {author} {\bibinfo {author} {\bibfnamefont {Klaus}\ \bibnamefont
  {Zollner}}, \bibinfo {author} {\bibfnamefont {Martin}\ \bibnamefont
  {Gmitra}}, \ and\ \bibinfo {author} {\bibfnamefont {Jaroslav}\ \bibnamefont
  {Fabian}},\ }\bibfield  {title} {\enquote {\bibinfo {title} {{Electrically
  tunable exchange splitting in bilayer graphene on monolayer Cr 2 X 2 Te 6
  with X = Ge, Si, and Sn}},}\ }\href {\doibase 10.1088/1367-2630/aace51}
  {\bibfield  {journal} {\bibinfo  {journal} {New J. Phys.}\ }\textbf {\bibinfo
  {volume} {20}},\ \bibinfo {pages} {073007} (\bibinfo {year}
  {2018})}\BibitemShut {NoStop}%
\bibitem [{\citenamefont {Zollner}\ \emph {et~al.}(2019)\citenamefont
  {Zollner}, \citenamefont {Faria~Junior},\ and\ \citenamefont
  {Fabian}}]{Zollner2019a:PRB}%
  \BibitemOpen
  \bibfield  {author} {\bibinfo {author} {\bibfnamefont {Klaus}\ \bibnamefont
  {Zollner}}, \bibinfo {author} {\bibfnamefont {Paulo~E.}\ \bibnamefont
  {Faria~Junior}}, \ and\ \bibinfo {author} {\bibfnamefont {Jaroslav}\
  \bibnamefont {Fabian}},\ }\bibfield  {title} {\enquote {\bibinfo {title}
  {Proximity exchange effects in ${\mathrm{mose}}_{2}$ and ${\mathrm{wse}}_{2}$
  heterostructures with ${\mathrm{cri}}_{3}$: Twist angle, layer, and gate
  dependence},}\ }\href {\doibase 10.1103/PhysRevB.100.085128} {\bibfield
  {journal} {\bibinfo  {journal} {Phys. Rev. B}\ }\textbf {\bibinfo {volume}
  {100}},\ \bibinfo {pages} {085128} (\bibinfo {year} {2019})}\BibitemShut
  {NoStop}%
\bibitem [{\citenamefont {Zollner}\ \emph
  {et~al.}(2020{\natexlab{a}})\citenamefont {Zollner}, \citenamefont
  {Faria~Junior},\ and\ \citenamefont {Fabian}}]{Zollner2020:PRB}%
  \BibitemOpen
  \bibfield  {author} {\bibinfo {author} {\bibfnamefont {Klaus}\ \bibnamefont
  {Zollner}}, \bibinfo {author} {\bibfnamefont {Paulo~E.}\ \bibnamefont
  {Faria~Junior}}, \ and\ \bibinfo {author} {\bibfnamefont {Jaroslav}\
  \bibnamefont {Fabian}},\ }\bibfield  {title} {\enquote {\bibinfo {title}
  {Giant proximity exchange and valley splitting in transition metal
  dichalcogenide/$h\mathrm{BN}$/(co, ni) heterostructures},}\ }\href {\doibase
  10.1103/PhysRevB.101.085112} {\bibfield  {journal} {\bibinfo  {journal}
  {Phys. Rev. B}\ }\textbf {\bibinfo {volume} {101}},\ \bibinfo {pages}
  {085112} (\bibinfo {year} {2020}{\natexlab{a}})}\BibitemShut {NoStop}%
\bibitem [{\citenamefont {Hallal}\ \emph {et~al.}(2017)\citenamefont {Hallal},
  \citenamefont {Ibrahim}, \citenamefont {Yang}, \citenamefont {Roche},\ and\
  \citenamefont {Chshiev}}]{Hallal2017:2DM}%
  \BibitemOpen
  \bibfield  {author} {\bibinfo {author} {\bibfnamefont {Ali}\ \bibnamefont
  {Hallal}}, \bibinfo {author} {\bibfnamefont {Fatima}\ \bibnamefont
  {Ibrahim}}, \bibinfo {author} {\bibfnamefont {Hongxin}\ \bibnamefont {Yang}},
  \bibinfo {author} {\bibfnamefont {Stephan}\ \bibnamefont {Roche}}, \ and\
  \bibinfo {author} {\bibfnamefont {Mairbek}\ \bibnamefont {Chshiev}},\
  }\bibfield  {title} {\enquote {\bibinfo {title} {{Tailoring magnetic
  insulator proximity effects in graphene: first-principles calculations}},}\
  }\href {\doibase 10.1088/2053-1583/aa6663} {\bibfield  {journal} {\bibinfo
  {journal} {2D Mater.}\ }\textbf {\bibinfo {volume} {4}},\ \bibinfo {pages}
  {025074} (\bibinfo {year} {2017})}\BibitemShut {NoStop}%
\bibitem [{\citenamefont {Zhang}\ \emph
  {et~al.}(2015{\natexlab{a}})\citenamefont {Zhang}, \citenamefont {Zhao},
  \citenamefont {Yao},\ and\ \citenamefont {Yang}}]{Zhang2015:PRB}%
  \BibitemOpen
  \bibfield  {author} {\bibinfo {author} {\bibfnamefont {Jiayong}\ \bibnamefont
  {Zhang}}, \bibinfo {author} {\bibfnamefont {Bao}\ \bibnamefont {Zhao}},
  \bibinfo {author} {\bibfnamefont {Yugui}\ \bibnamefont {Yao}}, \ and\
  \bibinfo {author} {\bibfnamefont {Zhongqin}\ \bibnamefont {Yang}},\
  }\bibfield  {title} {\enquote {\bibinfo {title} {Robust quantum anomalous
  hall effect in graphene-based van der waals heterostructures},}\ }\href
  {\doibase 10.1103/PhysRevB.92.165418} {\bibfield  {journal} {\bibinfo
  {journal} {Phys. Rev. B}\ }\textbf {\bibinfo {volume} {92}},\ \bibinfo
  {pages} {165418} (\bibinfo {year} {2015}{\natexlab{a}})}\BibitemShut
  {NoStop}%
\bibitem [{\citenamefont {Zhang}\ \emph {et~al.}(2018)\citenamefont {Zhang},
  \citenamefont {Zhao}, \citenamefont {Zhou}, \citenamefont {Xue},
  \citenamefont {Ma},\ and\ \citenamefont {Yang}}]{Zhang2018:PRB}%
  \BibitemOpen
  \bibfield  {author} {\bibinfo {author} {\bibfnamefont {Jiayong}\ \bibnamefont
  {Zhang}}, \bibinfo {author} {\bibfnamefont {Bao}\ \bibnamefont {Zhao}},
  \bibinfo {author} {\bibfnamefont {Tong}\ \bibnamefont {Zhou}}, \bibinfo
  {author} {\bibfnamefont {Yang}\ \bibnamefont {Xue}}, \bibinfo {author}
  {\bibfnamefont {Chunlan}\ \bibnamefont {Ma}}, \ and\ \bibinfo {author}
  {\bibfnamefont {Zhongqin}\ \bibnamefont {Yang}},\ }\bibfield  {title}
  {\enquote {\bibinfo {title} {{Strong magnetization and Chern insulators in
  compressed graphene/CrI$_3$ van der Waals heterostructures}},}\ }\href
  {\doibase 10.1103/PhysRevB.97.085401} {\bibfield  {journal} {\bibinfo
  {journal} {Phys. Rev. B}\ }\textbf {\bibinfo {volume} {97}},\ \bibinfo
  {pages} {085401} (\bibinfo {year} {2018})}\BibitemShut {NoStop}%
\bibitem [{\citenamefont {Yang}\ \emph {et~al.}(2013)\citenamefont {Yang},
  \citenamefont {Hallal}, \citenamefont {Terrade}, \citenamefont {Waintal},
  \citenamefont {Roche},\ and\ \citenamefont {Chshiev}}]{Yang2013:PRL}%
  \BibitemOpen
  \bibfield  {author} {\bibinfo {author} {\bibfnamefont {H.~X.}\ \bibnamefont
  {Yang}}, \bibinfo {author} {\bibfnamefont {A.}~\bibnamefont {Hallal}},
  \bibinfo {author} {\bibfnamefont {D.}~\bibnamefont {Terrade}}, \bibinfo
  {author} {\bibfnamefont {X.}~\bibnamefont {Waintal}}, \bibinfo {author}
  {\bibfnamefont {S.}~\bibnamefont {Roche}}, \ and\ \bibinfo {author}
  {\bibfnamefont {M.}~\bibnamefont {Chshiev}},\ }\bibfield  {title} {\enquote
  {\bibinfo {title} {Proximity effects induced in graphene by magnetic
  insulators: First-principles calculations on spin filtering and
  exchange-splitting gaps},}\ }\href {\doibase 10.1103/PhysRevLett.110.046603}
  {\bibfield  {journal} {\bibinfo  {journal} {Phys. Rev. Lett.}\ }\textbf
  {\bibinfo {volume} {110}},\ \bibinfo {pages} {046603} (\bibinfo {year}
  {2013})}\BibitemShut {NoStop}%
\bibitem [{\citenamefont {Dyrda{\l}}\ and\ \citenamefont
  {Barna{\'{s}}}(2017)}]{Dyrdal2017:2DM}%
  \BibitemOpen
  \bibfield  {author} {\bibinfo {author} {\bibfnamefont {A}~\bibnamefont
  {Dyrda{\l}}}\ and\ \bibinfo {author} {\bibfnamefont {J}~\bibnamefont
  {Barna{\'{s}}}},\ }\bibfield  {title} {\enquote {\bibinfo {title} {Anomalous,
  spin, and valley hall effects in graphene deposited on ferromagnetic
  substrates},}\ }\href {\doibase 10.1088/2053-1583/aa7bac} {\bibfield
  {journal} {\bibinfo  {journal} {2D Materials}\ }\textbf {\bibinfo {volume}
  {4}},\ \bibinfo {pages} {034003} (\bibinfo {year} {2017})}\BibitemShut
  {NoStop}%
\bibitem [{\citenamefont {Song}(2018)}]{Song2017:JPD}%
  \BibitemOpen
  \bibfield  {author} {\bibinfo {author} {\bibfnamefont {Yu}~\bibnamefont
  {Song}},\ }\bibfield  {title} {\enquote {\bibinfo {title}
  {{Electric-field-induced extremely large change in resistance in graphene
  ferromagnets}},}\ }\href {\doibase 10.1088/1361-6463/aa9b5e} {\bibfield
  {journal} {\bibinfo  {journal} {J. Phys. D: Appl. Phys.}\ }\textbf {\bibinfo
  {volume} {51}},\ \bibinfo {pages} {025002} (\bibinfo {year}
  {2018})}\BibitemShut {NoStop}%
\bibitem [{\citenamefont {Haugen}\ \emph {et~al.}(2008)\citenamefont {Haugen},
  \citenamefont {Huertas-Hernando},\ and\ \citenamefont
  {Brataas}}]{Haugen2008:PRB}%
  \BibitemOpen
  \bibfield  {author} {\bibinfo {author} {\bibfnamefont {H{\aa}vard}\
  \bibnamefont {Haugen}}, \bibinfo {author} {\bibfnamefont {Daniel}\
  \bibnamefont {Huertas-Hernando}}, \ and\ \bibinfo {author} {\bibfnamefont
  {Arne}\ \bibnamefont {Brataas}},\ }\bibfield  {title} {\enquote {\bibinfo
  {title} {{Spin transport in proximity-induced ferromagnetic graphene}},}\
  }\href {\doibase 10.1103/PhysRevB.77.115406} {\bibfield  {journal} {\bibinfo
  {journal} {Phys. Rev. B}\ }\textbf {\bibinfo {volume} {77}},\ \bibinfo
  {pages} {115406} (\bibinfo {year} {2008})}\BibitemShut {NoStop}%
\bibitem [{\citenamefont {Zhang}\ \emph
  {et~al.}(2015{\natexlab{b}})\citenamefont {Zhang}, \citenamefont {Zhao},
  \citenamefont {Yao},\ and\ \citenamefont {Yang}}]{Zhang2015:SR}%
  \BibitemOpen
  \bibfield  {author} {\bibinfo {author} {\bibfnamefont {Jiayong}\ \bibnamefont
  {Zhang}}, \bibinfo {author} {\bibfnamefont {Bao}\ \bibnamefont {Zhao}},
  \bibinfo {author} {\bibfnamefont {Yugui}\ \bibnamefont {Yao}}, \ and\
  \bibinfo {author} {\bibfnamefont {Zhongqin}\ \bibnamefont {Yang}},\
  }\bibfield  {title} {\enquote {\bibinfo {title} {{Quantum Anomalous Hall
  Effect in Graphene-based Heterostructure}},}\ }\href {\doibase
  10.1038/srep10629} {\bibfield  {journal} {\bibinfo  {journal} {Sci. Rep.}\
  }\textbf {\bibinfo {volume} {5}},\ \bibinfo {pages} {10629} (\bibinfo {year}
  {2015}{\natexlab{b}})}\BibitemShut {NoStop}%
\bibitem [{\citenamefont {Su}\ \emph {et~al.}(2017)\citenamefont {Su},
  \citenamefont {Barlas}, \citenamefont {Li}, \citenamefont {Shi},\ and\
  \citenamefont {Lake}}]{Su2017:PRB}%
  \BibitemOpen
  \bibfield  {author} {\bibinfo {author} {\bibfnamefont {Shanshan}\
  \bibnamefont {Su}}, \bibinfo {author} {\bibfnamefont {Yafis}\ \bibnamefont
  {Barlas}}, \bibinfo {author} {\bibfnamefont {Junxue}\ \bibnamefont {Li}},
  \bibinfo {author} {\bibfnamefont {Jing}\ \bibnamefont {Shi}}, \ and\ \bibinfo
  {author} {\bibfnamefont {Roger~K.}\ \bibnamefont {Lake}},\ }\bibfield
  {title} {\enquote {\bibinfo {title} {Effect of intervalley interaction on
  band topology of commensurate graphene/euo heterostructures},}\ }\href
  {\doibase 10.1103/PhysRevB.95.075418} {\bibfield  {journal} {\bibinfo
  {journal} {Phys. Rev. B}\ }\textbf {\bibinfo {volume} {95}},\ \bibinfo
  {pages} {075418} (\bibinfo {year} {2017})}\BibitemShut {NoStop}%
\bibitem [{\citenamefont {Gibertini}\ \emph {et~al.}(2019)\citenamefont
  {Gibertini}, \citenamefont {Koperski}, \citenamefont {Morpurgo},\ and\
  \citenamefont {Novoselov}}]{Gibertini2019:NN}%
  \BibitemOpen
  \bibfield  {author} {\bibinfo {author} {\bibfnamefont {M.}~\bibnamefont
  {Gibertini}}, \bibinfo {author} {\bibfnamefont {M.}~\bibnamefont {Koperski}},
  \bibinfo {author} {\bibfnamefont {A.~F.}\ \bibnamefont {Morpurgo}}, \ and\
  \bibinfo {author} {\bibfnamefont {K.~S.}\ \bibnamefont {Novoselov}},\
  }\bibfield  {title} {\enquote {\bibinfo {title} {{Magnetic 2D materials and
  heterostructures}},}\ }\href {\doibase 10.1038/s41565-019-0438-6} {\bibfield
  {journal} {\bibinfo  {journal} {Nat. Nanotechnol.}\ }\textbf {\bibinfo
  {volume} {14}},\ \bibinfo {pages} {408} (\bibinfo {year} {2019})}\BibitemShut
  {NoStop}%
\bibitem [{\citenamefont {Klein}\ \emph {et~al.}(2018)\citenamefont {Klein},
  \citenamefont {MacNeill}, \citenamefont {Lado}, \citenamefont {Soriano},
  \citenamefont {Navarro-Moratalla}, \citenamefont {Watanabe}, \citenamefont
  {Taniguchi}, \citenamefont {Manni}, \citenamefont {Canfield}, \citenamefont
  {Fern{\'{a}}ndez-Rossier},\ and\ \citenamefont
  {Jarillo-Herrero}}]{Klein2018:SC}%
  \BibitemOpen
  \bibfield  {author} {\bibinfo {author} {\bibfnamefont {D.~R.}\ \bibnamefont
  {Klein}}, \bibinfo {author} {\bibfnamefont {D.}~\bibnamefont {MacNeill}},
  \bibinfo {author} {\bibfnamefont {J.~L.}\ \bibnamefont {Lado}}, \bibinfo
  {author} {\bibfnamefont {D.}~\bibnamefont {Soriano}}, \bibinfo {author}
  {\bibfnamefont {E.}~\bibnamefont {Navarro-Moratalla}}, \bibinfo {author}
  {\bibfnamefont {K.}~\bibnamefont {Watanabe}}, \bibinfo {author}
  {\bibfnamefont {T.}~\bibnamefont {Taniguchi}}, \bibinfo {author}
  {\bibfnamefont {S.}~\bibnamefont {Manni}}, \bibinfo {author} {\bibfnamefont
  {P.}~\bibnamefont {Canfield}}, \bibinfo {author} {\bibfnamefont
  {J.}~\bibnamefont {Fern{\'{a}}ndez-Rossier}}, \ and\ \bibinfo {author}
  {\bibfnamefont {P.}~\bibnamefont {Jarillo-Herrero}},\ }\bibfield  {title}
  {\enquote {\bibinfo {title} {{Probing magnetism in 2D van der Waals
  crystalline insulators via electron tunneling}},}\ }\href {\doibase
  10.1126/science.aar3617} {\bibfield  {journal} {\bibinfo  {journal}
  {Science}\ }\textbf {\bibinfo {volume} {360}},\ \bibinfo {pages} {1218}
  (\bibinfo {year} {2018})}\BibitemShut {NoStop}%
\bibitem [{\citenamefont {Cardoso}\ \emph {et~al.}(2018)\citenamefont
  {Cardoso}, \citenamefont {Soriano}, \citenamefont
  {Garc\'{\i}a-Mart\'{\i}nez},\ and\ \citenamefont
  {Fern\'andez-Rossier}}]{Cardoso2018:PRL}%
  \BibitemOpen
  \bibfield  {author} {\bibinfo {author} {\bibfnamefont {C.}~\bibnamefont
  {Cardoso}}, \bibinfo {author} {\bibfnamefont {D.}~\bibnamefont {Soriano}},
  \bibinfo {author} {\bibfnamefont {N.~A.}\ \bibnamefont
  {Garc\'{\i}a-Mart\'{\i}nez}}, \ and\ \bibinfo {author} {\bibfnamefont
  {J.}~\bibnamefont {Fern\'andez-Rossier}},\ }\bibfield  {title} {\enquote
  {\bibinfo {title} {Van der waals spin valves},}\ }\href {\doibase
  10.1103/PhysRevLett.121.067701} {\bibfield  {journal} {\bibinfo  {journal}
  {Phys. Rev. Lett.}\ }\textbf {\bibinfo {volume} {121}},\ \bibinfo {pages}
  {067701} (\bibinfo {year} {2018})}\BibitemShut {NoStop}%
\bibitem [{\citenamefont {Henriques}\ \emph {et~al.}(2020)\citenamefont
  {Henriques}, \citenamefont {Catarina}, \citenamefont {Costa}, \citenamefont
  {Fern\'andez-Rossier},\ and\ \citenamefont {Peres}}]{Henriques2020:PRB}%
  \BibitemOpen
  \bibfield  {author} {\bibinfo {author} {\bibfnamefont {J.~C.~G.}\
  \bibnamefont {Henriques}}, \bibinfo {author} {\bibfnamefont {G.}~\bibnamefont
  {Catarina}}, \bibinfo {author} {\bibfnamefont {A.~T.}\ \bibnamefont {Costa}},
  \bibinfo {author} {\bibfnamefont {J.}~\bibnamefont {Fern\'andez-Rossier}}, \
  and\ \bibinfo {author} {\bibfnamefont {N.~M.~R.}\ \bibnamefont {Peres}},\
  }\bibfield  {title} {\enquote {\bibinfo {title} {Excitonic magneto-optical
  kerr effect in two-dimensional transition metal dichalcogenides induced by
  spin proximity},}\ }\href {\doibase 10.1103/PhysRevB.101.045408} {\bibfield
  {journal} {\bibinfo  {journal} {Phys. Rev. B}\ }\textbf {\bibinfo {volume}
  {101}},\ \bibinfo {pages} {045408} (\bibinfo {year} {2020})}\BibitemShut
  {NoStop}%
\bibitem [{\citenamefont {Singh}\ \emph {et~al.}(2017)\citenamefont {Singh},
  \citenamefont {Katoch}, \citenamefont {Zhu}, \citenamefont {Meng},
  \citenamefont {Liu}, \citenamefont {Brangham}, \citenamefont {Yang},
  \citenamefont {Flatt{\'{e}}},\ and\ \citenamefont
  {Kawakami}}]{Singh2017:PRL}%
  \BibitemOpen
  \bibfield  {author} {\bibinfo {author} {\bibfnamefont {Simranjeet}\
  \bibnamefont {Singh}}, \bibinfo {author} {\bibfnamefont {Jyoti}\ \bibnamefont
  {Katoch}}, \bibinfo {author} {\bibfnamefont {Tiancong}\ \bibnamefont {Zhu}},
  \bibinfo {author} {\bibfnamefont {Keng~Yuan}\ \bibnamefont {Meng}}, \bibinfo
  {author} {\bibfnamefont {Tianyu}\ \bibnamefont {Liu}}, \bibinfo {author}
  {\bibfnamefont {Jack~T.}\ \bibnamefont {Brangham}}, \bibinfo {author}
  {\bibfnamefont {Fengyuan}\ \bibnamefont {Yang}}, \bibinfo {author}
  {\bibfnamefont {Michael~E.}\ \bibnamefont {Flatt{\'{e}}}}, \ and\ \bibinfo
  {author} {\bibfnamefont {Roland~K.}\ \bibnamefont {Kawakami}},\ }\bibfield
  {title} {\enquote {\bibinfo {title} {{Strong Modulation of Spin Currents in
  Bilayer Graphene by Static and Fluctuating Proximity Exchange Fields}},}\
  }\href {\doibase 10.1103/PhysRevLett.118.187201} {\bibfield  {journal}
  {\bibinfo  {journal} {Phys. Rev. Lett.}\ }\textbf {\bibinfo {volume} {118}},\
  \bibinfo {pages} {187201} (\bibinfo {year} {2017})}\BibitemShut {NoStop}%
\bibitem [{\citenamefont {Swartz}\ \emph {et~al.}(2012)\citenamefont {Swartz},
  \citenamefont {Odenthal}, \citenamefont {Hao}, \citenamefont {Ruoff},\ and\
  \citenamefont {Kawakami}}]{Swartz2012:ACS}%
  \BibitemOpen
  \bibfield  {author} {\bibinfo {author} {\bibfnamefont {Adrian~G.}\
  \bibnamefont {Swartz}}, \bibinfo {author} {\bibfnamefont {Patrick~M.}\
  \bibnamefont {Odenthal}}, \bibinfo {author} {\bibfnamefont {Yufeng}\
  \bibnamefont {Hao}}, \bibinfo {author} {\bibfnamefont {Rodney~S.}\
  \bibnamefont {Ruoff}}, \ and\ \bibinfo {author} {\bibfnamefont {Roland~K.}\
  \bibnamefont {Kawakami}},\ }\bibfield  {title} {\enquote {\bibinfo {title}
  {{Integration of the ferromagnetic insulator EuO onto graphene}},}\ }\href
  {\doibase 10.1021/nn303771f} {\bibfield  {journal} {\bibinfo  {journal} {ACS
  Nano}\ }\textbf {\bibinfo {volume} {6}},\ \bibinfo {pages} {10063} (\bibinfo
  {year} {2012})}\BibitemShut {NoStop}%
\bibitem [{\citenamefont {Gmitra}\ and\ \citenamefont
  {Fabian}(2015)}]{Gmitra2015:PRB}%
  \BibitemOpen
  \bibfield  {author} {\bibinfo {author} {\bibfnamefont {Martin}\ \bibnamefont
  {Gmitra}}\ and\ \bibinfo {author} {\bibfnamefont {Jaroslav}\ \bibnamefont
  {Fabian}},\ }\bibfield  {title} {\enquote {\bibinfo {title} {Graphene on
  transition-metal dichalcogenides: A platform for proximity spin-orbit physics
  and optospintronics},}\ }\href {\doibase 10.1103/PhysRevB.92.155403}
  {\bibfield  {journal} {\bibinfo  {journal} {Phys. Rev. B}\ }\textbf {\bibinfo
  {volume} {92}},\ \bibinfo {pages} {155403} (\bibinfo {year}
  {2015})}\BibitemShut {NoStop}%
\bibitem [{\citenamefont {Gmitra}\ \emph {et~al.}(2016)\citenamefont {Gmitra},
  \citenamefont {Kochan}, \citenamefont {H\"ogl},\ and\ \citenamefont
  {Fabian}}]{Gmitra2016:PRB}%
  \BibitemOpen
  \bibfield  {author} {\bibinfo {author} {\bibfnamefont {Martin}\ \bibnamefont
  {Gmitra}}, \bibinfo {author} {\bibfnamefont {Denis}\ \bibnamefont {Kochan}},
  \bibinfo {author} {\bibfnamefont {Petra}\ \bibnamefont {H\"ogl}}, \ and\
  \bibinfo {author} {\bibfnamefont {Jaroslav}\ \bibnamefont {Fabian}},\
  }\bibfield  {title} {\enquote {\bibinfo {title} {Trivial and inverted dirac
  bands and the emergence of quantum spin hall states in graphene on
  transition-metal dichalcogenides},}\ }\href {\doibase
  10.1103/PhysRevB.93.155104} {\bibfield  {journal} {\bibinfo  {journal} {Phys.
  Rev. B}\ }\textbf {\bibinfo {volume} {93}},\ \bibinfo {pages} {155104}
  (\bibinfo {year} {2016})}\BibitemShut {NoStop}%
\bibitem [{\citenamefont {Zollner}\ and\ \citenamefont
  {Fabian}(2019)}]{Zollner2019b:PRB}%
  \BibitemOpen
  \bibfield  {author} {\bibinfo {author} {\bibfnamefont {Klaus}\ \bibnamefont
  {Zollner}}\ and\ \bibinfo {author} {\bibfnamefont {Jaroslav}\ \bibnamefont
  {Fabian}},\ }\bibfield  {title} {\enquote {\bibinfo {title} {Single and
  bilayer graphene on the topological insulator
  ${\mathrm{bi}}_{2}{\mathrm{se}}_{3}$: Electronic and spin-orbit properties
  from first principles},}\ }\href {\doibase 10.1103/PhysRevB.100.165141}
  {\bibfield  {journal} {\bibinfo  {journal} {Phys. Rev. B}\ }\textbf {\bibinfo
  {volume} {100}},\ \bibinfo {pages} {165141} (\bibinfo {year}
  {2019})}\BibitemShut {NoStop}%
\bibitem [{\citenamefont {Zollner}\ and\ \citenamefont
  {Fabian}(2021{\natexlab{a}})}]{Zollner2021:PSSB}%
  \BibitemOpen
  \bibfield  {author} {\bibinfo {author} {\bibfnamefont {Klaus}\ \bibnamefont
  {Zollner}}\ and\ \bibinfo {author} {\bibfnamefont {Jaroslav}\ \bibnamefont
  {Fabian}},\ }\bibfield  {title} {\enquote {\bibinfo {title} {Heterostructures
  of graphene and topological insulators bi2se3, bi2te3, and sb2te3},}\ }\href
  {\doibase https://doi.org/10.1002/pssb.202000081} {\bibfield  {journal}
  {\bibinfo  {journal} {physica status solidi (b)}\ }\textbf {\bibinfo {volume}
  {258}},\ \bibinfo {pages} {2000081} (\bibinfo {year}
  {2021}{\natexlab{a}})}\BibitemShut {NoStop}%
\bibitem [{\citenamefont {Khokhriakov}\ \emph {et~al.}(2020)\citenamefont
  {Khokhriakov}, \citenamefont {Hoque}, \citenamefont {Karpiak},\ and\
  \citenamefont {Dash}}]{Khokhriakov2020:NC}%
  \BibitemOpen
  \bibfield  {author} {\bibinfo {author} {\bibfnamefont {Dmitrii}\ \bibnamefont
  {Khokhriakov}}, \bibinfo {author} {\bibfnamefont {Anamul~Md.}\ \bibnamefont
  {Hoque}}, \bibinfo {author} {\bibfnamefont {Bogdan}\ \bibnamefont {Karpiak}},
  \ and\ \bibinfo {author} {\bibfnamefont {Saroj~P.}\ \bibnamefont {Dash}},\
  }\bibfield  {title} {\enquote {\bibinfo {title} {Gate-tunable spin-galvanic
  effect in graphene-topological insulator van der waals heterostructures at
  room temperature},}\ }\href {\doibase 10.1038/s41467-020-17481-1} {\bibfield
  {journal} {\bibinfo  {journal} {Nature Communications}\ }\textbf {\bibinfo
  {volume} {11}},\ \bibinfo {pages} {3657} (\bibinfo {year}
  {2020})}\BibitemShut {NoStop}%
\bibitem [{\citenamefont {Karpiak}\ \emph {et~al.}(2019)\citenamefont
  {Karpiak}, \citenamefont {Cummings}, \citenamefont {Zollner}, \citenamefont
  {Vila}, \citenamefont {Khokhriakov}, \citenamefont {Hoque}, \citenamefont
  {Dankert}, \citenamefont {Svedlindh}, \citenamefont {Fabian}, \citenamefont
  {Roche},\ and\ \citenamefont {Dash}}]{Karpiak2019:arxiv}%
  \BibitemOpen
  \bibfield  {author} {\bibinfo {author} {\bibfnamefont {Bogdan}\ \bibnamefont
  {Karpiak}}, \bibinfo {author} {\bibfnamefont {Aron~W.}\ \bibnamefont
  {Cummings}}, \bibinfo {author} {\bibfnamefont {Klaus}\ \bibnamefont
  {Zollner}}, \bibinfo {author} {\bibfnamefont {Marc}\ \bibnamefont {Vila}},
  \bibinfo {author} {\bibfnamefont {Dmitrii}\ \bibnamefont {Khokhriakov}},
  \bibinfo {author} {\bibfnamefont {Anamul~Md}\ \bibnamefont {Hoque}}, \bibinfo
  {author} {\bibfnamefont {Andr{\'{e}}}\ \bibnamefont {Dankert}}, \bibinfo
  {author} {\bibfnamefont {Peter}\ \bibnamefont {Svedlindh}}, \bibinfo {author}
  {\bibfnamefont {Jaroslav}\ \bibnamefont {Fabian}}, \bibinfo {author}
  {\bibfnamefont {Stephan}\ \bibnamefont {Roche}}, \ and\ \bibinfo {author}
  {\bibfnamefont {Saroj~P.}\ \bibnamefont {Dash}},\ }\bibfield  {title}
  {\enquote {\bibinfo {title} {{Magnetic proximity in a van der Waals
  heterostructure of magnetic insulator and graphene}},}\ }\href {\doibase
  10.1088/2053-1583/ab5915} {\bibfield  {journal} {\bibinfo  {journal} {2D
  Mater.}\ }\textbf {\bibinfo {volume} {7}},\ \bibinfo {pages} {015026}
  (\bibinfo {year} {2019})}\BibitemShut {NoStop}%
\bibitem [{\citenamefont {Zihlmann}\ \emph {et~al.}(2018)\citenamefont
  {Zihlmann}, \citenamefont {Cummings}, \citenamefont {Garcia}, \citenamefont
  {Kedves}, \citenamefont {Watanabe}, \citenamefont {Taniguchi}, \citenamefont
  {Sch\"onenberger},\ and\ \citenamefont {Makk}}]{Zihlmann2018:PRB}%
  \BibitemOpen
  \bibfield  {author} {\bibinfo {author} {\bibfnamefont {Simon}\ \bibnamefont
  {Zihlmann}}, \bibinfo {author} {\bibfnamefont {Aron~W.}\ \bibnamefont
  {Cummings}}, \bibinfo {author} {\bibfnamefont {Jose~H.}\ \bibnamefont
  {Garcia}}, \bibinfo {author} {\bibfnamefont {M\'at\'e}\ \bibnamefont
  {Kedves}}, \bibinfo {author} {\bibfnamefont {Kenji}\ \bibnamefont
  {Watanabe}}, \bibinfo {author} {\bibfnamefont {Takashi}\ \bibnamefont
  {Taniguchi}}, \bibinfo {author} {\bibfnamefont {Christian}\ \bibnamefont
  {Sch\"onenberger}}, \ and\ \bibinfo {author} {\bibfnamefont {P\'eter}\
  \bibnamefont {Makk}},\ }\bibfield  {title} {\enquote {\bibinfo {title} {Large
  spin relaxation anisotropy and valley-zeeman spin-orbit coupling in
  ${\mathrm{wse}}_{2}$/graphene/$h$-bn heterostructures},}\ }\href {\doibase
  10.1103/PhysRevB.97.075434} {\bibfield  {journal} {\bibinfo  {journal} {Phys.
  Rev. B}\ }\textbf {\bibinfo {volume} {97}},\ \bibinfo {pages} {075434}
  (\bibinfo {year} {2018})}\BibitemShut {NoStop}%
\bibitem [{\citenamefont {Song}\ \emph {et~al.}(2018)\citenamefont {Song},
  \citenamefont {Soriano}, \citenamefont {Cummings}, \citenamefont {Robles},
  \citenamefont {Ordej{\'{o}}n},\ and\ \citenamefont {Roche}}]{Song2018:NL}%
  \BibitemOpen
  \bibfield  {author} {\bibinfo {author} {\bibfnamefont {Kenan}\ \bibnamefont
  {Song}}, \bibinfo {author} {\bibfnamefont {David}\ \bibnamefont {Soriano}},
  \bibinfo {author} {\bibfnamefont {Aron~W.}\ \bibnamefont {Cummings}},
  \bibinfo {author} {\bibfnamefont {Roberto}\ \bibnamefont {Robles}}, \bibinfo
  {author} {\bibfnamefont {Pablo}\ \bibnamefont {Ordej{\'{o}}n}}, \ and\
  \bibinfo {author} {\bibfnamefont {Stephan}\ \bibnamefont {Roche}},\
  }\bibfield  {title} {\enquote {\bibinfo {title} {{Spin Proximity Effects in
  Graphene/Topological Insulator Heterostructures}},}\ }\href {\doibase
  10.1021/acs.nanolett.7b05482} {\bibfield  {journal} {\bibinfo  {journal}
  {Nano Lett.}\ }\textbf {\bibinfo {volume} {18}},\ \bibinfo {pages} {2033}
  (\bibinfo {year} {2018})}\BibitemShut {NoStop}%
\bibitem [{\citenamefont {Khokhriakov}\ \emph {et~al.}(2018)\citenamefont
  {Khokhriakov}, \citenamefont {Cummings}, \citenamefont {Song}, \citenamefont
  {Vila}, \citenamefont {Karpiak}, \citenamefont {Dankert}, \citenamefont
  {Roche},\ and\ \citenamefont {Dash}}]{Khokhriakov2018:SA}%
  \BibitemOpen
  \bibfield  {author} {\bibinfo {author} {\bibfnamefont {Dmitrii}\ \bibnamefont
  {Khokhriakov}}, \bibinfo {author} {\bibfnamefont {Aron~W.}\ \bibnamefont
  {Cummings}}, \bibinfo {author} {\bibfnamefont {Kenan}\ \bibnamefont {Song}},
  \bibinfo {author} {\bibfnamefont {Marc}\ \bibnamefont {Vila}}, \bibinfo
  {author} {\bibfnamefont {Bogdan}\ \bibnamefont {Karpiak}}, \bibinfo {author}
  {\bibfnamefont {Andr{\'{e}}}\ \bibnamefont {Dankert}}, \bibinfo {author}
  {\bibfnamefont {Stephan}\ \bibnamefont {Roche}}, \ and\ \bibinfo {author}
  {\bibfnamefont {Saroj~P.}\ \bibnamefont {Dash}},\ }\bibfield  {title}
  {\enquote {\bibinfo {title} {{Tailoring emergent spin phenomena in Dirac
  material heterostructures}},}\ }\href {\doibase 10.1126/sciadv.aat9349}
  {\bibfield  {journal} {\bibinfo  {journal} {Sci. Adv.}\ }\textbf {\bibinfo
  {volume} {4}},\ \bibinfo {pages} {eaat9349} (\bibinfo {year}
  {2018})}\BibitemShut {NoStop}%
\bibitem [{\citenamefont {Garcia}\ \emph {et~al.}(2018)\citenamefont {Garcia},
  \citenamefont {Vila}, \citenamefont {Cummings},\ and\ \citenamefont
  {Roche}}]{Garcia2018:CSR}%
  \BibitemOpen
  \bibfield  {author} {\bibinfo {author} {\bibfnamefont {Jose~H.}\ \bibnamefont
  {Garcia}}, \bibinfo {author} {\bibfnamefont {Marc}\ \bibnamefont {Vila}},
  \bibinfo {author} {\bibfnamefont {Aron~W.}\ \bibnamefont {Cummings}}, \ and\
  \bibinfo {author} {\bibfnamefont {Stephan}\ \bibnamefont {Roche}},\
  }\bibfield  {title} {\enquote {\bibinfo {title} {{Spin transport in
  graphene/transition metal dichalcogenide heterostructures}},}\ }\href
  {\doibase 10.1039/c7cs00864c} {\bibfield  {journal} {\bibinfo  {journal}
  {Chem. Soc. Rev.}\ }\textbf {\bibinfo {volume} {47}},\ \bibinfo {pages}
  {3359} (\bibinfo {year} {2018})}\BibitemShut {NoStop}%
\bibitem [{\citenamefont {Safeer}\ \emph {et~al.}(2019)\citenamefont {Safeer},
  \citenamefont {Ingla-Ayn{\'e}s}, \citenamefont {Herling}, \citenamefont
  {Garcia}, \citenamefont {Vila}, \citenamefont {Ontoso}, \citenamefont
  {Calvo}, \citenamefont {Roche}, \citenamefont {Hueso},\ and\ \citenamefont
  {Casanova}}]{Safeer2019:NL}%
  \BibitemOpen
  \bibfield  {author} {\bibinfo {author} {\bibfnamefont {CK}~\bibnamefont
  {Safeer}}, \bibinfo {author} {\bibfnamefont {Josep}\ \bibnamefont
  {Ingla-Ayn{\'e}s}}, \bibinfo {author} {\bibfnamefont {Franz}\ \bibnamefont
  {Herling}}, \bibinfo {author} {\bibfnamefont {Jos{\'e}~H}\ \bibnamefont
  {Garcia}}, \bibinfo {author} {\bibfnamefont {Marc}\ \bibnamefont {Vila}},
  \bibinfo {author} {\bibfnamefont {Nerea}\ \bibnamefont {Ontoso}}, \bibinfo
  {author} {\bibfnamefont {M~Reyes}\ \bibnamefont {Calvo}}, \bibinfo {author}
  {\bibfnamefont {Stephan}\ \bibnamefont {Roche}}, \bibinfo {author}
  {\bibfnamefont {Luis~E}\ \bibnamefont {Hueso}}, \ and\ \bibinfo {author}
  {\bibfnamefont {F{\`e}lix}\ \bibnamefont {Casanova}},\ }\bibfield  {title}
  {\enquote {\bibinfo {title} {{Room-temperature spin Hall effect in
  graphene/MoS$_2$ van der Waals heterostructures}},}\ }\href {\doibase
  10.1021/acs.nanolett.8b04368} {\bibfield  {journal} {\bibinfo  {journal}
  {Nano Lett.}\ }\textbf {\bibinfo {volume} {19}},\ \bibinfo {pages} {1074}
  (\bibinfo {year} {2019})}\BibitemShut {NoStop}%
\bibitem [{\citenamefont {Herling}\ \emph {et~al.}(2020)\citenamefont
  {Herling}, \citenamefont {Safeer}, \citenamefont {Ingla-Aynés},
  \citenamefont {Ontoso}, \citenamefont {Hueso},\ and\ \citenamefont
  {Casanova}}]{Herlin2020:APL}%
  \BibitemOpen
  \bibfield  {author} {\bibinfo {author} {\bibfnamefont {Franz}\ \bibnamefont
  {Herling}}, \bibinfo {author} {\bibfnamefont {C.~K.}\ \bibnamefont {Safeer}},
  \bibinfo {author} {\bibfnamefont {Josep}\ \bibnamefont {Ingla-Aynés}},
  \bibinfo {author} {\bibfnamefont {Nerea}\ \bibnamefont {Ontoso}}, \bibinfo
  {author} {\bibfnamefont {Luis~E.}\ \bibnamefont {Hueso}}, \ and\ \bibinfo
  {author} {\bibfnamefont {Fèlix}\ \bibnamefont {Casanova}},\ }\bibfield
  {title} {\enquote {\bibinfo {title} {Gate tunability of highly efficient
  spin-to-charge conversion by spin hall effect in graphene proximitized with
  wse2},}\ }\href {\doibase 10.1063/5.0006101} {\bibfield  {journal} {\bibinfo
  {journal} {APL Materials}\ }\textbf {\bibinfo {volume} {8}},\ \bibinfo
  {pages} {071103} (\bibinfo {year} {2020})}\BibitemShut {NoStop}%
\bibitem [{\citenamefont {Khoo}\ \emph {et~al.}(2017)\citenamefont {Khoo},
  \citenamefont {Morpurgo},\ and\ \citenamefont {Levitov}}]{Khoo2017:NL}%
  \BibitemOpen
  \bibfield  {author} {\bibinfo {author} {\bibfnamefont {Jun~Yong}\
  \bibnamefont {Khoo}}, \bibinfo {author} {\bibfnamefont {Alberto~F.}\
  \bibnamefont {Morpurgo}}, \ and\ \bibinfo {author} {\bibfnamefont {Leonid}\
  \bibnamefont {Levitov}},\ }\bibfield  {title} {\enquote {\bibinfo {title}
  {{On-Demand Spin-Orbit Interaction from Which-Layer Tunability in Bilayer
  Graphene}},}\ }\href {\doibase 10.1021/acs.nanolett.7b03604} {\bibfield
  {journal} {\bibinfo  {journal} {Nano Lett.}\ }\textbf {\bibinfo {volume}
  {17}},\ \bibinfo {pages} {7003} (\bibinfo {year} {2017})}\BibitemShut
  {NoStop}%
\bibitem [{\citenamefont {Wang}\ \emph
  {et~al.}(2015{\natexlab{a}})\citenamefont {Wang}, \citenamefont {Ki},
  \citenamefont {Chen}, \citenamefont {Berger}, \citenamefont {MacDonald},\
  and\ \citenamefont {Morpurgo}}]{Wang2015:NC}%
  \BibitemOpen
  \bibfield  {author} {\bibinfo {author} {\bibfnamefont {Zhe}\ \bibnamefont
  {Wang}}, \bibinfo {author} {\bibfnamefont {Dong-Keun}\ \bibnamefont {Ki}},
  \bibinfo {author} {\bibfnamefont {Hua}\ \bibnamefont {Chen}}, \bibinfo
  {author} {\bibfnamefont {Helmuth}\ \bibnamefont {Berger}}, \bibinfo {author}
  {\bibfnamefont {Allan~H}\ \bibnamefont {MacDonald}}, \ and\ \bibinfo {author}
  {\bibfnamefont {Alberto~F}\ \bibnamefont {Morpurgo}},\ }\bibfield  {title}
  {\enquote {\bibinfo {title} {{Strong interface-induced spin-orbit interaction
  in graphene on WS2}},}\ }\href {\doibase 10.1038/ncomms9339} {\bibfield
  {journal} {\bibinfo  {journal} {Nat. Commun.}\ }\textbf {\bibinfo {volume}
  {6}},\ \bibinfo {pages} {8339} (\bibinfo {year}
  {2015}{\natexlab{a}})}\BibitemShut {NoStop}%
\bibitem [{\citenamefont {Omar}\ and\ \citenamefont {van
  Wees}(2018)}]{Omar2018:PRB}%
  \BibitemOpen
  \bibfield  {author} {\bibinfo {author} {\bibfnamefont {S.}~\bibnamefont
  {Omar}}\ and\ \bibinfo {author} {\bibfnamefont {B.~J.}\ \bibnamefont {van
  Wees}},\ }\bibfield  {title} {\enquote {\bibinfo {title} {Spin transport in
  high-mobility graphene on ${\mathrm{ws}}_{2}$ substrate with electric-field
  tunable proximity spin-orbit interaction},}\ }\href {\doibase
  10.1103/PhysRevB.97.045414} {\bibfield  {journal} {\bibinfo  {journal} {Phys.
  Rev. B}\ }\textbf {\bibinfo {volume} {97}},\ \bibinfo {pages} {045414}
  (\bibinfo {year} {2018})}\BibitemShut {NoStop}%
\bibitem [{\citenamefont {Omar}\ and\ \citenamefont {van
  Wees}(2017)}]{Omar2017:PRB}%
  \BibitemOpen
  \bibfield  {author} {\bibinfo {author} {\bibfnamefont {S.}~\bibnamefont
  {Omar}}\ and\ \bibinfo {author} {\bibfnamefont {B.~J.}\ \bibnamefont {van
  Wees}},\ }\bibfield  {title} {\enquote {\bibinfo {title}
  {Graphene-${\mathrm{ws}}_{2}$ heterostructures for tunable spin injection and
  spin transport},}\ }\href {\doibase 10.1103/PhysRevB.95.081404} {\bibfield
  {journal} {\bibinfo  {journal} {Phys. Rev. B}\ }\textbf {\bibinfo {volume}
  {95}},\ \bibinfo {pages} {081404} (\bibinfo {year} {2017})}\BibitemShut
  {NoStop}%
\bibitem [{\citenamefont {Fülöp}\ \emph
  {et~al.}(2021{\natexlab{a}})\citenamefont {Fülöp}, \citenamefont {Márffy},
  \citenamefont {Zihlmann}, \citenamefont {Gmitra}, \citenamefont {Tóvári},
  \citenamefont {Szentpéteri}, \citenamefont {Kedves}, \citenamefont
  {Watanabe}, \citenamefont {Taniguchi}, \citenamefont {Fabian},\ and\
  \citenamefont {et~al.}}]{Fulop2021:arxiv}%
  \BibitemOpen
  \bibfield  {author} {\bibinfo {author} {\bibfnamefont {Bálint}\ \bibnamefont
  {Fülöp}}, \bibinfo {author} {\bibfnamefont {Albin}\ \bibnamefont
  {Márffy}}, \bibinfo {author} {\bibfnamefont {Simon}\ \bibnamefont
  {Zihlmann}}, \bibinfo {author} {\bibfnamefont {Martin}\ \bibnamefont
  {Gmitra}}, \bibinfo {author} {\bibfnamefont {Endre}\ \bibnamefont
  {Tóvári}}, \bibinfo {author} {\bibfnamefont {Bálint}\ \bibnamefont
  {Szentpéteri}}, \bibinfo {author} {\bibfnamefont {Máté}\ \bibnamefont
  {Kedves}}, \bibinfo {author} {\bibfnamefont {Kenji}\ \bibnamefont
  {Watanabe}}, \bibinfo {author} {\bibfnamefont {Takashi}\ \bibnamefont
  {Taniguchi}}, \bibinfo {author} {\bibfnamefont {Jaroslav}\ \bibnamefont
  {Fabian}}, \ and\ \bibinfo {author} {\bibnamefont {et~al.}},\ }\bibfield
  {title} {\enquote {\bibinfo {title} {Boosting proximity spin–orbit coupling
  in graphene/wse2 heterostructures via hydrostatic pressure},}\ }\href
  {\doibase 10.1038/s41699-021-00262-9} {\bibfield  {journal} {\bibinfo
  {journal} {npj 2D Materials and Applications}\ }\textbf {\bibinfo {volume}
  {5}},\ \bibinfo {pages} {82} (\bibinfo {year}
  {2021}{\natexlab{a}})}\BibitemShut {NoStop}%
\bibitem [{\citenamefont {Zollner}\ \emph
  {et~al.}(2020{\natexlab{b}})\citenamefont {Zollner}, \citenamefont
  {Petrovi\ifmmode~\acute{c}\else \'{c}\fi{}}, \citenamefont {Dolui},
  \citenamefont {Plech\'a\ifmmode~\check{c}\else \v{c}\fi{}}, \citenamefont
  {Nikoli\ifmmode~\acute{c}\else \'{c}\fi{}},\ and\ \citenamefont
  {Fabian}}]{Zollner2019:PRR}%
  \BibitemOpen
  \bibfield  {author} {\bibinfo {author} {\bibfnamefont {Klaus}\ \bibnamefont
  {Zollner}}, \bibinfo {author} {\bibfnamefont {Marko~D.}\ \bibnamefont
  {Petrovi\ifmmode~\acute{c}\else \'{c}\fi{}}}, \bibinfo {author}
  {\bibfnamefont {Kapildeb}\ \bibnamefont {Dolui}}, \bibinfo {author}
  {\bibfnamefont {Petr}\ \bibnamefont {Plech\'a\ifmmode~\check{c}\else
  \v{c}\fi{}}}, \bibinfo {author} {\bibfnamefont {Branislav~K.}\ \bibnamefont
  {Nikoli\ifmmode~\acute{c}\else \'{c}\fi{}}}, \ and\ \bibinfo {author}
  {\bibfnamefont {Jaroslav}\ \bibnamefont {Fabian}},\ }\bibfield  {title}
  {\enquote {\bibinfo {title} {Scattering-induced and highly tunable by gate
  damping-like spin-orbit torque in graphene doubly proximitized by
  two-dimensional magnet ${\mathrm{cr}}_{2}{\mathrm{ge}}_{2}{\mathrm{te}}_{6}$
  and monolayer ${\mathrm{ws}}_{2}$},}\ }\href {\doibase
  10.1103/PhysRevResearch.2.043057} {\bibfield  {journal} {\bibinfo  {journal}
  {Phys. Rev. Research}\ }\textbf {\bibinfo {volume} {2}},\ \bibinfo {pages}
  {043057} (\bibinfo {year} {2020}{\natexlab{b}})}\BibitemShut {NoStop}%
\bibitem [{\citenamefont {Farooq}\ and\ \citenamefont
  {Hong}(2019)}]{Farooq2019:NPJ}%
  \BibitemOpen
  \bibfield  {author} {\bibinfo {author} {\bibfnamefont {M~Umar}\ \bibnamefont
  {Farooq}}\ and\ \bibinfo {author} {\bibfnamefont {Jisang}\ \bibnamefont
  {Hong}},\ }\bibfield  {title} {\enquote {\bibinfo {title} {Switchable valley
  splitting by external electric field effect in graphene/cri 3
  heterostructures},}\ }\href {\doibase
  https://doi.org/10.1038/s41699-019-0086-6} {\bibfield  {journal} {\bibinfo
  {journal} {npj 2D Materials and Applications}\ }\textbf {\bibinfo {volume}
  {3}},\ \bibinfo {pages} {3} (\bibinfo {year} {2019})}\BibitemShut {NoStop}%
\bibitem [{\citenamefont {Seyler}\ \emph {et~al.}(2018)\citenamefont {Seyler},
  \citenamefont {Zhong}, \citenamefont {Huang}, \citenamefont {Linpeng},
  \citenamefont {Wilson}, \citenamefont {Taniguchi}, \citenamefont {Watanabe},
  \citenamefont {Yao}, \citenamefont {Xiao}, \citenamefont {McGuire},
  \citenamefont {Fu},\ and\ \citenamefont {Xu}}]{Seyler2018:NL}%
  \BibitemOpen
  \bibfield  {author} {\bibinfo {author} {\bibfnamefont {Kyle~L.}\ \bibnamefont
  {Seyler}}, \bibinfo {author} {\bibfnamefont {Ding}\ \bibnamefont {Zhong}},
  \bibinfo {author} {\bibfnamefont {Bevin}\ \bibnamefont {Huang}}, \bibinfo
  {author} {\bibfnamefont {Xiayu}\ \bibnamefont {Linpeng}}, \bibinfo {author}
  {\bibfnamefont {Nathan~P.}\ \bibnamefont {Wilson}}, \bibinfo {author}
  {\bibfnamefont {Takashi}\ \bibnamefont {Taniguchi}}, \bibinfo {author}
  {\bibfnamefont {Kenji}\ \bibnamefont {Watanabe}}, \bibinfo {author}
  {\bibfnamefont {Wang}\ \bibnamefont {Yao}}, \bibinfo {author} {\bibfnamefont
  {Di}~\bibnamefont {Xiao}}, \bibinfo {author} {\bibfnamefont {Michael~A.}\
  \bibnamefont {McGuire}}, \bibinfo {author} {\bibfnamefont {Kai-Mei~C.}\
  \bibnamefont {Fu}}, \ and\ \bibinfo {author} {\bibfnamefont {Xiaodong}\
  \bibnamefont {Xu}},\ }\bibfield  {title} {\enquote {\bibinfo {title} {{Valley
  Manipulation by Optically Tuning the Magnetic Proximity Effect in WSe 2 /CrI
  3 Heterostructures}},}\ }\href {\doibase 10.1021/acs.nanolett.8b01105}
  {\bibfield  {journal} {\bibinfo  {journal} {Nano Lett.}\ }\textbf {\bibinfo
  {volume} {18}},\ \bibinfo {pages} {3823} (\bibinfo {year}
  {2018})}\BibitemShut {NoStop}%
\bibitem [{\citenamefont {Wang}\ \emph
  {et~al.}(2015{\natexlab{b}})\citenamefont {Wang}, \citenamefont {Tang},
  \citenamefont {Sachs}, \citenamefont {Barlas},\ and\ \citenamefont
  {Shi}}]{Wang2015:PRL}%
  \BibitemOpen
  \bibfield  {author} {\bibinfo {author} {\bibfnamefont {Zhiyong}\ \bibnamefont
  {Wang}}, \bibinfo {author} {\bibfnamefont {Chi}\ \bibnamefont {Tang}},
  \bibinfo {author} {\bibfnamefont {Raymond}\ \bibnamefont {Sachs}}, \bibinfo
  {author} {\bibfnamefont {Yafis}\ \bibnamefont {Barlas}}, \ and\ \bibinfo
  {author} {\bibfnamefont {Jing}\ \bibnamefont {Shi}},\ }\bibfield  {title}
  {\enquote {\bibinfo {title} {Proximity-induced ferromagnetism in graphene
  revealed by the anomalous hall effect},}\ }\href {\doibase
  10.1103/PhysRevLett.114.016603} {\bibfield  {journal} {\bibinfo  {journal}
  {Phys. Rev. Lett.}\ }\textbf {\bibinfo {volume} {114}},\ \bibinfo {pages}
  {016603} (\bibinfo {year} {2015}{\natexlab{b}})}\BibitemShut {NoStop}%
\bibitem [{\citenamefont {Mendes}\ \emph {et~al.}(2015)\citenamefont {Mendes},
  \citenamefont {{Alves Santos}}, \citenamefont {Meireles}, \citenamefont
  {Lacerda}, \citenamefont {Vilela-Le{\~{a}}o}, \citenamefont {Machado},
  \citenamefont {Rodr{\'{i}}guez-Su{\'{a}}rez}, \citenamefont {Azevedo},\ and\
  \citenamefont {Rezende}}]{Mendes2015:PRL}%
  \BibitemOpen
  \bibfield  {author} {\bibinfo {author} {\bibfnamefont {J.~B.~S.}\
  \bibnamefont {Mendes}}, \bibinfo {author} {\bibfnamefont {O.}~\bibnamefont
  {{Alves Santos}}}, \bibinfo {author} {\bibfnamefont {L.~M.}\ \bibnamefont
  {Meireles}}, \bibinfo {author} {\bibfnamefont {R.~G.}\ \bibnamefont
  {Lacerda}}, \bibinfo {author} {\bibfnamefont {L.~H.}\ \bibnamefont
  {Vilela-Le{\~{a}}o}}, \bibinfo {author} {\bibfnamefont {F.~L.~A.}\
  \bibnamefont {Machado}}, \bibinfo {author} {\bibfnamefont {R.~L.}\
  \bibnamefont {Rodr{\'{i}}guez-Su{\'{a}}rez}}, \bibinfo {author}
  {\bibfnamefont {A.}~\bibnamefont {Azevedo}}, \ and\ \bibinfo {author}
  {\bibfnamefont {S.~M.}\ \bibnamefont {Rezende}},\ }\bibfield  {title}
  {\enquote {\bibinfo {title} {{Spin-Current to Charge-Current Conversion and
  Magnetoresistance in a Hybrid Structure of Graphene and Yttrium Iron
  Garnet}},}\ }\href {\doibase 10.1103/PhysRevLett.115.226601} {\bibfield
  {journal} {\bibinfo  {journal} {Phys. Rev. Lett.}\ }\textbf {\bibinfo
  {volume} {115}},\ \bibinfo {pages} {226601} (\bibinfo {year}
  {2015})}\BibitemShut {NoStop}%
\bibitem [{\citenamefont {Leutenantsmeyer}\ \emph {et~al.}(2016)\citenamefont
  {Leutenantsmeyer}, \citenamefont {Kaverzin}, \citenamefont {Wojtaszek},\ and\
  \citenamefont {van Wees}}]{Leutenantsmeyer2016:2DM}%
  \BibitemOpen
  \bibfield  {author} {\bibinfo {author} {\bibfnamefont {Johannes~Christian}\
  \bibnamefont {Leutenantsmeyer}}, \bibinfo {author} {\bibfnamefont {Alexey~A}\
  \bibnamefont {Kaverzin}}, \bibinfo {author} {\bibfnamefont {Magdalena}\
  \bibnamefont {Wojtaszek}}, \ and\ \bibinfo {author} {\bibfnamefont {Bart~J}\
  \bibnamefont {van Wees}},\ }\bibfield  {title} {\enquote {\bibinfo {title}
  {Proximity induced room temperature ferromagnetism in graphene probed with
  spin currents},}\ }\href {\doibase 10.1088/2053-1583/4/1/014001} {\bibfield
  {journal} {\bibinfo  {journal} {2D Mater.}\ }\textbf {\bibinfo {volume}
  {4}},\ \bibinfo {pages} {014001} (\bibinfo {year} {2016})}\BibitemShut
  {NoStop}%
\bibitem [{\citenamefont {Zollner}\ \emph
  {et~al.}(2020{\natexlab{c}})\citenamefont {Zollner}, \citenamefont {Gmitra},\
  and\ \citenamefont {Fabian}}]{Zollner2020:PRL}%
  \BibitemOpen
  \bibfield  {author} {\bibinfo {author} {\bibfnamefont {Klaus}\ \bibnamefont
  {Zollner}}, \bibinfo {author} {\bibfnamefont {Martin}\ \bibnamefont
  {Gmitra}}, \ and\ \bibinfo {author} {\bibfnamefont {Jaroslav}\ \bibnamefont
  {Fabian}},\ }\bibfield  {title} {\enquote {\bibinfo {title} {Swapping
  exchange and spin-orbit coupling in 2d van der waals heterostructures},}\
  }\href {\doibase 10.1103/PhysRevLett.125.196402} {\bibfield  {journal}
  {\bibinfo  {journal} {Phys. Rev. Lett.}\ }\textbf {\bibinfo {volume} {125}},\
  \bibinfo {pages} {196402} (\bibinfo {year} {2020}{\natexlab{c}})}\BibitemShut
  {NoStop}%
\bibitem [{\citenamefont {Dolui}\ \emph {et~al.}(2020)\citenamefont {Dolui},
  \citenamefont {Petrovic}, \citenamefont {Zollner}, \citenamefont {Plechac},
  \citenamefont {Fabian},\ and\ \citenamefont {Nikolic}}]{Dolui2020:NL}%
  \BibitemOpen
  \bibfield  {author} {\bibinfo {author} {\bibfnamefont {Kapildeb}\
  \bibnamefont {Dolui}}, \bibinfo {author} {\bibfnamefont {Marko~D}\
  \bibnamefont {Petrovic}}, \bibinfo {author} {\bibfnamefont {Klaus}\
  \bibnamefont {Zollner}}, \bibinfo {author} {\bibfnamefont {Petr}\
  \bibnamefont {Plechac}}, \bibinfo {author} {\bibfnamefont {Jaroslav}\
  \bibnamefont {Fabian}}, \ and\ \bibinfo {author} {\bibfnamefont
  {Branislav~K}\ \bibnamefont {Nikolic}},\ }\bibfield  {title} {\enquote
  {\bibinfo {title} {Proximity spin--orbit torque on a two-dimensional magnet
  within van der waals heterostructure: current-driven
  antiferromagnet-to-ferromagnet reversible nonequilibrium phase transition in
  bilayer cri3},}\ }\href {\doibase
  https://doi.org/10.1021/acs.nanolett.9b04556} {\bibfield  {journal} {\bibinfo
   {journal} {Nano letters}\ }\textbf {\bibinfo {volume} {20}},\ \bibinfo
  {pages} {2288--2295} (\bibinfo {year} {2020})}\BibitemShut {NoStop}%
\bibitem [{\citenamefont {Alghamdi}\ \emph {et~al.}(2019)\citenamefont
  {Alghamdi}, \citenamefont {Lohmann}, \citenamefont {Li}, \citenamefont
  {Jothi}, \citenamefont {Shao}, \citenamefont {Aldosary}, \citenamefont {Su},
  \citenamefont {Fokwa},\ and\ \citenamefont {Shi}}]{Alghamdi2019:NL}%
  \BibitemOpen
  \bibfield  {author} {\bibinfo {author} {\bibfnamefont {Mohammed}\
  \bibnamefont {Alghamdi}}, \bibinfo {author} {\bibfnamefont {Mark}\
  \bibnamefont {Lohmann}}, \bibinfo {author} {\bibfnamefont {Junxue}\
  \bibnamefont {Li}}, \bibinfo {author} {\bibfnamefont {Palani~R}\ \bibnamefont
  {Jothi}}, \bibinfo {author} {\bibfnamefont {Qiming}\ \bibnamefont {Shao}},
  \bibinfo {author} {\bibfnamefont {Mohammed}\ \bibnamefont {Aldosary}},
  \bibinfo {author} {\bibfnamefont {Tang}\ \bibnamefont {Su}}, \bibinfo
  {author} {\bibfnamefont {Boniface P.~T.}\ \bibnamefont {Fokwa}}, \ and\
  \bibinfo {author} {\bibfnamefont {Jing}\ \bibnamefont {Shi}},\ }\bibfield
  {title} {\enquote {\bibinfo {title} {{Highly Efficient Spin–Orbit Torque
  and Switching of Layered Ferromagnet Fe 3 GeTe 2}},}\ }\href {\doibase
  10.1021/acs.nanolett.9b01043} {\bibfield  {journal} {\bibinfo  {journal}
  {Nano Lett.}\ }\textbf {\bibinfo {volume} {19}},\ \bibinfo {pages} {4400}
  (\bibinfo {year} {2019})}\BibitemShut {NoStop}%
\bibitem [{\citenamefont {Wang}\ \emph {et~al.}(2019)\citenamefont {Wang},
  \citenamefont {Tang}, \citenamefont {Xia}, \citenamefont {He}, \citenamefont
  {Zhang}, \citenamefont {Liu}, \citenamefont {Wan}, \citenamefont {Fang},
  \citenamefont {Guo}, \citenamefont {Yang}, \citenamefont {Guang},
  \citenamefont {Zhang}, \citenamefont {Xu}, \citenamefont {Wei}, \citenamefont
  {Liao}, \citenamefont {Lu}, \citenamefont {Feng}, \citenamefont {Li},
  \citenamefont {Peng}, \citenamefont {Wei}, \citenamefont {Yang},
  \citenamefont {Shi}, \citenamefont {Zhang}, \citenamefont {Han},
  \citenamefont {Zhang}, \citenamefont {Zhang}, \citenamefont {Yu},\ and\
  \citenamefont {Han}}]{Wang2019:SA}%
  \BibitemOpen
  \bibfield  {author} {\bibinfo {author} {\bibfnamefont {Xiao}\ \bibnamefont
  {Wang}}, \bibinfo {author} {\bibfnamefont {Jian}\ \bibnamefont {Tang}},
  \bibinfo {author} {\bibfnamefont {Xiuxin}\ \bibnamefont {Xia}}, \bibinfo
  {author} {\bibfnamefont {Congli}\ \bibnamefont {He}}, \bibinfo {author}
  {\bibfnamefont {Junwei}\ \bibnamefont {Zhang}}, \bibinfo {author}
  {\bibfnamefont {Yizhou}\ \bibnamefont {Liu}}, \bibinfo {author}
  {\bibfnamefont {Caihua}\ \bibnamefont {Wan}}, \bibinfo {author}
  {\bibfnamefont {Chi}\ \bibnamefont {Fang}}, \bibinfo {author} {\bibfnamefont
  {Chenyang}\ \bibnamefont {Guo}}, \bibinfo {author} {\bibfnamefont {Wenlong}\
  \bibnamefont {Yang}}, \bibinfo {author} {\bibfnamefont {Yao}\ \bibnamefont
  {Guang}}, \bibinfo {author} {\bibfnamefont {Xiaomin}\ \bibnamefont {Zhang}},
  \bibinfo {author} {\bibfnamefont {Hongjun}\ \bibnamefont {Xu}}, \bibinfo
  {author} {\bibfnamefont {Jinwu}\ \bibnamefont {Wei}}, \bibinfo {author}
  {\bibfnamefont {Mengzhou}\ \bibnamefont {Liao}}, \bibinfo {author}
  {\bibfnamefont {Xiaobo}\ \bibnamefont {Lu}}, \bibinfo {author} {\bibfnamefont
  {Jiafeng}\ \bibnamefont {Feng}}, \bibinfo {author} {\bibfnamefont {Xiaoxi}\
  \bibnamefont {Li}}, \bibinfo {author} {\bibfnamefont {Yong}\ \bibnamefont
  {Peng}}, \bibinfo {author} {\bibfnamefont {Hongxiang}\ \bibnamefont {Wei}},
  \bibinfo {author} {\bibfnamefont {Rong}\ \bibnamefont {Yang}}, \bibinfo
  {author} {\bibfnamefont {Dongxia}\ \bibnamefont {Shi}}, \bibinfo {author}
  {\bibfnamefont {Xixiang}\ \bibnamefont {Zhang}}, \bibinfo {author}
  {\bibfnamefont {Zheng}\ \bibnamefont {Han}}, \bibinfo {author} {\bibfnamefont
  {Zhidong}\ \bibnamefont {Zhang}}, \bibinfo {author} {\bibfnamefont {Guangyu}\
  \bibnamefont {Zhang}}, \bibinfo {author} {\bibfnamefont {Guoqiang}\
  \bibnamefont {Yu}}, \ and\ \bibinfo {author} {\bibfnamefont {Xiufeng}\
  \bibnamefont {Han}},\ }\bibfield  {title} {\enquote {\bibinfo {title}
  {Current-driven magnetization switching in a van der waals ferromagnet
  fe3gete2},}\ }\href {\doibase 10.1126/sciadv.aaw8904} {\bibfield  {journal}
  {\bibinfo  {journal} {Science Advances}\ }\textbf {\bibinfo {volume} {5}},\
  \bibinfo {pages} {eaaw8904} (\bibinfo {year} {2019})}\BibitemShut {NoStop}%
\bibitem [{\citenamefont {Lazi{\'{c}}}\ \emph {et~al.}(2016)\citenamefont
  {Lazi{\'{c}}}, \citenamefont {Belashchenko},\ and\ \citenamefont
  {{\v{Z}}uti{\'{c}}}}]{Lazic2016:PRB}%
  \BibitemOpen
  \bibfield  {author} {\bibinfo {author} {\bibfnamefont {Predrag}\ \bibnamefont
  {Lazi{\'{c}}}}, \bibinfo {author} {\bibfnamefont {K.~D.}\ \bibnamefont
  {Belashchenko}}, \ and\ \bibinfo {author} {\bibfnamefont {Igor}\ \bibnamefont
  {{\v{Z}}uti{\'{c}}}},\ }\bibfield  {title} {\enquote {\bibinfo {title}
  {{Effective gating and tunable magnetic proximity effects in two-dimensional
  heterostructures}},}\ }\href {\doibase 10.1103/PhysRevB.93.241401} {\bibfield
   {journal} {\bibinfo  {journal} {Phys. Rev. B}\ }\textbf {\bibinfo {volume}
  {93}},\ \bibinfo {pages} {241401} (\bibinfo {year} {2016})}\BibitemShut
  {NoStop}%
\bibitem [{\citenamefont {Yan}\ \emph {et~al.}(2021)\citenamefont {Yan},
  \citenamefont {Qi}, \citenamefont {Wang},\ and\ \citenamefont
  {Mi}}]{Yan2021:PE}%
  \BibitemOpen
  \bibfield  {author} {\bibinfo {author} {\bibfnamefont {Shiming}\ \bibnamefont
  {Yan}}, \bibinfo {author} {\bibfnamefont {Shengmei}\ \bibnamefont {Qi}},
  \bibinfo {author} {\bibfnamefont {Dunhui}\ \bibnamefont {Wang}}, \ and\
  \bibinfo {author} {\bibfnamefont {Wenbo}\ \bibnamefont {Mi}},\ }\bibfield
  {title} {\enquote {\bibinfo {title} {Novel electronic structures and magnetic
  properties in twisted two-dimensional graphene/janus 2h–vsete
  heterostructures},}\ }\href {\doibase
  https://doi.org/10.1016/j.physe.2021.114854} {\bibfield  {journal} {\bibinfo
  {journal} {Physica E: Low-dimensional Systems and Nanostructures}\ }\textbf
  {\bibinfo {volume} {134}},\ \bibinfo {pages} {114854} (\bibinfo {year}
  {2021})}\BibitemShut {NoStop}%
\bibitem [{\citenamefont {David}\ \emph {et~al.}(2019)\citenamefont {David},
  \citenamefont {Rakyta}, \citenamefont {Korm\'anyos},\ and\ \citenamefont
  {Burkard}}]{David2019:arxiv}%
  \BibitemOpen
  \bibfield  {author} {\bibinfo {author} {\bibfnamefont {Alessandro}\
  \bibnamefont {David}}, \bibinfo {author} {\bibfnamefont {P\'eter}\
  \bibnamefont {Rakyta}}, \bibinfo {author} {\bibfnamefont {Andor}\
  \bibnamefont {Korm\'anyos}}, \ and\ \bibinfo {author} {\bibfnamefont {Guido}\
  \bibnamefont {Burkard}},\ }\bibfield  {title} {\enquote {\bibinfo {title}
  {Induced spin-orbit coupling in twisted graphene--transition metal
  dichalcogenide heterobilayers: Twistronics meets spintronics},}\ }\href
  {\doibase 10.1103/PhysRevB.100.085412} {\bibfield  {journal} {\bibinfo
  {journal} {Phys. Rev. B}\ }\textbf {\bibinfo {volume} {100}},\ \bibinfo
  {pages} {085412} (\bibinfo {year} {2019})}\BibitemShut {NoStop}%
\bibitem [{\citenamefont {Li}\ and\ \citenamefont
  {Koshino}(2019)}]{Li2019:PRB}%
  \BibitemOpen
  \bibfield  {author} {\bibinfo {author} {\bibfnamefont {Yang}\ \bibnamefont
  {Li}}\ and\ \bibinfo {author} {\bibfnamefont {Mikito}\ \bibnamefont
  {Koshino}},\ }\bibfield  {title} {\enquote {\bibinfo {title} {Twist-angle
  dependence of the proximity spin-orbit coupling in graphene on
  transition-metal dichalcogenides},}\ }\href {\doibase
  10.1103/PhysRevB.99.075438} {\bibfield  {journal} {\bibinfo  {journal} {Phys.
  Rev. B}\ }\textbf {\bibinfo {volume} {99}},\ \bibinfo {pages} {075438}
  (\bibinfo {year} {2019})}\BibitemShut {NoStop}%
\bibitem [{Note1()}]{Note1}%
  \BibitemOpen
  \bibinfo {note} {See Supplemental Material [url] where we show a more
  extended summary of results, which includes Refs.~\cite
  {ASE,Lazic2015:CPC,Koda2016:JPCC,Carr2020:NRM,Baskin1955:PR,Carteaux1995:JPCM,
  Hohenberg1964:PRB,Giannozzi2009:JPCM,Gong2017:Nat,Kresse1999:PRB,Perdew1996:PRL,Zhang2015:PRB,
  Zollner2019:PRR,Karpiak2019:arxiv,Grimme2006:JCC,Grimme2010:JCP,Barone2009:JCC,Bengtsson1999:PRB,
  Li2014:JMCC,Chen2015:PRB,Liu2021:N,Leutenantsmeyer2016:2DM,Hogl2020:PRL,
  Ghiasi2021:NN,Rosenberger2020:ACS,Weston2020:NN,Fulop2021:arxiv,Fulop2021:arxiv2,Tribhuwan2016:S,
  Li2019:PRB,David2019:arxiv,Zollner2021:arxiv,Zollner2016:PRB}}\BibitemShut
  {NoStop}%
\bibitem [{\citenamefont {Kochan}\ \emph {et~al.}(2017)\citenamefont {Kochan},
  \citenamefont {Irmer},\ and\ \citenamefont {Fabian}}]{Kochan2017:PRB}%
  \BibitemOpen
  \bibfield  {author} {\bibinfo {author} {\bibfnamefont {Denis}\ \bibnamefont
  {Kochan}}, \bibinfo {author} {\bibfnamefont {Susanne}\ \bibnamefont {Irmer}},
  \ and\ \bibinfo {author} {\bibfnamefont {Jaroslav}\ \bibnamefont {Fabian}},\
  }\bibfield  {title} {\enquote {\bibinfo {title} {Model spin-orbit coupling
  hamiltonians for graphene systems},}\ }\href {\doibase
  10.1103/PhysRevB.95.165415} {\bibfield  {journal} {\bibinfo  {journal} {Phys.
  Rev. B}\ }\textbf {\bibinfo {volume} {95}},\ \bibinfo {pages} {165415}
  (\bibinfo {year} {2017})}\BibitemShut {NoStop}%
\bibitem [{\citenamefont {Phong}\ \emph {et~al.}(2017)\citenamefont {Phong},
  \citenamefont {Walet},\ and\ \citenamefont {Guinea}}]{Phong2017:2DM}%
  \BibitemOpen
  \bibfield  {author} {\bibinfo {author} {\bibfnamefont {V{\~{o}}~Tien}\
  \bibnamefont {Phong}}, \bibinfo {author} {\bibfnamefont {Niels~R}\
  \bibnamefont {Walet}}, \ and\ \bibinfo {author} {\bibfnamefont {Francisco}\
  \bibnamefont {Guinea}},\ }\bibfield  {title} {\enquote {\bibinfo {title}
  {{Effective interactions in a graphene layer induced by the proximity to a
  ferromagnet}},}\ }\href {\doibase 10.1088/2053-1583/aa9fca} {\bibfield
  {journal} {\bibinfo  {journal} {2D Mater.}\ }\textbf {\bibinfo {volume}
  {5}},\ \bibinfo {pages} {014004} (\bibinfo {year} {2017})}\BibitemShut
  {NoStop}%
\bibitem [{\citenamefont {Fülöp}\ \emph
  {et~al.}(2021{\natexlab{b}})\citenamefont {Fülöp}, \citenamefont {Márffy},
  \citenamefont {Tóvári}, \citenamefont {Kedves}, \citenamefont {Zihlmann},
  \citenamefont {Indolese}, \citenamefont {Kovács-Krausz}, \citenamefont
  {Watanabe}, \citenamefont {Taniguchi}, \citenamefont {Schönenberger},\ and\
  \citenamefont {et~al.}}]{Fulop2021:arxiv2}%
  \BibitemOpen
  \bibfield  {author} {\bibinfo {author} {\bibfnamefont {Bálint}\ \bibnamefont
  {Fülöp}}, \bibinfo {author} {\bibfnamefont {Albin}\ \bibnamefont
  {Márffy}}, \bibinfo {author} {\bibfnamefont {Endre}\ \bibnamefont
  {Tóvári}}, \bibinfo {author} {\bibfnamefont {Máté}\ \bibnamefont
  {Kedves}}, \bibinfo {author} {\bibfnamefont {Simon}\ \bibnamefont
  {Zihlmann}}, \bibinfo {author} {\bibfnamefont {David}\ \bibnamefont
  {Indolese}}, \bibinfo {author} {\bibfnamefont {Zoltán}\ \bibnamefont
  {Kovács-Krausz}}, \bibinfo {author} {\bibfnamefont {Kenji}\ \bibnamefont
  {Watanabe}}, \bibinfo {author} {\bibfnamefont {Takashi}\ \bibnamefont
  {Taniguchi}}, \bibinfo {author} {\bibfnamefont {Christian}\ \bibnamefont
  {Schönenberger}}, \ and\ \bibinfo {author} {\bibnamefont {et~al.}},\
  }\bibfield  {title} {\enquote {\bibinfo {title} {New method of transport
  measurements on van der waals heterostructures under pressure},}\ }\href
  {\doibase 10.1063/5.0058583} {\bibfield  {journal} {\bibinfo  {journal}
  {Journal of Applied Physics}\ }\textbf {\bibinfo {volume} {130}},\ \bibinfo
  {pages} {064303} (\bibinfo {year} {2021}{\natexlab{b}})}\BibitemShut
  {NoStop}%
\bibitem [{\citenamefont {Naimer}\ \emph {et~al.}(2021)\citenamefont {Naimer},
  \citenamefont {Zollner}, \citenamefont {Gmitra},\ and\ \citenamefont
  {Fabian}}]{Naimer2021:arXiv}%
  \BibitemOpen
  \bibfield  {author} {\bibinfo {author} {\bibfnamefont {Thomas}\ \bibnamefont
  {Naimer}}, \bibinfo {author} {\bibfnamefont {Klaus}\ \bibnamefont {Zollner}},
  \bibinfo {author} {\bibfnamefont {Martin}\ \bibnamefont {Gmitra}}, \ and\
  \bibinfo {author} {\bibfnamefont {Jaroslav}\ \bibnamefont {Fabian}},\
  }\bibfield  {title} {\enquote {\bibinfo {title} {Twist-angle dependent
  proximity induced spin-orbit coupling in graphene/transition metal
  dichalcogenide heterostructures},}\ }\href {\doibase
  10.1103/PhysRevB.104.195156} {\bibfield  {journal} {\bibinfo  {journal}
  {Phys. Rev. B}\ }\textbf {\bibinfo {volume} {104}},\ \bibinfo {pages}
  {195156} (\bibinfo {year} {2021})}\BibitemShut {NoStop}%
\bibitem [{\citenamefont {Pezo}\ \emph {et~al.}(2021)\citenamefont {Pezo},
  \citenamefont {Zanolli}, \citenamefont {Wittemeier}, \citenamefont
  {Ordej{\'{o}}n}, \citenamefont {Fazzio}, \citenamefont {Roche},\ and\
  \citenamefont {Garcia}}]{Pezo2021:arXiv}%
  \BibitemOpen
  \bibfield  {author} {\bibinfo {author} {\bibfnamefont {Armando}\ \bibnamefont
  {Pezo}}, \bibinfo {author} {\bibfnamefont {Zeila}\ \bibnamefont {Zanolli}},
  \bibinfo {author} {\bibfnamefont {Nils}\ \bibnamefont {Wittemeier}}, \bibinfo
  {author} {\bibfnamefont {Pablo}\ \bibnamefont {Ordej{\'{o}}n}}, \bibinfo
  {author} {\bibfnamefont {Adalberto}\ \bibnamefont {Fazzio}}, \bibinfo
  {author} {\bibfnamefont {Stephan}\ \bibnamefont {Roche}}, \ and\ \bibinfo
  {author} {\bibfnamefont {Jose~H}\ \bibnamefont {Garcia}},\ }\bibfield
  {title} {\enquote {\bibinfo {title} {Manipulation of spin transport in
  graphene/transition metal dichalcogenide heterobilayers upon twisting},}\
  }\href {\doibase 10.1088/2053-1583/ac3378} {\bibfield  {journal} {\bibinfo
  {journal} {2D Materials}\ }\textbf {\bibinfo {volume} {9}},\ \bibinfo {pages}
  {015008} (\bibinfo {year} {2021})}\BibitemShut {NoStop}%
\bibitem [{\citenamefont {H\"ogl}\ \emph {et~al.}(2020)\citenamefont {H\"ogl},
  \citenamefont {Frank}, \citenamefont {Zollner}, \citenamefont {Kochan},
  \citenamefont {Gmitra},\ and\ \citenamefont {Fabian}}]{Hogl2020:PRL}%
  \BibitemOpen
  \bibfield  {author} {\bibinfo {author} {\bibfnamefont {Petra}\ \bibnamefont
  {H\"ogl}}, \bibinfo {author} {\bibfnamefont {Tobias}\ \bibnamefont {Frank}},
  \bibinfo {author} {\bibfnamefont {Klaus}\ \bibnamefont {Zollner}}, \bibinfo
  {author} {\bibfnamefont {Denis}\ \bibnamefont {Kochan}}, \bibinfo {author}
  {\bibfnamefont {Martin}\ \bibnamefont {Gmitra}}, \ and\ \bibinfo {author}
  {\bibfnamefont {Jaroslav}\ \bibnamefont {Fabian}},\ }\bibfield  {title}
  {\enquote {\bibinfo {title} {Quantum anomalous hall effects in graphene from
  proximity-induced uniform and staggered spin-orbit and exchange coupling},}\
  }\href {\doibase 10.1103/PhysRevLett.124.136403} {\bibfield  {journal}
  {\bibinfo  {journal} {Phys. Rev. Lett.}\ }\textbf {\bibinfo {volume} {124}},\
  \bibinfo {pages} {136403} (\bibinfo {year} {2020})}\BibitemShut {NoStop}%
\bibitem [{\citenamefont {Vila}\ \emph {et~al.}(2021)\citenamefont {Vila},
  \citenamefont {Garcia},\ and\ \citenamefont {Roche}}]{Vila2021:arxiv}%
  \BibitemOpen
  \bibfield  {author} {\bibinfo {author} {\bibfnamefont {Marc}\ \bibnamefont
  {Vila}}, \bibinfo {author} {\bibfnamefont {Jose~H.}\ \bibnamefont {Garcia}},
  \ and\ \bibinfo {author} {\bibfnamefont {Stephan}\ \bibnamefont {Roche}},\
  }\bibfield  {title} {\enquote {\bibinfo {title} {Valley-polarized quantum
  anomalous hall phase in bilayer graphene with layer-dependent proximity
  effects},}\ }\href {\doibase 10.1103/PhysRevB.104.L161113} {\bibfield
  {journal} {\bibinfo  {journal} {Phys. Rev. B}\ }\textbf {\bibinfo {volume}
  {104}},\ \bibinfo {pages} {L161113} (\bibinfo {year} {2021})}\BibitemShut
  {NoStop}%
\bibitem [{\citenamefont {Bahn}\ and\ \citenamefont {Jacobsen}(2002)}]{ASE}%
  \BibitemOpen
  \bibfield  {author} {\bibinfo {author} {\bibfnamefont {S.~R.}\ \bibnamefont
  {Bahn}}\ and\ \bibinfo {author} {\bibfnamefont {K.~W.}\ \bibnamefont
  {Jacobsen}},\ }\bibfield  {title} {\enquote {\bibinfo {title} {An
  object-oriented scripting interface to a legacy electronic structure code},}\
  }\href {\doibase 10.1109/5992.998641} {\bibfield  {journal} {\bibinfo
  {journal} {Comput. Sci. Eng.}\ }\textbf {\bibinfo {volume} {4}},\ \bibinfo
  {pages} {56} (\bibinfo {year} {2002})}\BibitemShut {NoStop}%
\bibitem [{\citenamefont {Lazic}(2015)}]{Lazic2015:CPC}%
  \BibitemOpen
  \bibfield  {author} {\bibinfo {author} {\bibfnamefont {Predrag}\ \bibnamefont
  {Lazic}},\ }\bibfield  {title} {\enquote {\bibinfo {title} {Cellmatch:
  Combining two unit cells into a common supercell with minimal strain},}\
  }\href {\doibase https://doi.org/10.1016/j.cpc.2015.08.038} {\bibfield
  {journal} {\bibinfo  {journal} {Computer Physics Communications}\ }\textbf
  {\bibinfo {volume} {197}},\ \bibinfo {pages} {324 -- 334} (\bibinfo {year}
  {2015})}\BibitemShut {NoStop}%
\bibitem [{\citenamefont {Koda}\ \emph {et~al.}(2016)\citenamefont {Koda},
  \citenamefont {Bechstedt}, \citenamefont {Marques},\ and\ \citenamefont
  {Teles}}]{Koda2016:JPCC}%
  \BibitemOpen
  \bibfield  {author} {\bibinfo {author} {\bibfnamefont {Daniel~S}\
  \bibnamefont {Koda}}, \bibinfo {author} {\bibfnamefont {Friedhelm}\
  \bibnamefont {Bechstedt}}, \bibinfo {author} {\bibfnamefont {Marcelo}\
  \bibnamefont {Marques}}, \ and\ \bibinfo {author} {\bibfnamefont {Lara~K}\
  \bibnamefont {Teles}},\ }\bibfield  {title} {\enquote {\bibinfo {title}
  {Coincidence lattices of 2d crystals: heterostructure predictions and
  applications},}\ }\href {\doibase 10.1021/acs.jpcc.6b01496} {\bibfield
  {journal} {\bibinfo  {journal} {The Journal of Physical Chemistry C}\
  }\textbf {\bibinfo {volume} {120}},\ \bibinfo {pages} {10895--10908}
  (\bibinfo {year} {2016})}\BibitemShut {NoStop}%
\bibitem [{\citenamefont {Baskin}\ and\ \citenamefont
  {Meyer}(1955)}]{Baskin1955:PR}%
  \BibitemOpen
  \bibfield  {author} {\bibinfo {author} {\bibfnamefont {Y.}~\bibnamefont
  {Baskin}}\ and\ \bibinfo {author} {\bibfnamefont {L.}~\bibnamefont {Meyer}},\
  }\bibfield  {title} {\enquote {\bibinfo {title} {Lattice constants of
  graphite at low temperatures},}\ }\href {\doibase 10.1103/PhysRev.100.544}
  {\bibfield  {journal} {\bibinfo  {journal} {Phys. Rev.}\ }\textbf {\bibinfo
  {volume} {100}},\ \bibinfo {pages} {544} (\bibinfo {year}
  {1955})}\BibitemShut {NoStop}%
\bibitem [{\citenamefont {Carteaux}\ \emph {et~al.}(1995)\citenamefont
  {Carteaux}, \citenamefont {Brunet}, \citenamefont {Ouvrard},\ and\
  \citenamefont {Andre}}]{Carteaux1995:JPCM}%
  \BibitemOpen
  \bibfield  {author} {\bibinfo {author} {\bibfnamefont {V}~\bibnamefont
  {Carteaux}}, \bibinfo {author} {\bibfnamefont {D}~\bibnamefont {Brunet}},
  \bibinfo {author} {\bibfnamefont {G}~\bibnamefont {Ouvrard}}, \ and\ \bibinfo
  {author} {\bibfnamefont {G}~\bibnamefont {Andre}},\ }\bibfield  {title}
  {\enquote {\bibinfo {title} {{Crystallographic, magnetic and electronic
  structures of a new layered ferromagnetic compound Cr2Ge2Te6}},}\ }\href
  {\doibase 10.1088/0953-8984/7/1/008} {\bibfield  {journal} {\bibinfo
  {journal} {J. Phys.: Condens. Mat.}\ }\textbf {\bibinfo {volume} {7}},\
  \bibinfo {pages} {69} (\bibinfo {year} {1995})}\BibitemShut {NoStop}%
\bibitem [{\citenamefont {Hohenberg}\ and\ \citenamefont
  {Kohn}(1964)}]{Hohenberg1964:PRB}%
  \BibitemOpen
  \bibfield  {author} {\bibinfo {author} {\bibfnamefont {P.}~\bibnamefont
  {Hohenberg}}\ and\ \bibinfo {author} {\bibfnamefont {W.}~\bibnamefont
  {Kohn}},\ }\bibfield  {title} {\enquote {\bibinfo {title} {Inhomogeneous
  electron gas},}\ }\href {\doibase 10.1103/PhysRev.136.B864} {\bibfield
  {journal} {\bibinfo  {journal} {Phys. Rev.}\ }\textbf {\bibinfo {volume}
  {136}},\ \bibinfo {pages} {B864} (\bibinfo {year} {1964})}\BibitemShut
  {NoStop}%
\bibitem [{\citenamefont {Giannozzi}\ and\ \citenamefont
  {et~al.}(2009)}]{Giannozzi2009:JPCM}%
  \BibitemOpen
  \bibfield  {author} {\bibinfo {author} {\bibfnamefont {Paolo}\ \bibnamefont
  {Giannozzi}}\ and\ \bibinfo {author} {\bibnamefont {et~al.}},\ }\bibfield
  {title} {\enquote {\bibinfo {title} {Quantum espresso: a modular and
  open-source software project for quantum simulations of materials},}\
  }\href@noop {} {\bibfield  {journal} {\bibinfo  {journal} {J. Phys.: Cond.
  Mat.}\ }\textbf {\bibinfo {volume} {21}},\ \bibinfo {pages} {395502}
  (\bibinfo {year} {2009})}\BibitemShut {NoStop}%
\bibitem [{\citenamefont {Gong}\ \emph {et~al.}(2017)\citenamefont {Gong},
  \citenamefont {Li}, \citenamefont {Li}, \citenamefont {Ji}, \citenamefont
  {Stern}, \citenamefont {Xia}, \citenamefont {Cao}, \citenamefont {Bao},
  \citenamefont {Wang}, \citenamefont {Wang}, \citenamefont {Qiu},
  \citenamefont {Cava}, \citenamefont {Louie}, \citenamefont {Xia},\ and\
  \citenamefont {Zhang}}]{Gong2017:Nat}%
  \BibitemOpen
  \bibfield  {author} {\bibinfo {author} {\bibfnamefont {Cheng}\ \bibnamefont
  {Gong}}, \bibinfo {author} {\bibfnamefont {Lin}\ \bibnamefont {Li}}, \bibinfo
  {author} {\bibfnamefont {Zhenglu}\ \bibnamefont {Li}}, \bibinfo {author}
  {\bibfnamefont {Huiwen}\ \bibnamefont {Ji}}, \bibinfo {author} {\bibfnamefont
  {Alex}\ \bibnamefont {Stern}}, \bibinfo {author} {\bibfnamefont {Yang}\
  \bibnamefont {Xia}}, \bibinfo {author} {\bibfnamefont {Ting}\ \bibnamefont
  {Cao}}, \bibinfo {author} {\bibfnamefont {Wei}\ \bibnamefont {Bao}}, \bibinfo
  {author} {\bibfnamefont {Chenzhe}\ \bibnamefont {Wang}}, \bibinfo {author}
  {\bibfnamefont {Yuan}\ \bibnamefont {Wang}}, \bibinfo {author} {\bibfnamefont
  {Z.~Q.}\ \bibnamefont {Qiu}}, \bibinfo {author} {\bibfnamefont {R.~J.}\
  \bibnamefont {Cava}}, \bibinfo {author} {\bibfnamefont {Steven~G.}\
  \bibnamefont {Louie}}, \bibinfo {author} {\bibfnamefont {Jing}\ \bibnamefont
  {Xia}}, \ and\ \bibinfo {author} {\bibfnamefont {Xiang}\ \bibnamefont
  {Zhang}},\ }\bibfield  {title} {\enquote {\bibinfo {title} {{Discovery of
  intrinsic ferromagnetism in two-dimensional van der Waals crystals}},}\
  }\href {\doibase 10.1038/nature22060} {\bibfield  {journal} {\bibinfo
  {journal} {Nature}\ }\textbf {\bibinfo {volume} {546}},\ \bibinfo {pages}
  {265} (\bibinfo {year} {2017})}\BibitemShut {NoStop}%
\bibitem [{\citenamefont {Kresse}\ and\ \citenamefont
  {Joubert}(1999)}]{Kresse1999:PRB}%
  \BibitemOpen
  \bibfield  {author} {\bibinfo {author} {\bibfnamefont {G.}~\bibnamefont
  {Kresse}}\ and\ \bibinfo {author} {\bibfnamefont {D.}~\bibnamefont
  {Joubert}},\ }\bibfield  {title} {\enquote {\bibinfo {title} {From ultrasoft
  pseudopotentials to the projector augmented-wave method},}\ }\href {\doibase
  10.1103/PhysRevB.59.1758} {\bibfield  {journal} {\bibinfo  {journal} {Phys.
  Rev. B}\ }\textbf {\bibinfo {volume} {59}},\ \bibinfo {pages} {1758}
  (\bibinfo {year} {1999})}\BibitemShut {NoStop}%
\bibitem [{\citenamefont {Perdew}\ \emph {et~al.}(1996)\citenamefont {Perdew},
  \citenamefont {Burke},\ and\ \citenamefont {Ernzerhof}}]{Perdew1996:PRL}%
  \BibitemOpen
  \bibfield  {author} {\bibinfo {author} {\bibfnamefont {John~P.}\ \bibnamefont
  {Perdew}}, \bibinfo {author} {\bibfnamefont {Kieron}\ \bibnamefont {Burke}},
  \ and\ \bibinfo {author} {\bibfnamefont {Matthias}\ \bibnamefont
  {Ernzerhof}},\ }\bibfield  {title} {\enquote {\bibinfo {title} {Generalized
  gradient approximation made simple},}\ }\href {\doibase
  10.1103/PhysRevLett.77.3865} {\bibfield  {journal} {\bibinfo  {journal}
  {Phys. Rev. Lett.}\ }\textbf {\bibinfo {volume} {77}},\ \bibinfo {pages}
  {3865} (\bibinfo {year} {1996})}\BibitemShut {NoStop}%
\bibitem [{\citenamefont {Grimme}(2006)}]{Grimme2006:JCC}%
  \BibitemOpen
  \bibfield  {author} {\bibinfo {author} {\bibfnamefont {Stefan}\ \bibnamefont
  {Grimme}},\ }\bibfield  {title} {\enquote {\bibinfo {title} {Semiempirical
  gga-type density functional constructed with a long-range dispersion
  correction},}\ }\href {\doibase 10.1002/jcc.20495} {\bibfield  {journal}
  {\bibinfo  {journal} {J. Comput. Chem.}\ }\textbf {\bibinfo {volume} {27}},\
  \bibinfo {pages} {1787} (\bibinfo {year} {2006})}\BibitemShut {NoStop}%
\bibitem [{\citenamefont {Grimme}\ \emph {et~al.}(2010)\citenamefont {Grimme},
  \citenamefont {Antony}, \citenamefont {Ehrlich},\ and\ \citenamefont
  {Krieg}}]{Grimme2010:JCP}%
  \BibitemOpen
  \bibfield  {author} {\bibinfo {author} {\bibfnamefont {Stefan}\ \bibnamefont
  {Grimme}}, \bibinfo {author} {\bibfnamefont {Jens}\ \bibnamefont {Antony}},
  \bibinfo {author} {\bibfnamefont {Stephan}\ \bibnamefont {Ehrlich}}, \ and\
  \bibinfo {author} {\bibfnamefont {Helge}\ \bibnamefont {Krieg}},\ }\bibfield
  {title} {\enquote {\bibinfo {title} {{A consistent and accurate ab initio
  parametrization of density functional dispersion correction (DFT-D) for the
  94 elements H-Pu}},}\ }\href {\doibase 10.1063/1.3382344} {\bibfield
  {journal} {\bibinfo  {journal} {J. Chem. Phys.}\ }\textbf {\bibinfo {volume}
  {132}},\ \bibinfo {pages} {154104} (\bibinfo {year} {2010})}\BibitemShut
  {NoStop}%
\bibitem [{\citenamefont {Barone}\ \emph {et~al.}(2009)\citenamefont {Barone},
  \citenamefont {Casarin}, \citenamefont {Forrer}, \citenamefont {Pavone},
  \citenamefont {Sambi},\ and\ \citenamefont {Vittadini}}]{Barone2009:JCC}%
  \BibitemOpen
  \bibfield  {author} {\bibinfo {author} {\bibfnamefont {Vincenzo}\
  \bibnamefont {Barone}}, \bibinfo {author} {\bibfnamefont {Maurizio}\
  \bibnamefont {Casarin}}, \bibinfo {author} {\bibfnamefont {Daniel}\
  \bibnamefont {Forrer}}, \bibinfo {author} {\bibfnamefont {Michele}\
  \bibnamefont {Pavone}}, \bibinfo {author} {\bibfnamefont {Mauro}\
  \bibnamefont {Sambi}}, \ and\ \bibinfo {author} {\bibfnamefont {Andrea}\
  \bibnamefont {Vittadini}},\ }\bibfield  {title} {\enquote {\bibinfo {title}
  {Role and effective treatment of dispersive forces in materials: Polyethylene
  and graphite crystals as test cases},}\ }\href {\doibase 10.1002/jcc.21112}
  {\bibfield  {journal} {\bibinfo  {journal} {J. Comput. Chem.}\ }\textbf
  {\bibinfo {volume} {30}},\ \bibinfo {pages} {934} (\bibinfo {year}
  {2009})}\BibitemShut {NoStop}%
\bibitem [{\citenamefont {Bengtsson}(1999)}]{Bengtsson1999:PRB}%
  \BibitemOpen
  \bibfield  {author} {\bibinfo {author} {\bibfnamefont {Lennart}\ \bibnamefont
  {Bengtsson}},\ }\bibfield  {title} {\enquote {\bibinfo {title} {Dipole
  correction for surface supercell calculations},}\ }\href {\doibase
  10.1103/PhysRevB.59.12301} {\bibfield  {journal} {\bibinfo  {journal} {Phys.
  Rev. B}\ }\textbf {\bibinfo {volume} {59}},\ \bibinfo {pages} {12301}
  (\bibinfo {year} {1999})}\BibitemShut {NoStop}%
\bibitem [{\citenamefont {Li}\ and\ \citenamefont {Yang}(2014)}]{Li2014:JMCC}%
  \BibitemOpen
  \bibfield  {author} {\bibinfo {author} {\bibfnamefont {Xingxing}\
  \bibnamefont {Li}}\ and\ \bibinfo {author} {\bibfnamefont {Jinlong}\
  \bibnamefont {Yang}},\ }\bibfield  {title} {\enquote {\bibinfo {title}
  {{CrXTe 3 (X = Si, Ge) nanosheets: two dimensional intrinsic ferromagnetic
  semiconductors}},}\ }\href {\doibase 10.1039/C4TC01193G} {\bibfield
  {journal} {\bibinfo  {journal} {J. Mater. Chem. C}\ }\textbf {\bibinfo
  {volume} {2}},\ \bibinfo {pages} {7071} (\bibinfo {year} {2014})}\BibitemShut
  {NoStop}%
\bibitem [{\citenamefont {Chen}\ \emph {et~al.}(2015)\citenamefont {Chen},
  \citenamefont {Qi},\ and\ \citenamefont {Shi}}]{Chen2015:PRB}%
  \BibitemOpen
  \bibfield  {author} {\bibinfo {author} {\bibfnamefont {Xiaofang}\
  \bibnamefont {Chen}}, \bibinfo {author} {\bibfnamefont {Jingshan}\
  \bibnamefont {Qi}}, \ and\ \bibinfo {author} {\bibfnamefont {Daning}\
  \bibnamefont {Shi}},\ }\bibfield  {title} {\enquote {\bibinfo {title}
  {Strain-engineering of magnetic coupling in two-dimensional magnetic
  semiconductor crsite3: Competition of direct exchange interaction and
  superexchange interaction},}\ }\href {\doibase
  https://doi.org/10.1016/j.physleta.2014.10.042} {\bibfield  {journal}
  {\bibinfo  {journal} {Phys. Letters A}\ }\textbf {\bibinfo {volume} {379}},\
  \bibinfo {pages} {60} (\bibinfo {year} {2015})}\BibitemShut {NoStop}%
\bibitem [{\citenamefont {Liu}\ \emph {et~al.}(2021)\citenamefont {Liu},
  \citenamefont {Hu}, \citenamefont {Wang}, \citenamefont {Krasheninnikov},
  \citenamefont {Chen},\ and\ \citenamefont {Sun}}]{Liu2021:N}%
  \BibitemOpen
  \bibfield  {author} {\bibinfo {author} {\bibfnamefont {Lifei}\ \bibnamefont
  {Liu}}, \bibinfo {author} {\bibfnamefont {Xiaohui}\ \bibnamefont {Hu}},
  \bibinfo {author} {\bibfnamefont {Yifeng}\ \bibnamefont {Wang}}, \bibinfo
  {author} {\bibfnamefont {Arkady~V}\ \bibnamefont {Krasheninnikov}}, \bibinfo
  {author} {\bibfnamefont {Zhongfang}\ \bibnamefont {Chen}}, \ and\ \bibinfo
  {author} {\bibfnamefont {Litao}\ \bibnamefont {Sun}},\ }\bibfield  {title}
  {\enquote {\bibinfo {title} {Tunable electronic properties and enhanced
  ferromagnetism in cr2ge2te6 monolayer by strain engineering},}\ }\href
  {\doibase 10.1088/1361-6528/ac1a94} {\bibfield  {journal} {\bibinfo
  {journal} {Nanotechnology}\ }\textbf {\bibinfo {volume} {32}},\ \bibinfo
  {pages} {485408} (\bibinfo {year} {2021})}\BibitemShut {NoStop}%
\bibitem [{\citenamefont {Ghiasi}\ \emph {et~al.}(2021)\citenamefont {Ghiasi},
  \citenamefont {Kaverzin}, \citenamefont {Dismukes}, \citenamefont {de~Wal},
  \citenamefont {Roy},\ and\ \citenamefont {van Wees}}]{Ghiasi2021:NN}%
  \BibitemOpen
  \bibfield  {author} {\bibinfo {author} {\bibfnamefont {Talieh~S.}\
  \bibnamefont {Ghiasi}}, \bibinfo {author} {\bibfnamefont {Alexey~A.}\
  \bibnamefont {Kaverzin}}, \bibinfo {author} {\bibfnamefont {Avalon~H.}\
  \bibnamefont {Dismukes}}, \bibinfo {author} {\bibfnamefont {Dennis~K.}\
  \bibnamefont {de~Wal}}, \bibinfo {author} {\bibfnamefont {Xavier}\
  \bibnamefont {Roy}}, \ and\ \bibinfo {author} {\bibfnamefont {Bart~J.}\
  \bibnamefont {van Wees}},\ }\bibfield  {title} {\enquote {\bibinfo {title}
  {Electrical and thermal generation of spin currents by magnetic bilayer
  graphene},}\ }\href {\doibase 10.1038/s41565-021-00887-3} {\bibfield
  {journal} {\bibinfo  {journal} {Nature Nanotechnology}\ }\textbf {\bibinfo
  {volume} {16}},\ \bibinfo {pages} {788--794} (\bibinfo {year}
  {2021})}\BibitemShut {NoStop}%
\bibitem [{\citenamefont {Rosenberger}\ \emph {et~al.}(2020)\citenamefont
  {Rosenberger}, \citenamefont {Chuang}, \citenamefont {Phillips},
  \citenamefont {Oleshko}, \citenamefont {McCreary}, \citenamefont {Sivaram},
  \citenamefont {Hellberg},\ and\ \citenamefont
  {Jonker}}]{Rosenberger2020:ACS}%
  \BibitemOpen
  \bibfield  {author} {\bibinfo {author} {\bibfnamefont {Matthew~R}\
  \bibnamefont {Rosenberger}}, \bibinfo {author} {\bibfnamefont {Hsun-Jen}\
  \bibnamefont {Chuang}}, \bibinfo {author} {\bibfnamefont {Madeleine}\
  \bibnamefont {Phillips}}, \bibinfo {author} {\bibfnamefont {Vladimir~P}\
  \bibnamefont {Oleshko}}, \bibinfo {author} {\bibfnamefont {Kathleen~M}\
  \bibnamefont {McCreary}}, \bibinfo {author} {\bibfnamefont {Saujan~V}\
  \bibnamefont {Sivaram}}, \bibinfo {author} {\bibfnamefont {C~Stephen}\
  \bibnamefont {Hellberg}}, \ and\ \bibinfo {author} {\bibfnamefont {Berend~T}\
  \bibnamefont {Jonker}},\ }\bibfield  {title} {\enquote {\bibinfo {title}
  {Twist angle-dependent atomic reconstruction and moir{\'e} patterns in
  transition metal dichalcogenide heterostructures},}\ }\href {\doibase
  https://doi.org/10.1021/acsnano.0c00088} {\bibfield  {journal} {\bibinfo
  {journal} {ACS nano}\ }\textbf {\bibinfo {volume} {14}},\ \bibinfo {pages}
  {4550--4558} (\bibinfo {year} {2020})}\BibitemShut {NoStop}%
\bibitem [{\citenamefont {Weston}\ \emph {et~al.}(2020)\citenamefont {Weston},
  \citenamefont {Zou}, \citenamefont {Enaldiev}, \citenamefont {Summerfield},
  \citenamefont {Clark}, \citenamefont {Z{\'o}lyomi}, \citenamefont {Graham},
  \citenamefont {Yelgel}, \citenamefont {Magorrian}, \citenamefont {Zhou} \emph
  {et~al.}}]{Weston2020:NN}%
  \BibitemOpen
  \bibfield  {author} {\bibinfo {author} {\bibfnamefont {Astrid}\ \bibnamefont
  {Weston}}, \bibinfo {author} {\bibfnamefont {Yichao}\ \bibnamefont {Zou}},
  \bibinfo {author} {\bibfnamefont {Vladimir}\ \bibnamefont {Enaldiev}},
  \bibinfo {author} {\bibfnamefont {Alex}\ \bibnamefont {Summerfield}},
  \bibinfo {author} {\bibfnamefont {Nicholas}\ \bibnamefont {Clark}}, \bibinfo
  {author} {\bibfnamefont {Viktor}\ \bibnamefont {Z{\'o}lyomi}}, \bibinfo
  {author} {\bibfnamefont {Abigail}\ \bibnamefont {Graham}}, \bibinfo {author}
  {\bibfnamefont {Celal}\ \bibnamefont {Yelgel}}, \bibinfo {author}
  {\bibfnamefont {Samuel}\ \bibnamefont {Magorrian}}, \bibinfo {author}
  {\bibfnamefont {Mingwei}\ \bibnamefont {Zhou}},  \emph {et~al.},\ }\bibfield
  {title} {\enquote {\bibinfo {title} {Atomic reconstruction in twisted
  bilayers of transition metal dichalcogenides},}\ }\href {\doibase
  https://doi.org/10.1038/s41565-020-0682-9} {\bibfield  {journal} {\bibinfo
  {journal} {Nature Nanotechnology}\ }\textbf {\bibinfo {volume} {15}},\
  \bibinfo {pages} {592--597} (\bibinfo {year} {2020})}\BibitemShut {NoStop}%
\bibitem [{\citenamefont {Pandey}\ \emph {et~al.}(2016)\citenamefont {Pandey},
  \citenamefont {Nayak}, \citenamefont {Liu}, \citenamefont {Moran},
  \citenamefont {Kim}, \citenamefont {Li}, \citenamefont {Lin}, \citenamefont
  {Akinwande},\ and\ \citenamefont {Singh}}]{Tribhuwan2016:S}%
  \BibitemOpen
  \bibfield  {author} {\bibinfo {author} {\bibfnamefont {Tribhuwan}\
  \bibnamefont {Pandey}}, \bibinfo {author} {\bibfnamefont {Avinash~P.}\
  \bibnamefont {Nayak}}, \bibinfo {author} {\bibfnamefont {Jin}\ \bibnamefont
  {Liu}}, \bibinfo {author} {\bibfnamefont {Samuel~T.}\ \bibnamefont {Moran}},
  \bibinfo {author} {\bibfnamefont {Joon-Seok}\ \bibnamefont {Kim}}, \bibinfo
  {author} {\bibfnamefont {Lain-Jong}\ \bibnamefont {Li}}, \bibinfo {author}
  {\bibfnamefont {Jung-Fu}\ \bibnamefont {Lin}}, \bibinfo {author}
  {\bibfnamefont {Deji}\ \bibnamefont {Akinwande}}, \ and\ \bibinfo {author}
  {\bibfnamefont {Abhishek~K.}\ \bibnamefont {Singh}},\ }\bibfield  {title}
  {\enquote {\bibinfo {title} {Pressure-induced charge transfer doping of
  monolayer graphene/mos2 heterostructure},}\ }\href {\doibase
  https://doi.org/10.1002/smll.201600808} {\bibfield  {journal} {\bibinfo
  {journal} {Small}\ }\textbf {\bibinfo {volume} {12}},\ \bibinfo {pages}
  {4063--4069} (\bibinfo {year} {2016})}\BibitemShut {NoStop}%
\bibitem [{\citenamefont {Zollner}\ and\ \citenamefont
  {Fabian}(2021{\natexlab{b}})}]{Zollner2021:arxiv}%
  \BibitemOpen
  \bibfield  {author} {\bibinfo {author} {\bibfnamefont {Klaus}\ \bibnamefont
  {Zollner}}\ and\ \bibinfo {author} {\bibfnamefont {Jaroslav}\ \bibnamefont
  {Fabian}},\ }\bibfield  {title} {\enquote {\bibinfo {title} {Bilayer graphene
  encapsulated within monolayers of ${\mathrm{ws}}_{2}$ or
  ${\mathrm{cr}}_{2}{\mathrm{ge}}_{2}{\mathrm{te}}_{6}$: Tunable proximity
  spin-orbit or exchange coupling},}\ }\href {\doibase
  10.1103/PhysRevB.104.075126} {\bibfield  {journal} {\bibinfo  {journal}
  {Phys. Rev. B}\ }\textbf {\bibinfo {volume} {104}},\ \bibinfo {pages}
  {075126} (\bibinfo {year} {2021}{\natexlab{b}})}\BibitemShut {NoStop}%
\end{thebibliography}%

\cleardoublepage


\section*{Supplementary material}

# Engineering proximity exchange by twisting: Reversal of ferromagnetic and emergence of antiferromagnetic Dirac bands in graphene/Cr₂Ge₂Te₆

Klaus Zollner[1] and Jaroslav Fabian[1]

[1]*Institute for Theoretical Physics, University of Regensburg, 93040 Regensburg, Germany*

In the Supplemental Material, we give a more extended summary of our results from the main text. First, we give a description of the computational details, the geometry setup, and structural information. Next, we show the band structure and fit results for three most relevant twist angles, 0, 19.1, and 30 degrees. For all angles, we summarize the fit parameters in tabular form. After that, we discuss the influence of in-plane strain on the results. For this second set of structures, where Cr₂Ge₂Te₆ (CGT) is unstrained, we also summarize the structural information, fitted parameters, and show band structures and detailed fits. For the electric field study, we summarize all fit parameters in tabular form, and plot the two sublattice-resolved proximity exchange parameters as function of twist angle and gate field. Similarly, the results for laterally and vertically shifting the graphene with respect to the CGT are also summarized in tabular form. In addition, we briefly discuss CGT encapsulated heterostructures and how one can further tailor proximity exchange in graphene. Finally, we extensively discuss the origin of the different proximity exchange couplings, from the calculated density of states, spin polarizations, projected band structures, as well as from the backfolding of the Dirac states into the CGT Brillouin zone, for different twist angles.

## I. COMPUTATIONAL DETAILS AND GEOMETRY

The heterostructures of graphene and CGT, where we consider several twist angles between the two monolayers, are set-up with the `atomic simulation environment (ASE)` [1] and the `CellMatch` code [2], implementing the Coincidence lattice method [3, 4]. Monolayers of graphene and CGT are based on hexagonal unit cells, with lattice constants of $a = 2.46$ Å [5] and $6.8275$ Å [6], which need to be strained in the twisted heterostructures, in order to form commensurate supercells for periodic density functional theory (DFT) calculations. In Table S1 we summarize the main structural information for the twist angles we consider. In total, we investigate 10 different angles between 0° and 30°. Especially these angles are suitable for DFT calculations, since strain applied to the monolayers is at maximum 5% and the number of atoms in the supercells is still manageable. Otherwise, also other angles could be investigated, but beyond reasonable strain limits and above 300 atoms in the structure.

TABLE S1. Structural information for the graphene/CGT heterostructures. We list the twist angle between the layers, the number of atoms (NoA) in the heterostructure supercell, the number $n_k$ for the $k$-point sampling, the lattice constants of graphene and CGT, the calculated dipole of the structure, the relaxed interlayer distance $d_{int}$, the rippling of the graphene layer $\Delta z_{grp}$, and the averaged calculated magnetic moments of Cr, Ge, Te, and C atoms.

| twist angle [°] | NoA | $n_k$ | $a_{grp}$ [Å] | $a_{CGT}$ [Å] | dipole [debye] | $d_{int}$ [Å] | $\Delta z_{grp}$ [pm] | Cr [$\mu_B$] | Ge [$\mu_B$] | Te [$\mu_B$] | C [$10^{-3}\mu_B$] |
|---|---|---|---|---|---|---|---|---|---|---|---|
| 0.0000 | 218 | 9 | 2.52 | 6.718 | -2.5453 | 3.5685 | 0.554 | 3.110 | 0.0213 | -0.0817 | -0.247 |
| 3.0045 | 302 | 6 | 2.48 | 6.826 | -4.4864 | 3.5303 | 0.545 | 3.139 | 0.0247 | -0.0870 | -0.132 |
| 5.8175 | 236 | 9 | 2.45 | 6.974 | -3.5160 | 3.5575 | 0.400 | 3.191 | 0.0282 | -0.0967 | -0.137 |
| 8.9483 | 102 | 24 | 2.46 | 6.848 | -1.5072 | 3.5473 | 0.613 | 3.148 | 0.0251 | -0.0886 | -0.200 |
| 12.2163 | 224 | 9 | 2.50 | 6.821 | -3.1757 | 3.5619 | 0.561 | 3.140 | 0.0244 | -0.0869 | -0.214 |
| 14.7047 | 314 | 6 | 2.45 | 6.963 | -4.9073 | 3.5435 | 0.838 | 3.186 | 0.0279 | -0.0957 | -0.279 |
| 19.1066 | 24 | 42 | 2.52 | 6.663 | -0.2338 | 3.5735 | 0.392 | 3.097 | 0.0194 | -0.0797 | -0.200 |
| 23.4132 | 242 | 9 | 2.45 | 7.108 | -4.0642 | 3.5372 | 0.653 | 3.236 | 0.0310 | -0.1058 | -0.240 |
| 26.9955 | 174 | 12 | 2.50 | 6.814 | -2.5319 | 3.5632 | 0.654 | 3.138 | 0.0241 | -0.0864 | -0.236 |
| 30.0000 | 80 | 30 | 2.45 | 7.073 | -1.3044 | 3.5450 | 0.918 | 3.224 | 0.0301 | -0.1033 | -0.184 |

The electronic structure calculations and structural relaxation of graphene/CGT heterostructures are performed by DFT [7] with `Quantum ESPRESSO` [8]. Self-consistent calculations are carried out with a $k$-point sampling of $n_k \times n_k \times 1$. The number $n_k$ is listed in Table S1 for all twist angles and depends on the number of atoms in the heterostructure. In addition, $n_k$ is limited by our computational power. Nevertheless, for large supercells the heterostructure Brillouin Zone is small and only few $k$-points are necessary to get converged results.

We perform open shell calculations that provide the spin-polarized ground state of the CGT monolayer. A Hubbard parameter of $U = 1.0$ eV is used for Cr $d$-orbitals, being in the range of proposed $U$ values for CGT [9]. We use

an energy cutoff for charge density of 520 Ry and the kinetic energy cutoff for wavefunctions is 65 Ry for the scalar relativistic pseudopotential with the projector augmented wave method [10] with the Perdew-Burke-Ernzerhof exchange correlation functional [11]. Spin-orbit coupling (SOC) is neglected, since we are mainly interested in the twist-angle dependent proximity-induced exchange coupling. Moreover, recent calculations have already shown that proximity-induced SOC in graphene, originating from the CGT, is small compared to the exchange coupling [12–14]. For the relaxation of the heterostructures, we add van der Waals (vdW) corrections [15–17] and use quasi-Newton algorithm based on trust radius procedure. Dipole corrections [18] are also included to get correct band offsets and internal electric fields. In order to simulate quasi-2D systems, we add a vacuum of about 24 Å to avoid interactions between periodic images in our slab geometry. To determine the interlayer distances, the atoms of graphene are allowed to relax only along $z$-direction (vertical to the layers) and the atoms of CGT are allowed to move in all directions, until every component of each force is reduced below $2 \times 10^{-4}$ [Ry/$a_0$], where $a_0$ is the Bohr radius.

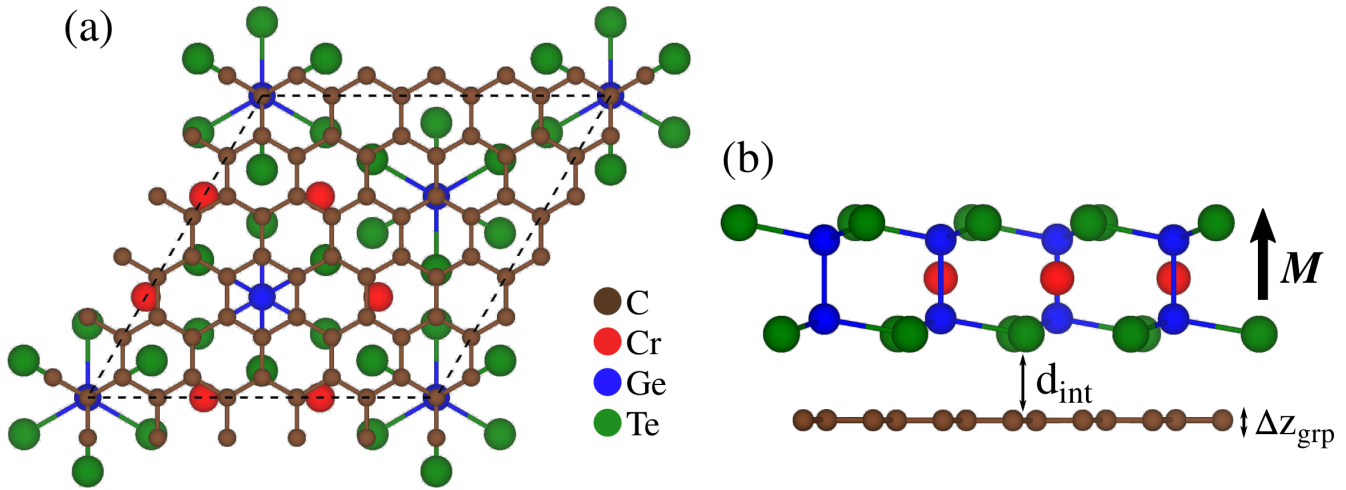

FIG. S1.   (a,b) Bottom and side view of the graphene/CGT heterostructure with a twist angle of 30°. Dashed line in (a) defines the heterostructure unit cell and in (b) we define the interlayer distance $d_{int}$ and the rippling of the graphene layer $\Delta z_{grp}$. Different colors correspond to different atom types. In all our heterostructures, the CGT layer (with magnetization $\boldsymbol{M}$ along $z$ direction) is above graphene.

After relaxation of the graphene/CGT heterostructures, we calculate the mean interlayer distances, $d_{int}$, and the standard deviations, $\Delta z_{grp}$, from the $z$ coordinates of the C atoms of graphene. The standard deviations represent the amount of rippling of graphene. The results are summarized in Table S1. The interlayer distance is nearly independent of the twist angle and is about 3.55 Å. The graphene itself stays nearly flat, as the rippling is always below 1 pm. In Fig. S1, we show an exemplary case of the graphene/CGT heterostructure for a twist angle of 30°. All our heterostructures have the CGT layer above graphene, with the CGT magnetization $\boldsymbol{M}$ along $z$ direction, specifying the spin quantization axis (spin up = $z$, spin down = $-z$). When we apply the transverse electric field (modeled by a zigzag potential), a positive field also points along $z$ direction, from graphene to CGT (see also Fig. 1 in the main text).

The averaged calculated magnetic moments for all atom types and all twist angles are also summarized in Table S1. The magnetic moments of CGT are directly proportional to its lattice constant, as we find. In addition, the magnetic moments of the C atoms already give us a first hint that graphene experiences some proximity-induced magnetism from the CGT layer. For all twist angles, the C atoms have a small magnetic moment parallel to the Te atoms, but antiparallel to Cr and Ge atoms. However, we do not notice an obvious correspondence between the induced magnetism in graphene and the CGT magnetic moments. Here, the atomic registry should play a major role, i. e., even though the layers are twisted with respect to each other, at the moment we consider only a specific stacking configuration, with a certain amount of strain applied to the monolayers.

## II. BAND STRUCTURE AND FIT RESULTS

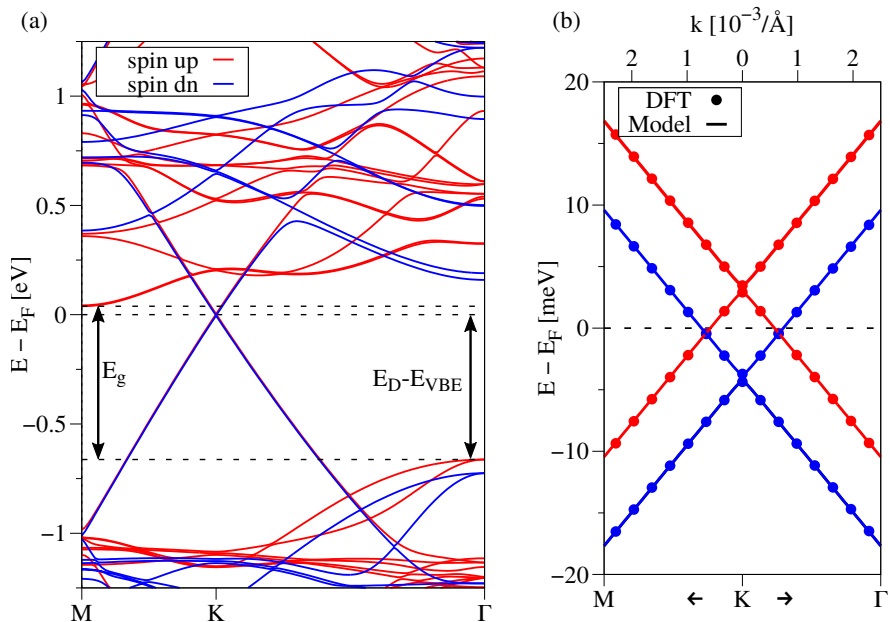

FIG. S2. (a) DFT-calculated band structure of the graphene/CGT heterostructure along the high-symmetry path M-K-$\Gamma$ for a twist angle of 30°. Red (blue) solid lines correspond to spin up (spin down). We also define the CGT band gap, $E_g$, and the position of the Dirac point with respect to the valence band edge of CGT at the $\Gamma$ point, $E_D - E_{VBE}$. (b) Zoom to the DFT-calculated (symbols) low energy Dirac bands near the K point with a fit to the model Hamiltonian (solid lines).

TABLE S2. Fit parameters of Hamiltonian $\mathcal{H}$ for the graphene/CGT heterostructures for different twist angles. We summarize the Fermi velocity $v_F$, the staggered potential $\Delta$, proximity exchange parameters $\lambda_{ex}^A$ and $\lambda_{ex}^B$, the Dirac point energy with respect to the Fermi level $E_D$, the band gap of CGT $E_g$, and the position of the Dirac point with respect to the valence band edge of CGT at the $\Gamma$ point $E_D - E_{VBE}$.

| twist angle [°] | $v_F$ [$10^5$ m/s] | $\Delta$ [meV] | $\lambda_{ex}^A$ [meV] | $\lambda_{ex}^B$ [meV] | $E_D$ [meV] | $E_g$ [eV] | $E_D - E_{VBE}$ [eV] |
|---|---|---|---|---|---|---|---|
| 0.0000 | 7.8555 | 0.000 | 4.201 | 4.199 | -71.904 | 0.300 | -0.080 |
| 3.0045 | 7.9972 | 0.054 | 4.079 | 4.116 | -0.323 | 0.483 | 0.203 |
| 5.8175 | 8.0349 | 1.584 | 2.402 | 2.297 | -0.520 | 0.629 | 0.554 |
| 8.9483 | 8.1209 | 0.048 | 1.181 | 1.133 | -0.191 | 0.519 | 0.299 |
| 12.2163 | 7.9355 | 0.062 | 0.089 | 0.104 | -0.054 | 0.474 | 0.129 |
| 14.7047 | 8.1996 | 0.172 | -3.146 | -3.167 | -0.029 | 0.632 | 0.513 |
| 19.1066 | 7.8378 | 5.450 | -2.050 | 1.331 | -143.350 | 0.197 | -0.168 |
| 23.4132 | 8.2224 | 0.128 | -4.289 | -4.200 | 1.669 | 0.717 | 0.716 |
| 26.9955 | 7.9568 | 0.149 | -4.084 | -3.957 | 0.349 | 0.460 | 0.095 |
| 30.0000 | 8.2856 | 0.298 | -3.630 | -3.596 | -0.428 | 0.705 | 0.662 |

In Fig. S2, we show the results for the graphene/CGT heterostructure for a twist angle of 30°, corresponding to the supercell shown in Fig. S1. In agreement with recent calculations [12, 13], we find the Dirac cone located at the Fermi level and close to the conduction band edge of the CGT at the M point. The band gap of the CGT is about 0.7 eV and indirect ($\Gamma \to M$), consistent with the tensile-strained lattice constant in the 30° case and the strain-tunable band gap of CGT [19–21]. However, for different twist angles the situation can be rather different. Of importance for us is the location of the Dirac point with respect to the Fermi level, i. e., the doping of graphene, that we measure with the parameter $E_D$. Another important quantity is the band gap of the CGT, $E_g$, which is related to strain. Finally we extract the position of the Dirac point within the CGT band gap, by measuring the energy difference to the CGT valence band edge at the $\Gamma$ point, i. e., $E_D - E_{VBE}$, as defined in Fig. S2(a). All those quantities are summarized in Table S2 for the different twist angles.

In Fig. S2(b), we show a zoom to the Dirac point of graphene with a fit to the model Hamiltonian from the main text. The results are in perfect agreement, employing the parameters in Table S2. For the 30° case, we find that

the spin down (spin up) Dirac cone is shifted towards lower (higher) energies. Here, proximity-induced exchange is similar to the Zeeman splitting from an external magnetic field [22]. The corresponding sublattice-dependent exchange parameters have the same sign and are nearly equal in value, $\lambda_{ex}^{A} \approx \lambda_{ex}^{B} \approx -3.6$ meV, hence we speak of uniform (ferromagnetic) proximity exchange.

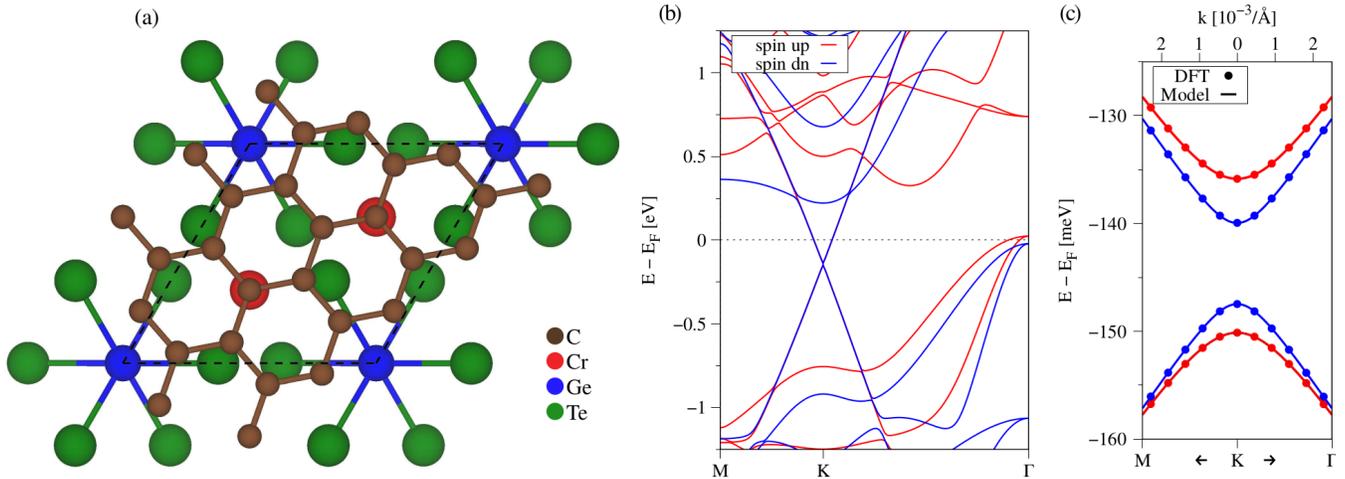

FIG. S3. (a) Bottom view of the graphene/CGT heterostructure with a twist angle of 19.1°. (b) DFT-calculated band structure along the high-symmetry path M-K-Γ. Red (blue) solid lines correspond to spin up (spin down). (c) Zoom to the DFT-calculated (symbols) low energy Dirac bands near the K point with a fit to the model Hamiltonian (solid lines).

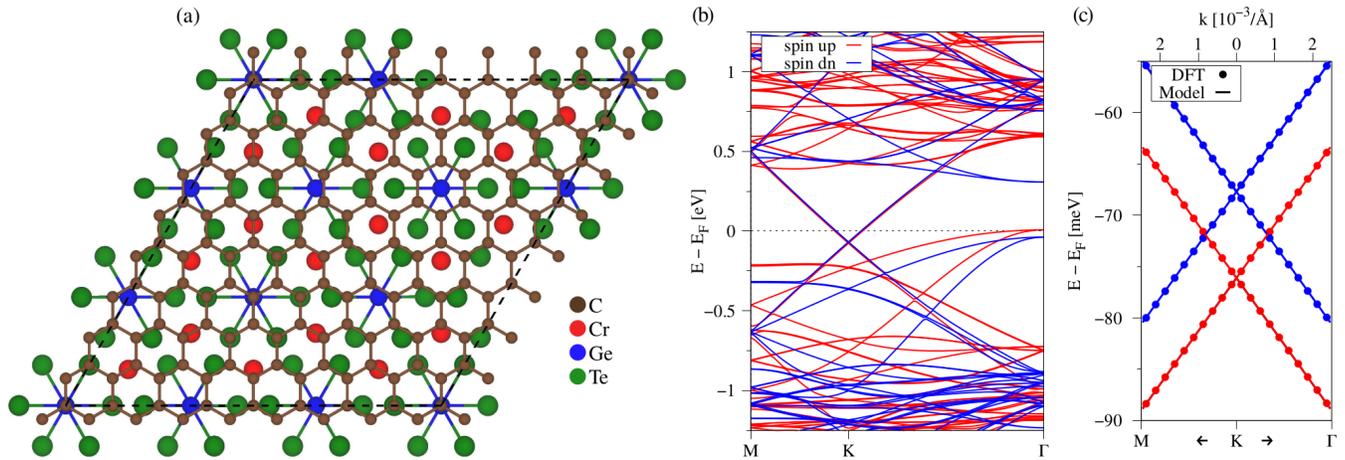

FIG. S4. (a) Bottom view of the graphene/CGT heterostructure with a twist angle of 0°. (b) DFT-calculated band structure along the high-symmetry path M-K-Γ. Red (blue) solid lines correspond to spin up (spin down). (c) Zoom to the DFT-calculated (symbols) low energy Dirac bands near the K point with a fit to the model Hamiltonian (solid lines).

Comparing the different twist angles, see Table S2, we find a nearly continuous tunability of proximity exchange parameters from about 4 to −4 meV, when twisting from 0° to 30°. In addition, we find some very surprising results. For 30°, we have uniform proximity exchange as already mentioned. Similarly, for 0° we also have uniform proximity exchange (4.2 meV), but the parameters have the opposite sign compared to the 30° case (−3.6 meV). Even more surprising, at an angle of 19.1° we find $\lambda_{ex}^{A} < 0$ and $\lambda_{ex}^{B} > 0$, i. e., staggered (antiferromagnetic) proximity-induced exchange, which is highly important for the realization of topological edge states in graphene [23]. Moreover, in some cases graphene experiences significant doping effects (0° and 19.1°) since $E_D$ is below the Fermi level and also below the CGT valence band edge. In Fig. S3 and Fig. S4, we show the geometries, the calculated band structures, and fit results for the 19.1° and 0° cases, also perfectly matching with the model.

## III. THE INFLUENCE OF STRAIN

TABLE S3. Structural information for the graphene/CGT heterostructures, when the CGT lattice constant is fixed at roughly 6.85 Å. We list the twist angle between the layers, the lattice constants of graphene and CGT, the calculated dipole of the structure, and the relaxed interlayer distance $d_{\text{int}}$.

| twist angle [°] | $a_{\text{grp}}$ [Å] | $a_{\text{CGT}}$ [Å] | dipole [debye] | $d_{\text{int}}$ [Å] |
|---|---|---|---|---|
| 0.0000 | 2.5679 | 6.8475 | -2.7303 | 3.6005 |
| 3.0045 | 2.4873 | 6.8506 | -4.4405 | 3.5382 |
| 5.8175 | 2.4050 | 6.8480 | -3.6099 | 3.5300 |
| 8.9483 | 2.4600 | 6.8480 | -1.5072 | 3.5473 |
| 12.2163 | 2.5105 | 6.8500 | -3.1492 | 3.5688 |
| 14.7047 | 2.4086 | 6.8480 | -5.0438 | 3.5180 |
| 19.1066 | 2.5873 | 6.8454 | -0.2760 | 3.6187 |
| 23.4132 | 2.3537 | 6.8400 | -4.4003 | 3.5028 |
| 26.9955 | 2.5125 | 6.8475 | -2.5066 | 3.5713 |
| 30.0000 | 2.3702 | 6.8423 | -1.4053 | 3.4940 |

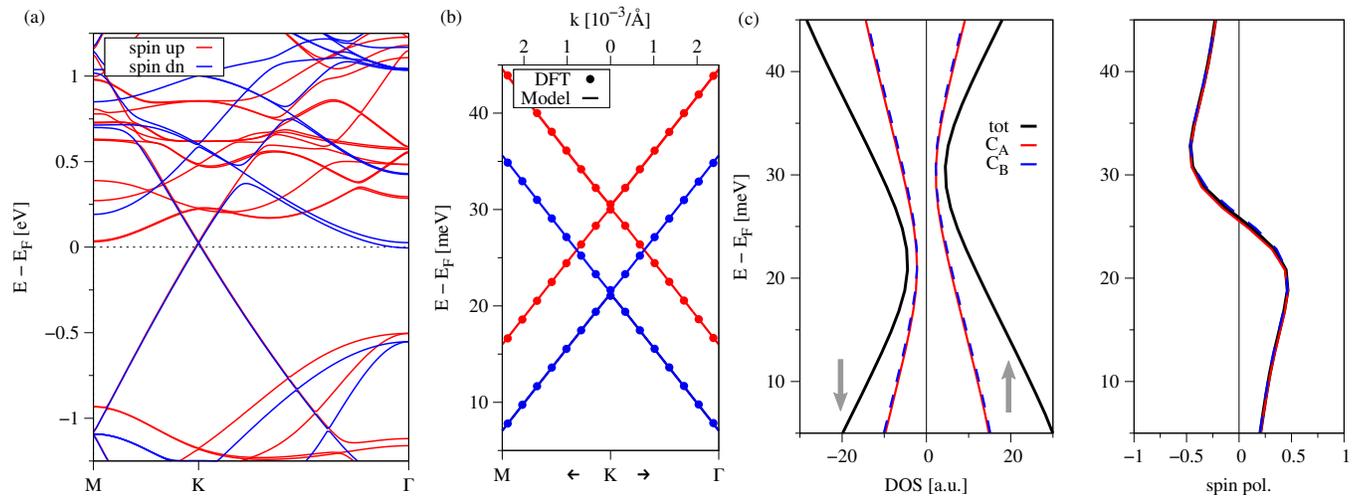

FIG. S5. (a) DFT-calculated band structure of the graphene/CGT heterostructure along the high-symmetry path M-K-Γ for a twist angle of 30° and when the CGT lattice constant is about 6.85 Å. Red (blue) solid lines correspond to spin up (spin down). (b) Zoom to the DFT-calculated (symbols) low energy Dirac bands near the K point with a fit to the model Hamiltonian (solid lines). (c) The sublattice (red = $C_A$, blue = $C_B$, black = total) and spin-resolved (positive = ↑, negative = ↓) density of states (DOS) of graphene near the Dirac point. The spin polarizations, calculated from the density of states as $(N_{\uparrow} - N_{\downarrow})/(N_{\uparrow} + N_{\downarrow})$.

So far, for each twist angle we distributed the strain to both materials (graphene and CGT) to form commensurate heterostructure supercells. Thus, we cannot a priori distinguish which results are due to straining, twisting, or a combination of both. Since the exchange coupling in graphene originates from CGT, we reconsider all structures with the different twist angles, but strain the heterostructure supercells such that the CGT lattice constant is always at about 6.85 Å, close to the experimental value. As a consequence, the strain applied to graphene increases, which should mainly influence the nearest neighbor hopping constant, reflected in the Fermi velocity. In Table S3, we again summarize the individual lattice constants of graphene and CGT for the different twist angles. Similar to before, we again allow the CGT atoms to relax their positions, while keeping the graphene layer fixed. Therefore, the interlayer distance may change, while the rippling of graphene stays fixed. Comparing Table S1 and Table S3, we can see that the interlayer distances change by at maximum 2%, which will certainly influence the magnitude, but not the type (uniform or staggered) of proximity-induced exchange coupling [12, 23]. In addition, the dipole (internal electric field) of the structure is slightly influenced, which will have an impact on the relevant band offsets.

In the same manner as above, we calculate the electronic band structure and apply our model Hamiltonian to the proximitized low energy Dirac bands. In Fig. S5, we again show the band structure of the graphene/CGT heterostructure for a twist angle of 30°, but now with a CGT lattice constant of about 6.85 Å. The overall band structure features remain the same, but the CGT now has a direct band gap at the Γ point, while the Dirac point is

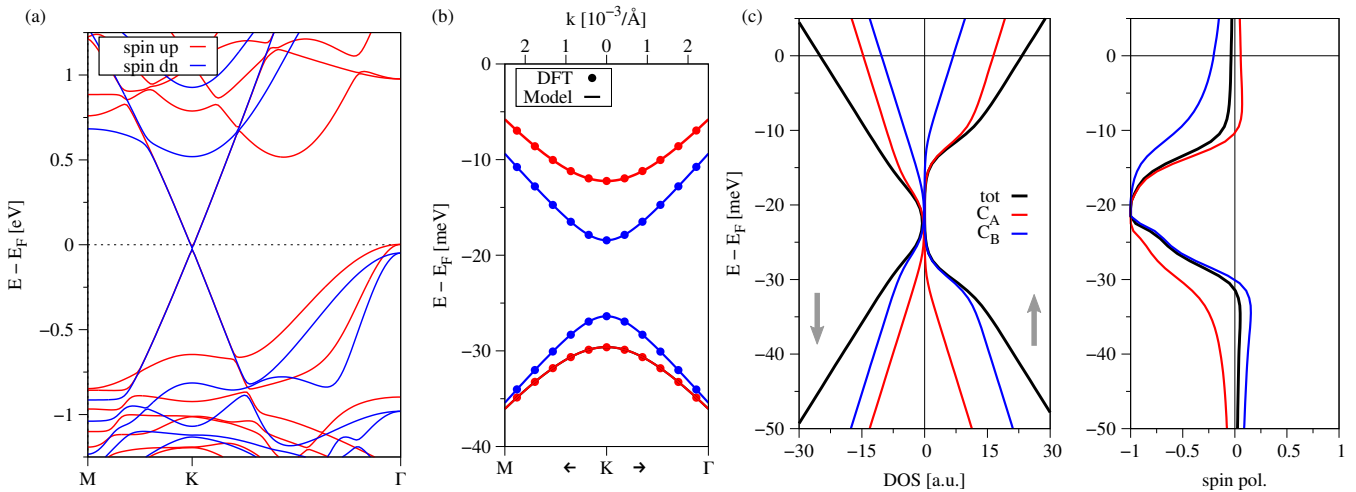

FIG. S6. (a) DFT-calculated band structure of the graphene/CGT heterostructure along the high-symmetry path M-K-Γ for a twist angle of 19.1° and when the CGT lattice constant is about 6.85 Å. Red (blue) solid lines correspond to spin up (spin down). (b) Zoom to the DFT-calculated (symbols) low energy Dirac bands near the K point with a fit to the model Hamiltonian (solid lines). (c) The sublattice (red = $C_A$, blue = $C_B$, black = total) and spin-resolved (positive = ↑, negative = ↓) density of states (DOS) of graphene near the Dirac point. The spin polarizations, calculated from the density of states as $(N_\uparrow - N_\downarrow)/(N_\uparrow + N_\downarrow)$.

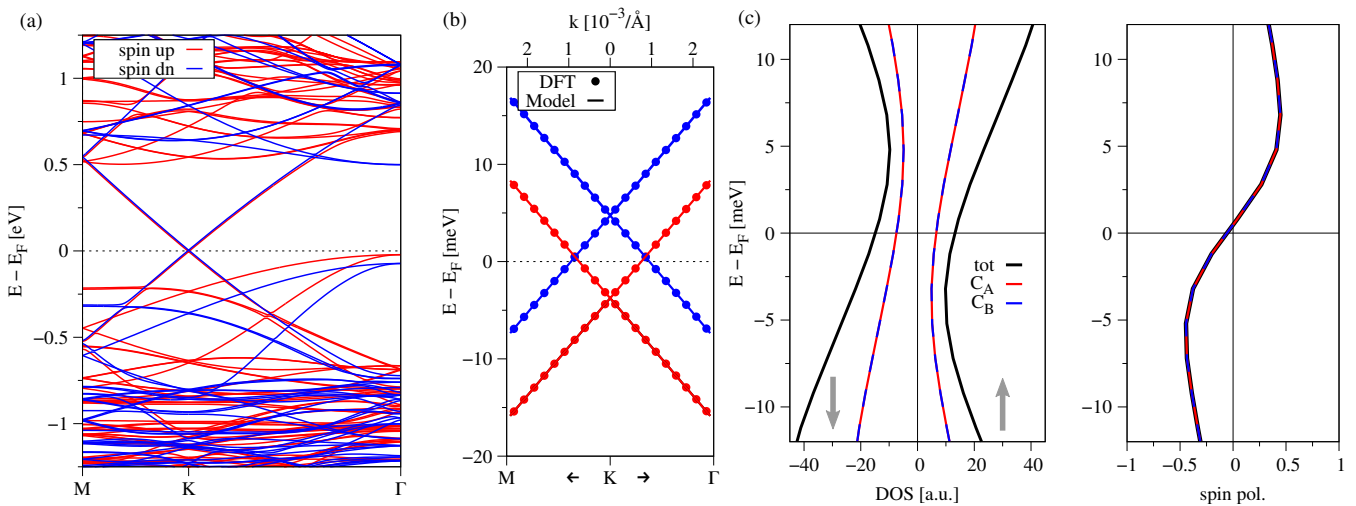

FIG. S7. (a) DFT-calculated band structure of the graphene/CGT heterostructure along the high-symmetry path M-K-Γ for a twist angle of 0° and when the CGT lattice constant is about 6.85 Å. Red (blue) solid lines correspond to spin up (spin down). (b) Zoom to the DFT-calculated (symbols) low energy Dirac bands near the K point with a fit to the model Hamiltonian (solid lines). (c) The sublattice (red = $C_A$, blue = $C_B$, black = total) and spin-resolved (positive = ↑, negative = ↓) density of states (DOS) of graphene near the Dirac point. The spin polarizations, calculated from the density of states as $(N_\uparrow - N_\downarrow)/(N_\uparrow + N_\downarrow)$.

located about 25 meV above the Fermi level. The calculated spin- and sublattice-resolved density of states together with the corresponding spin polarization of the Dirac states is shown in Fig. S5 (c). At about 40 meV above the Fermi level, the $C_A$ ($C_B$) sublattice contributes with 51% (49%) and at 10 meV the contributions are switched, which is consistent with the tiny sublattice symmetry breaking Δ for the 30° structure. Above (below) the Dirac point, both sublattice spin polarizations are almost equal and negative (positive). The most important quantity for Hanle spin relaxation experiments [24] is the total spin polarization on graphene, which switches sign at the Dirac point, since the two spin densities become equal. At 40 meV (10 meV), the total spin polarization equals −29% (26%).

In Table S4, we summarize the same information as in Table S2, but for the heterostructures when CGT is nearly unstrained. First of all, we notice that the CGT band gap, $E_g$, stays nearly constant, due to almost similar CGT lattice constants for all twist angles. Moreover, the averaged magnetic moments of Cr, Ge, and Te atoms are 3.145, 0.025, and −0.089 $\mu_B$ and are nearly constant now. Thus, in Table S4, we list only the calculated magnetic moments

TABLE S4. Fit parameters of Hamiltonian $\mathcal{H}$ for the graphene/CGT heterostructures for different twist angles when the CGT lattice constant is fixed at roughly 6.85 Å. Parameters have the same meaning as in Table S2. We also list the averaged calculated magnetic moment of C atoms.

| twist angle [°] | $v_F$ [$10^5$ m/s] | $\Delta$ [meV] | $\lambda_{ex}^A$ [meV] | $\lambda_{ex}^B$ [meV] | $E_D$ [meV] | $E_g$ [eV] | $E_D - E_{VBE}$ [eV] | C [$10^{-3}\mu_B$] |
|---|---|---|---|---|---|---|---|---|
| 0.0000 | 7.6032 | 0.002 | 4.252 | 4.249 | 0.488 | 0.522 | 0.022 | -0.184 |
| 3.0045 | 7.9587 | 0.055 | 4.045 | 4.079 | -0.265 | 0.522 | 0.224 | -0.136 |
| 5.8175 | 8.2500 | 1.666 | 2.768 | 2.608 | -0.358 | 0.522 | 0.478 | -0.126 |
| 8.9483 | 8.1209 | 0.048 | 1.181 | 1.133 | -0.191 | 0.519 | 0.299 | -0.200 |
| 12.2163 | 7.8825 | 0.062 | 0.028 | 0.044 | 0.090 | 0.521 | 0.155 | -0.221 |
| 14.7047 | 8.3861 | 0.217 | -3.617 | -3.630 | -0.031 | 0.528 | 0.439 | -0.271 |
| 19.1066 | 7.5375 | 6.329 | -3.092 | 1.613 | -21.664 | 0.514 | -0.025 | -0.243 |
| 23.4132 | 8.9630 | 0.145 | -5.819 | -5.560 | 55.749 | 0.516 | 0.576 | -0.126 |
| 26.9955 | 7.8970 | 0.146 | -3.878 | -3.753 | 0.449 | 0.515 | 0.126 | -0.242 |
| 30.0000 | 8.6758 | 0.288 | -4.510 | -4.427 | 25.804 | 0.500 | 0.530 | -0.166 |

of C atoms. Second, the type and magnitude of proximity-induced exchange coupling stays nearly the same, i. e., straining does not dictate the proximity exchange, but the twisting does. Third, the Fermi velocity adapts itself to the new graphene lattice constant. Finally, the band offsets are drastically different for certain twist angles. For example, comparing the 0° results, we now find the Dirac point at the Fermi level and about 20 meV above the CGT valence band edge, see Table S4. In contrast, before we had the Dirac point at about −70 meV below the Fermi level and −80 meV below the CGT valence band edge, see Table S2. Similar scenarios hold for the other twist angles, but sometimes not as drastic as for the 0° case.

In Fig. S6 and Fig. S7, we show the band structure and density of states results for the 19.1° and 0° cases, when CGT is unstrained. For 19.1°, the conduction (valence) band Dirac states are mainly formed by $C_A$ ($C_B$) $p_z$ orbitals, which supports our sublattice resolved proximity exchange Hamiltonian with broken sublattice symmetry for arbitrary twist angles. At (−45 meV below) the Fermi level, the $C_A$ atom contribution to the total graphene DOS is 65% (37%), while the $C_B$ atom contribution is 35% (63%), see Fig. S6(c). At the Fermi level, the total spin polarization is negative (−3.6%), consistent with the conduction band Dirac dispersion. Nevertheless, the individual sublattice spin polarizations are opposite, positive for $C_A$ and negative for $C_B$, supporting the antiferromagnetic proximity exchange that we have found by fitting the dispersion. For the valence band Dirac states, similar things hold, but spin polarizations and sublattice characters are reversed, compared to the conduction band states. At −45 meV below the Fermi level, the total spin polarization is 3.1%.

For 0°, see Fig. S7, we find similar results as for 30°, but the ferromagnetic proximity exchange has the opposite sign. Since the sublattice symmetry breaking $\Delta$ is nearly zero Dirac bands are equally formed by $C_A$ and $C_B$ sublattices. At about 10 meV (−10 meV) above (below) the Dirac point, the total spin polarization is positive (negative) and about 37% (−39%).

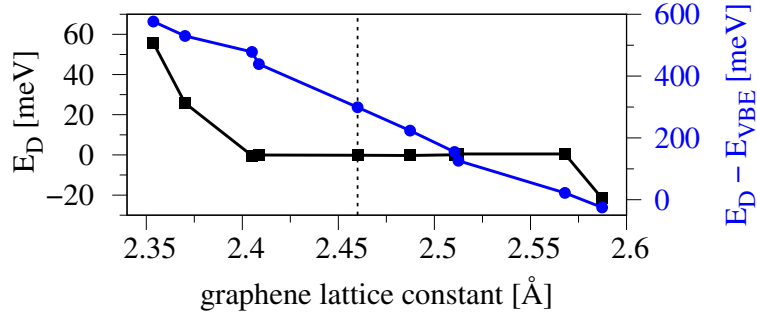

FIG. S8. The Dirac point energy $E_D$ (black) and the band offset, $E_D - E_{VBE}$ (blue), as function of the graphene lattice constant. The information are listed in Tab. S3 and Tab. S4. The vertical dashed line indicates the experimental graphene lattice constant.

In Fig. S8, we plot the Dirac point energy, $E_D$, and the band offset, $E_D - E_{VBE}$, as function of the graphene lattice constant, as listed in Table S3 and Table S4. Astonishingly, we find a linear relationship between the band offset and the graphene lattice constant, which tells us that the strain in graphene is responsible for the different band offsets. Most important, we can now read off the *intrinsic* band offset of the heterostructure, that can be expected in experiment for all twist angles. The Dirac point should be located at the Fermi level and about 300 meV above

the CGT valence band edge for unstrained layers. In addition, if the graphene Dirac point is shifted out of the CGT band gap, i. e., if $E_g < (E_D - E_{VBE}) < 0$, graphene gets doped ($E_D \neq 0$). In our scenario, straining is responsible for the doping, but experimentally a gate voltage can be employed to tune the Dirac point within the CGT band gap, certainly influencing the proximity-induced exchange.

## IV. MOIRÉ RECONSTRUCTIONS

In the following, we also want to comment on possible moiré reconstructions, as observed in twisted transition metal dichalcogenide heterostructures [25, 26]. As mentioned above, during relaxation we only allowed graphene atoms to move along $z$-direction, for proper determination of the interlayer distances. What about the in-plane forces and related deformations of the graphene lattice? By inspecting the output files of the relaxed structures for all twist angles when both layers are strained, we find that all (in-plane and out-of-plane) forces acting on graphene and CGT are below $10^{-3}$ [Ry/$a_0$], even with the mentioned constraint. Therefore moiré reconstructions are not likely to be observed, at least not in our supercell geometries. The only exception is for the twist angle of 5.8°, where we find significant ($\approx 10^{-2}$ [Ry/$a_0$]) in-plane forces acting on C atoms. We thus performed a subsequent relaxation without constraining the graphene atoms, and recalculated the electronic structure for this angle. In addition, and to crosscheck results, we perform a full relaxation for our smallest supercells (19.1° and 30°). The results are summarized in Table S5 and are barely different to the original results in Table S2. Especially the proximity exchange parameters are nearly identical. Only for 5.8°, the staggered potential $\Delta$ is now much smaller and proximity exchange parameters are slightly enhanced due to the relaxation.

TABLE S5.  Fit parameters of Hamiltonian $\mathcal{H}$ for the fully relaxed graphene/CGT heterostructures for selected twist angles. Parameters have the same meaning as in Table S2.

| twist angle [°] | $v_F$ [$10^5$ m/s] | $\Delta$ [meV] | $\lambda_{ex}^A$ [meV] | $\lambda_{ex}^B$ [meV] | $E_D$ [meV] | $E_D - E_{VBE}$ [eV] | dipole [debye] |
|---|---|---|---|---|---|---|---|
| 5.8175 | 8.1528 | 0.070 | 2.685 | 2.591 | -0.415 | 0.552 | -3.5466 |
| 19.1066 | 7.8378 | 5.446 | -2.053 | 1.332 | -143.870 | -0.169 | -0.2537 |
| 30.0000 | 8.2856 | 0.299 | -3.635 | -3.600 | -0.414 | 0.663 | -1.3042 |

# V. ELECTRIC FIELD TUNABILITY

We have found that the proximity exchange depends on the twist angle. In addition, strain controls the band offset, also influencing proximity exchange. We now consider the first heterostructure set corresponding to Table S1 (graphene and CGT strained) with different twist angles and apply a transverse electric field between $\pm 1.5$ V/nm. Can we tune the proximity-induced exchange coupling by gating?

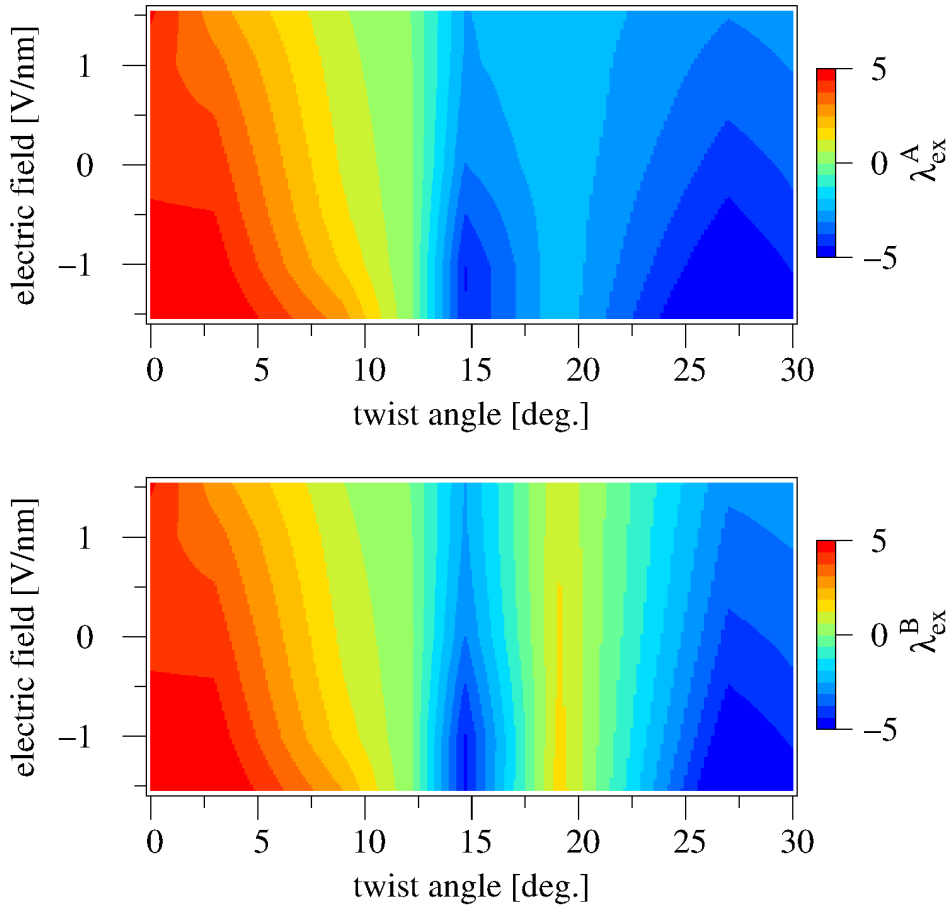

FIG. S9. Electric field and twist angle dependence of the two proximity-induced exchange parameters $\lambda_{ex}^A$ (top) and $\lambda_{ex}^B$ (bottom).

In Fig. S9, we summarize the electric field and twist angle dependence of the two proximity-induced exchange parameters $\lambda_{ex}^A$ and $\lambda_{ex}^B$. In Table S6, we summarize the fit parameters employed for Fig. S9. We find that gating and twisting are two efficient control knobs to tailor the sign and magnitude of the proximity-induced exchange coupling in graphene/CGT heterostructures. In agreement with Table S2, one can nearly continuously tune the sign of the proximity exchange from positive to negative by twisting. In addition, an external electric field can be employed to tune the magnitude. For some twist angles, a relative tunability of more than 100% can be reached within our electric field limits. Most important, at a twist angle of $12.2°$, proximity exchange parameters are small, but can be even tuned from positive to negative values by the gate field. Not only the proximity exchange can be tuned, but also the doping level, $E_D$, and the band offset, $E_D - E_{VBE}$, see Table S6. Unfortunately, our results always take into account the strain of the layers, which does not reflect the situation in experiments. Therefore, we can give at most predictions about the gate- and twist-tunability of the proximity exchange coupling. Note that for the electric field calculations, we have neglected two angles, $5.8°$ and $23.4°$. The reason is that already small negative gate fields would shift the Dirac cone into the CGT conduction bands, leading to significant band hybridization and spoiling of the bare proximity exchange and the Dirac dispersion. However, the overall trends of the gate-tunable proximity exchange should still hold.

TABLE S6. Electric field dependence of the fit parameters of Hamiltonian $\mathcal{H}$ for the graphene/CGT heterostructures for different twist angles. We summarize the Fermi velocity $v_F$, the staggered potential $\Delta$, proximity exchange parameters $\lambda_{ex}^A$ and $\lambda_{ex}^B$, the Dirac point energy with respect to the Fermi level $E_D$, the position of the Dirac point with respect to the valence band edge of CGT at the $\Gamma$ point $E_D - E_{VBE}$, and the calculated dipole of the structure.

| el. field [V/nm] | twist angle [°] | $v_F$ [$10^5$ m/s] | $\Delta$ [meV] | $\lambda_{ex}^A$ [meV] | $\lambda_{ex}^B$ [meV] | $E_D$ [meV] | $E_D - E_{VBE}$ [eV] | dipole [debye] |
|---|---|---|---|---|---|---|---|---|
| -1.543 | 0.0000 | 7.8596 | 0.0006 | 5.2511 | 5.2508 | -0.3157 | 0.0734 | -14.7934 |
| -1.028 | 0.0000 | 7.8606 | 0.0002 | 4.9167 | 4.9172 | 0.0422 | 0.0115 | -10.8131 |
| -0.514 | 0.0000 | 7.8596 | 0.0008 | 4.4663 | 4.4662 | -13.0826 | -0.0209 | -6.5423 |
| 0.000 | 0.0000 | 7.8555 | 0.0001 | 4.2005 | 4.1994 | -71.9042 | -0.0798 | -2.5453 |
| 0.514 | 0.0000 | 7.8477 | 0.0002 | 4.0489 | 4.0494 | -131.5339 | -0.1392 | 1.4452 |
| 1.028 | 0.0000 | 7.8274 | 0.0002 | 4.0131 | 4.0127 | -192.3786 | -0.2001 | 5.4117 |
| 1.543 | 0.0000 | 7.6683 | 0.0003 | 4.5233 | 4.5210 | -199.8656 | -0.2225 | 9.7786 |
| -1.543 | 3.0045 | 8.0256 | 0.0589 | 5.1937 | 5.2226 | 0.4909 | 0.3815 | -20.7561 |
| -1.028 | 3.0045 | 8.0227 | 0.0567 | 4.7636 | 4.7939 | -0.0705 | 0.3215 | -15.3389 |
| -0.514 | 3.0045 | 8.0135 | 0.0557 | 4.4071 | 4.4382 | -0.2486 | 0.2621 | -9.9093 |
| 0.000 | 3.0045 | 7.9972 | 0.0541 | 4.0791 | 4.1157 | -0.3232 | 0.2026 | -4.4864 |
| 0.514 | 3.0045 | 7.9713 | 0.0587 | 3.7419 | 3.7800 | 0.0305 | 0.1447 | 0.9421 |
| 1.028 | 3.0045 | 7.9302 | 0.0635 | 3.3389 | 3.3777 | -0.1628 | 0.0868 | 6.3719 |
| 1.543 | 3.0045 | 7.8621 | 0.0687 | 2.7703 | 2.8025 | 0.0489 | 0.0300 | 11.8043 |
| -1.543 | 8.9483 | 7.8864 | 0.1548 | 2.8482 | 2.5391 | -0.0516 | 0.4841 | -6.9494 |
| -1.028 | 8.9483 | 8.0810 | 0.0877 | 1.8394 | 1.6477 | 0.5584 | 0.4235 | -5.1303 |
| -0.514 | 8.9483 | 8.1155 | 0.0607 | 1.4062 | 1.3342 | -0.2144 | 0.3611 | -3.3179 |
| 0.000 | 8.9483 | 8.1209 | 0.0475 | 1.1806 | 1.1331 | -0.1914 | 0.2992 | -1.5072 |
| 0.514 | 8.9483 | 8.1159 | 0.0416 | 1.0094 | 0.9692 | 0.3067 | 0.2369 | 0.3070 |
| 1.028 | 8.9483 | 8.1040 | 0.0370 | 0.8580 | 0.8160 | -0.1417 | 0.1755 | 2.1149 |
| 1.543 | 8.9483 | 8.0855 | 0.0355 | 0.6986 | 0.6495 | -0.0463 | 0.1137 | 3.9268 |
| -1.543 | 12.2163 | 7.9494 | 0.0891 | 0.0484 | 0.0654 | -0.0307 | 0.3165 | -15.3657 |
| -1.028 | 12.2163 | 7.9479 | 0.0784 | 0.0832 | 0.0965 | 0.6230 | 0.2548 | -11.2889 |
| -0.514 | 12.2163 | 7.9432 | 0.0704 | 0.0965 | 0.1067 | -0.1465 | 0.1916 | -7.2409 |
| 0.000 | 12.2163 | 7.9356 | 0.0619 | 0.0916 | 0.1005 | -0.0538 | 0.1295 | -3.1757 |
| 0.514 | 12.2163 | 7.9248 | 0.0565 | 0.0687 | 0.0723 | -0.2459 | 0.0669 | 0.8752 |
| 1.028 | 12.2163 | 7.9110 | 0.0464 | 0.0275 | 0.0229 | -0.0879 | 0.0064 | 4.9510 |
| 1.543 | 12.2163 | 7.9018 | 0.0442 | -0.0192 | -0.0248 | -19.4284 | -0.0276 | 9.2931 |
| -1.543 | 14.7047 | 8.1085 | 0.2480 | -4.3889 | -4.4529 | 26.3551 | 0.6529 | -22.3183 |
| -1.028 | 14.7047 | 8.1398 | 0.2102 | -4.4075 | -4.4966 | 2.9413 | 0.6243 | -16.2363 |
| -0.514 | 14.7047 | 8.1745 | 0.1926 | -3.7977 | -3.8629 | 0.7328 | 0.5754 | -10.4690 |
| 0.000 | 14.7047 | 8.1996 | 0.1727 | -3.1462 | -3.1669 | -0.0287 | 0.5127 | -4.9073 |
| 0.514 | 14.7047 | 8.2127 | 0.1428 | -2.8358 | -2.8646 | -0.1019 | 0.4510 | 0.6701 |
| 1.028 | 14.7047 | 8.2189 | 0.1355 | -2.5872 | -2.6134 | -0.0568 | 0.3890 | 6.2426 |
| 1.543 | 14.7047 | 8.2203 | 0.1355 | -2.5296 | -2.5499 | 0.4126 | 0.3274 | 11.8259 |
| -1.543 | 19.1066 | 7.8418 | 5.3188 | -2.0396 | 1.6232 | -33.8931 | -0.0392 | -1.6063 |
| -1.028 | 19.1066 | 7.8400 | 5.3783 | -2.0338 | 1.5096 | -97.2511 | -0.0993 | -1.1669 |
| -0.514 | 19.1066 | 7.8384 | 5.4110 | -2.0734 | 1.3874 | -132.8669 | -0.1461 | -0.7137 |
| 0.000 | 19.1066 | 7.8378 | 5.4503 | -2.0498 | 1.3314 | -143.3497 | -0.1680 | -0.2338 |
| 0.514 | 19.1066 | 7.8350 | 5.5173 | -2.0321 | 1.2987 | -190.5737 | -0.2169 | 0.2187 |
| 1.028 | 19.1066 | 7.8320 | 5.5899 | -2.0911 | 1.1936 | -230.1734 | -0.2629 | 0.6699 |
| 1.543 | 19.1066 | 7.8307 | 5.6359 | -2.0944 | 1.1650 | -238.8009 | -0.2867 | 1.1496 |
| -1.543 | 26.9955 | 7.9018 | 0.1907 | -5.9643 | -5.7640 | 0.2573 | 0.2799 | -12.0090 |
| -1.028 | 26.9955 | 7.9275 | 0.1725 | -5.1814 | -5.0074 | 1.1925 | 0.2197 | -8.8315 |
| -0.514 | 26.9955 | 7.9450 | 0.1648 | -4.5646 | -4.4206 | 1.1464 | 0.1582 | -5.6726 |
| 0.000 | 26.9955 | 7.9568 | 0.1496 | -4.0845 | -3.9566 | 0.3491 | 0.0952 | -2.5319 |
| 0.514 | 26.9955 | 7.9643 | 0.1398 | -3.7125 | -3.5897 | -0.2120 | 0.0323 | 0.6153 |
| 1.028 | 26.9955 | 7.9677 | 0.1345 | -3.3923 | -3.2929 | -9.2685 | -0.0143 | 3.8887 |
| 1.543 | 26.9955 | 7.9697 | 0.1342 | -3.0823 | -2.9970 | -65.4473 | -0.0710 | 7.0871 |
| -1.543 | 30.0000 | 8.2439 | 0.3676 | -4.8155 | -4.7776 | 112.1088 | 0.8114 | -5.7332 |
| -1.028 | 30.0000 | 8.2572 | 0.3521 | -4.3315 | -4.2944 | 86.2641 | 0.7787 | -4.1952 |
| -0.514 | 30.0000 | 8.2745 | 0.3222 | -3.8842 | -3.8495 | 23.6695 | 0.7172 | -2.7585 |
| 0.000 | 30.0000 | 8.2856 | 0.2983 | -3.6302 | -3.5964 | -0.4279 | 0.6621 | -1.3044 |
| 0.514 | 30.0000 | 8.2945 | 0.2756 | -3.3269 | -3.2935 | -0.3483 | 0.6001 | 0.1250 |
| 1.028 | 30.0000 | 8.3005 | 0.2603 | -3.0822 | -3.0535 | -0.1490 | 0.5377 | 1.5561 |
| 1.543 | 30.0000 | 8.3045 | 0.2488 | -2.8847 | -2.8606 | 0.3343 | 0.4766 | 2.9846 |

## VI. LATERALLY AND VERTICALLY SHIFTING THE LAYERS

How sensitive is the proximity-induced exchange coupling with respect to the atomic registry and the interlayer distance? For this purpose, we consider only the most important twist angles, i. e., $0°$, $19.1°$, and $30°$, in order to cover all types of proximity exchange (uniform and staggered) and shift the graphene layer with respect to the CGT substrate. Again, we consider only the heterostructures where both layers are strained. We start from the relaxed structures as summarized in Table S1 and apply vertical shifts ($z$ direction) of graphene, i. e., we tune the interlayer distance by $\Delta d$. For the lateral shifts ($x$ and $y$), we use crystal coordinate notation, i. e., we shift graphene by fractions $x$ and $y$ of the supercell lattice vectors. In Table S7 we summarize the fit results.

TABLE S7. Fit parameters of Hamiltonian $\mathcal{H}$ for the graphene/CGT heterostructures for selected twist angles and for lateral ($x$ and $y$) and vertical ($z$) shifts. The vertical shifts tune the interlayer distance by $\Delta d$, while lateral shifts $x$ and $y$ are in fractions of the supercell lattice vectors. Other parameters have the same meaning as in Table S2. We also list the averaged calculated magnetic moment of C atoms.

| twist angle [°] | $(x,y)$ | $\Delta d$ [Å] | $v_F$ [$10^5$ m/s] | $\Delta$ [meV] | $\lambda_{ex}^A$ [meV] | $\lambda_{ex}^B$ [meV] | $E_D$ [meV] | $E_D - E_{VBE}$ [eV] | C [$10^{-3}\mu_B$] |
|---|---|---|---|---|---|---|---|---|---|
| 0.0000 | (0,0) | -0.1 | 7.7801 | 0.000 | 5.752 | 5.751 | -119.931 | -0.127 | -0.308 |
| 0.0000 | (0,0) | 0.1 | 7.9063 | 0.002 | 3.066 | 3.064 | -22.279 | -0.031 | -0.194 |
| 0.0000 | (1/6,1/6) | 0.0 | 7.8549 | 0.025 | 4.272 | 4.274 | -71.123 | -0.079 | -0.249 |
| 0.0000 | (1/9,2/9) | 0.0 | 7.8537 | 0.117 | 4.280 | 4.253 | -72.503 | -0.080 | -0.238 |
| 19.1066 | (0,0) | -0.1 | 7.7793 | 7.862 | -2.853 | 1.842 | -192.029 | -0.226 | -0.264 |
| 19.1066 | (0,0) | 0.1 | 7.8764 | 3.794 | -1.462 | 0.947 | -134.069 | -0.148 | -0.157 |
| 19.1066 | (2/3,1/3) | 0.0 | 7.6762 | 3.924 | -2.240 | -4.033 | -149.732 | -0.177 | -0.214 |
| 19.1066 | (1/3,2/3) | 0.0 | 7.6631 | 7.026 | -2.407 | -6.621 | -148.355 | -0.177 | -0.193 |
| 30.0000 | (0,0) | -0.1 | 8.2472 | 0.382 | -4.593 | -4.555 | 0.015 | 0.585 | -0.248 |
| 30.0000 | (0,0) | 0.1 | 8.3140 | 0.233 | -2.791 | -2.761 | 33.910 | 0.724 | -0.144 |
| 30.0000 | (1/6,1/3) | 0.0 | 8.2814 | 0.070 | -3.578 | -3.592 | -0.149 | 0.662 | -0.202 |
| 30.0000 | (1/3,1/3) | 0.0 | 8.2852 | 0.137 | -3.607 | -3.646 | -0.131 | 0.663 | -0.206 |

We find that the interlayer distance strongly influences the proximity exchange. Tuning $\Delta d$ by $\pm 0.1$ Å, which equals about $\pm 3\%$ of the interlayer distance only, the exchange parameters can be tuned by on average $\mp 30\%$. Such tunability has been recently reported for the proximity SOC in graphene/WSe$_2$ heterostructures, where a special experimental setup allows to apply hydrostatic pressure and tune the vdW gap between the layers [27, 28]. In addition, the interlayer distance also influences the doping and band offsets, as we can see from $E_D$ and $E_D - E_{VBE}$. Increasing the distance, decreases the charge transfer between the layers and vice versa, also in agreement with experimental findings for graphene/MoS$_2$ heterostructures [29].

What about the lateral shifts? In the two cases, $0°$ and $30°$, when the heterostructure supercell is large, the atomic registry does not play a role for the results, especially for proximity exchange couplings. The reason is, that on average the two C sublattices always feel roughly the same influence from the CGT layer, independent of the shift. However, for the $19.1°$ case, there is a strong dependence on the lateral shift, since the heterostructure supercell is small, see Fig. S3(a). Our original result of staggered exchange parameters, see Table S2, is not valid anymore when shifting the layers relative to each other, see Table S7. This makes the $19.1°$ case a special one, since we can achieve uniform and staggered exchange couplings, important for tailoring the edge transport properties in proximitized graphene [23].

## VII. ENCAPSULATED STRUCTURES

Assume graphene to be encapsulated between two CGT layers. Depending on the twist angles ($\vartheta_b$ and $\vartheta_t$) of bottom and top CGT layers with respect to graphene, and their individual magnetizatons ($\boldsymbol{M}_b$ and $\boldsymbol{M}_t$) we can further tailor proximity exchange coupling in graphene.

Case 1: When the magnetizations are parallel (antiparallel) and twist angles are the same, proximity exchange couplings are additive (effectively cancel) [12, 30].

Case 2: When the magnetizations are parallel and twist angles are different (say 0° and 30°), proximity exchange couplings can also vanish, based on the above results in Table S4.

Case 3: The magnetizations are parallel and both twist angles are 19.1°. Depending on the individual stacking configurations, one should be able to create a scenario of ferromagnetic (antiferromagnetic) proximity exchange from, say, bottom (top) CGT, based on Table S7. Consequently, the layer-dependent proximity exchange couplings could add up for one graphene sublattice, while for the other sublattice the couplings cancel each other.

In Fig. S10, we summarize the results for three exemplary structures of CGT encapsulated graphene, where both twist angles are 19.1° and parallel magnetizations, but with different encapsulation configurations. Similar to our non-encapsulated structures, we first performed a relaxation to obtain reasonable band structure results. Here, we also used structures where the CGT lattice constant is unstrained. Comparing the three cases, we find the global band structure to be barely different. However, the low energy graphene Dirac bands differ significantly. In the first case, the sublattice symmetry breaking $\Delta$ is large, opening a gap in the Dirac spectrum, and proximity exchange couplings are ferromagnetic. In the second case, $\Delta$ is also large, but proximity exchange couplings are antiferromagnetic. In the third case, we also find ferromagnetic couplings, but $\Delta$ is small leading to band overlap at the K point. For all three cases, proximity exchange parameters are also much larger than for the non-encapsulated structure. This demonstrates, that one can boost and further tailor proximity exchange in graphene by CGT encapsulation.

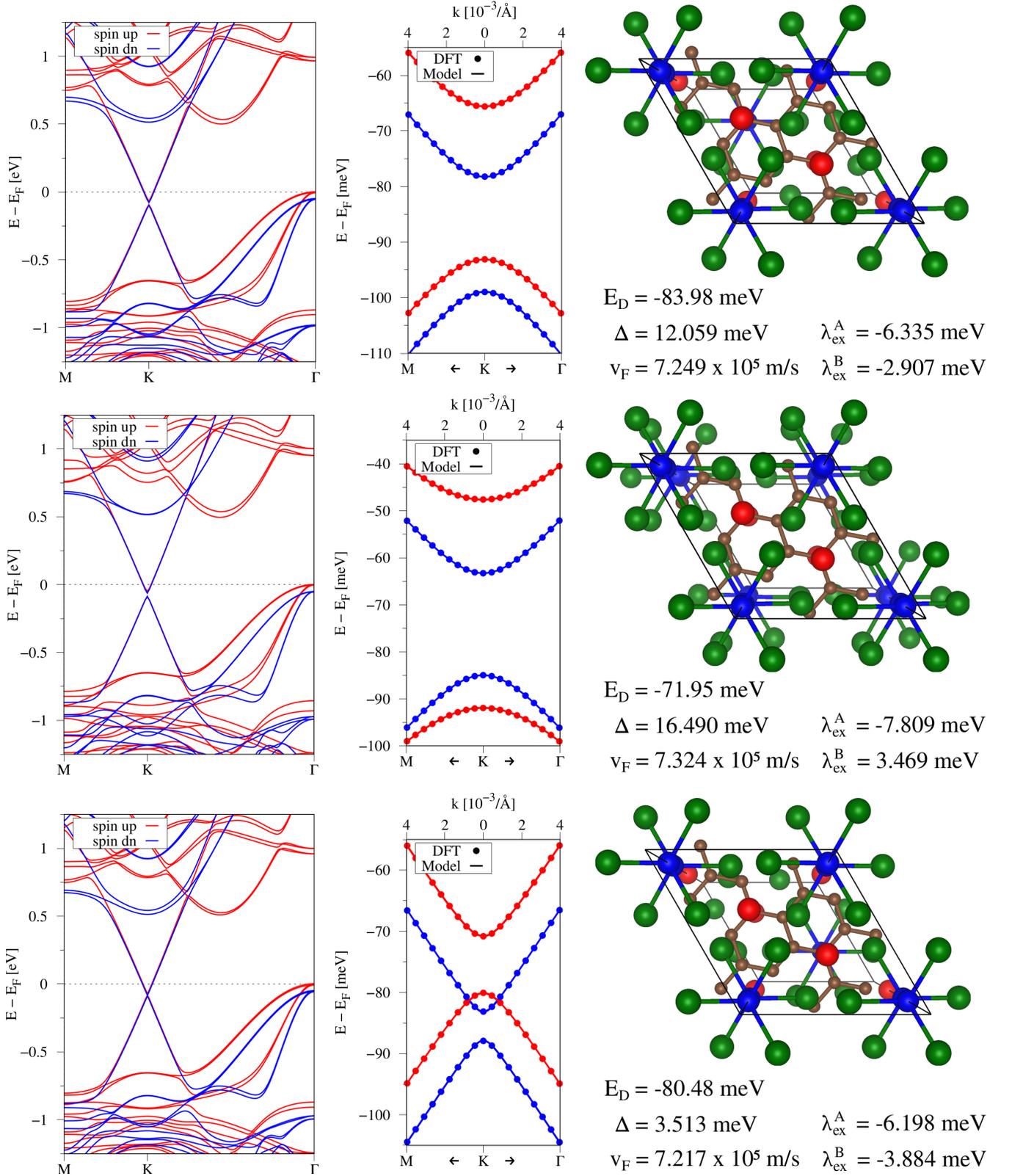

FIG. S10. Left: DFT-calculated band structures of the CGT/graphene/CGT heterostructures along the high-symmetry path M-K-Γ for twist angles of 19.1° and when the CGT lattice constant is about 6.85 Å. Red (blue) solid lines correspond to spin up (spin down). Middle: Zooms to the DFT-calculated (symbols) low energy Dirac bands near the K point with a fit to the model Hamiltonian (solid lines). Right: Top views of the corresponding geometries and fit parameters.

# VIII.   ORIGIN OF PROXIMITY EXCHANGE

One of the main questions is: What causes the different proximity exchange couplings?

## S1.   Analyzing the local Density of States

We consider three important angles, 19.1° with antiferromagnetic proximity exchange, and 0° and 30° with opposite ferromagnetic proximity exchange. Here, we consider structures from Table S3 where CGT is unstrained. In Fig. S11, we again show the calculated dispersion, the spin and atom resolved density of states (corresponding to the Dirac bands) and the spin polarization of Cr, Ge, and Te atoms. We have also calculated the spin polarizations in real space (Fig. S12, Fig. S13, and Fig. S14), $\Delta\rho = \rho_\uparrow - \rho_\downarrow$, where the spin densities $\rho_{\uparrow/\downarrow}(\boldsymbol{r}) = \sum_{n,\boldsymbol{k}} |\phi^{\boldsymbol{k}}_{n,\uparrow/\downarrow}(\boldsymbol{r})|^2$ are sums over eigenstates with the corresponding spin. To find out about the proximity exchange couplings, we take into account the Dirac states within the shaded regions in Fig. S11 for the calculation of spin densities (conduction and valence bands individually).

For 30°, we find an overall negative spin polarization on CGT that contributes to the Dirac spectrum, see Fig. S11. Looking at the decomposition of the DOS at about 40 meV above the Fermi level, the spin up (down) Dirac bands are formed by about 98.86% C, 0.35% Cr, 0.18% Ge, and 0.61% Te (97.27% C, 0.89% Cr, 0.82% Ge, 1.02% Te). At about 10 meV above the Fermi level, the spin up (down) Dirac bands are formed by about 98.90% C, 0.33% Cr, 0.18% Ge, and 0.60% Te (97.43% C, 0.83% Cr, 0.78% Ge, and 0.96% Te). From Fig. S12, we find a negative (positive) spin polarization on graphene, when taking into account conduction (valence) band states only, consistent with the calculated density of states and the low energy bands, see Fig. S5. From Fig. S12 it is evident that there is mainly a coupling of graphene $p_z$ orbitals to a spin down density on CGT (hence a significant negative spin polarization from CGT). From a simple perturbation theory point of view, coupling to a spin down density could lower the energy of spin down Dirac states, which is consistent with the Dirac spectrum for 30°.

For 0°, looking at the decomposition of the DOS at 10 meV above the Fermi level, the spin up (down) Dirac bands are formed by about 97.83% C, 0.66% Cr, 0.30% Ge, and 1.20% Te (98.03% C, 0.34% Cr, 0.38% Ge, and 1.25% Te). At about −10 meV below the Fermi level, we find that spin up (down) Dirac bands are formed by about 97.82% C, 0.66% Cr, 0.31% Ge, and 1.22% Te (97.95% C, 0.35% Cr, 0.39% Ge, and 1.31% Te). From this it is evident that there is mainly a coupling between graphene $p_z$ to Te $p$ orbitals. From the real space spin polarization, see Fig. S13, we find that conduction band Dirac states couple mainly to a positive spin density on Cr and non-interfacial Te atoms, while for the valence band there is a coupling to a negative spin density on interfacial Te atoms. Consistent with the DOS, see Fig. S7(c), we find positive (negative) spin polarization on graphene when considering conduction (valence) band Dirac states only. The spin polarizations are exactly opposite as for the 30° case.

For 19.1°, looking at the decomposition of the DOS at the Fermi level, the spin up (down) Dirac bands are formed by about 98.56% C, 0.21% Cr, 0.11% Ge, and 1.12% Te (98.98% C, 0.17% Cr, 0.12% Ge, and 0.73% Te). At about −45 meV below the Fermi level, we find that spin up (down) Dirac bands are formed by about 98.77% C, 0.28% Cr, 0.15% Ge, and 0.79% Te (99.00% C, 0.17% Cr, 0.16% Ge, and 0.67% Te). From Fig. S14, we observe opposite spin polarizations on the graphene sublattices, which supports our finding of antiferromagnetic exchange coupling. Consistent with the calculated density of states, see Fig. S6(c), we also see a change of the spin polarization on the graphene sublattices, when considering valence or conduction band states only.

15

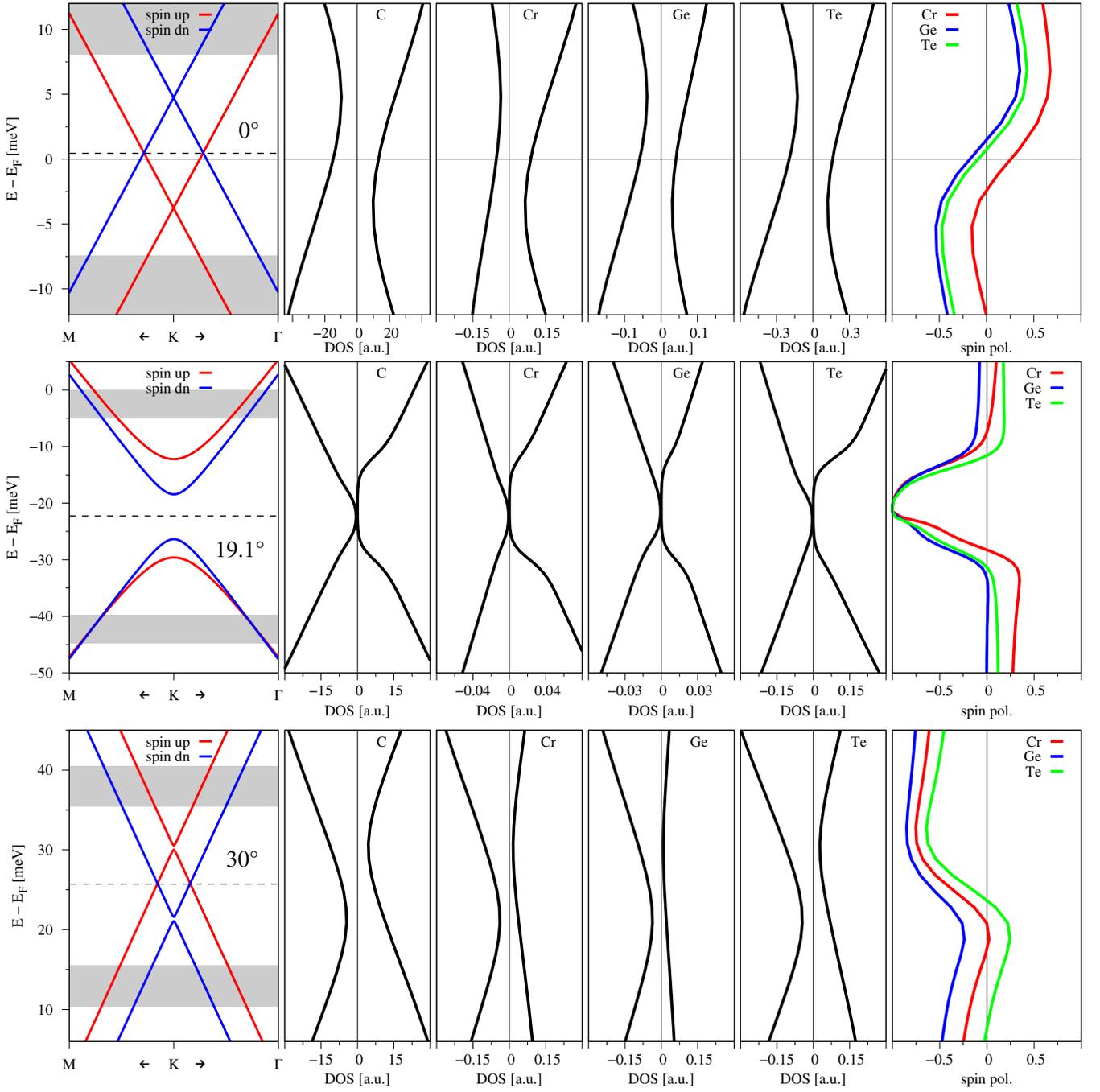

FIG. S11. For 0°, 19.1°, and 30°, we again show the low energy Dirac band structure. The dashed line indicates the Dirac point energy $E_D$. In addition, we show the corresponding atom and spin resolved local density of states, and the spin polarization calculated as $(N_\uparrow - N_\downarrow)/(N_\uparrow + N_\downarrow)$ for Cr, Ge, and Te atoms. The C density of states and spin polarization is investigated in more detail in Figs. S5, S6, and S7. The shaded regions are employed for the calculation of the spin densities in real space.

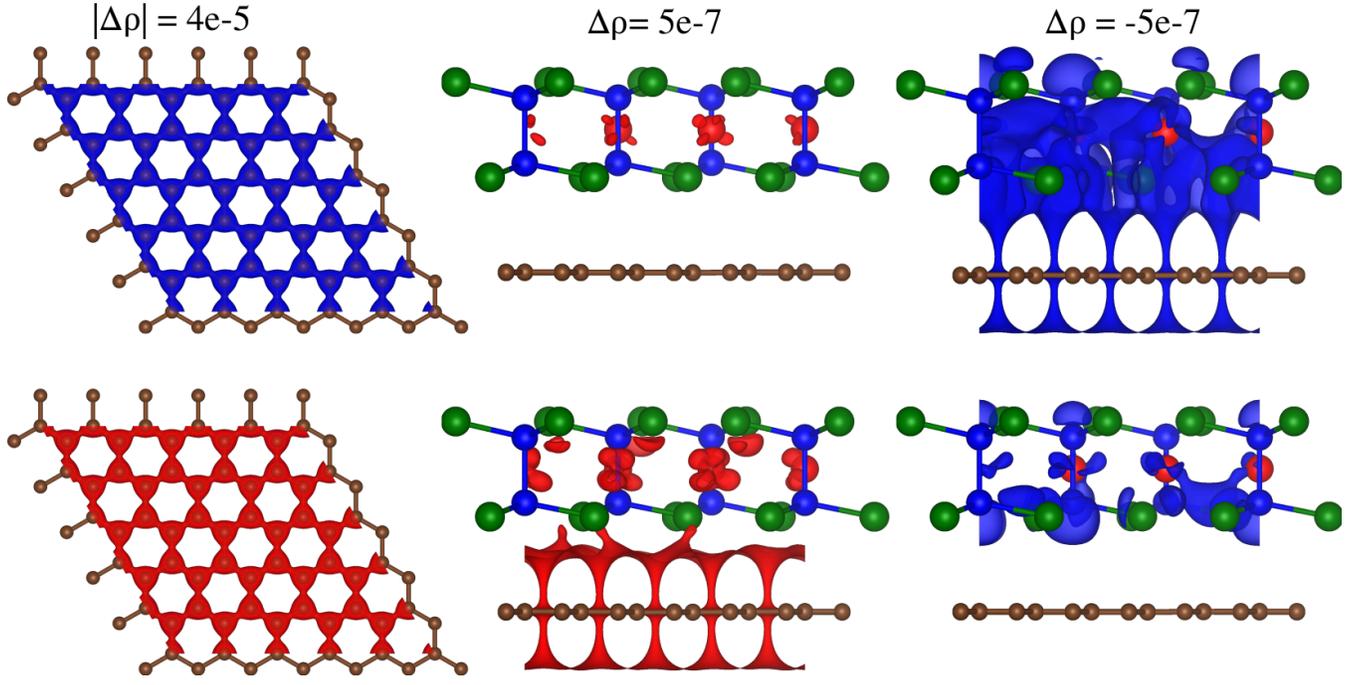

FIG. S12. Calculated spin polarization, $\Delta\rho = \rho_\uparrow - \rho_\downarrow$, for the graphene/CGT heterostructure with a twist angle of $30°$. The color red (blue) corresponds to $\Delta\rho > 0$ ($\Delta\rho < 0$). The isosurfaces correspond to isovalues (units $Å^{-3}$) as indicated by the labels above the figures. Left: Top view on graphene only. Middle and Right: Side views of the full heterostructure. Upper (lower) row takes into account conduction band (valence band) states only.

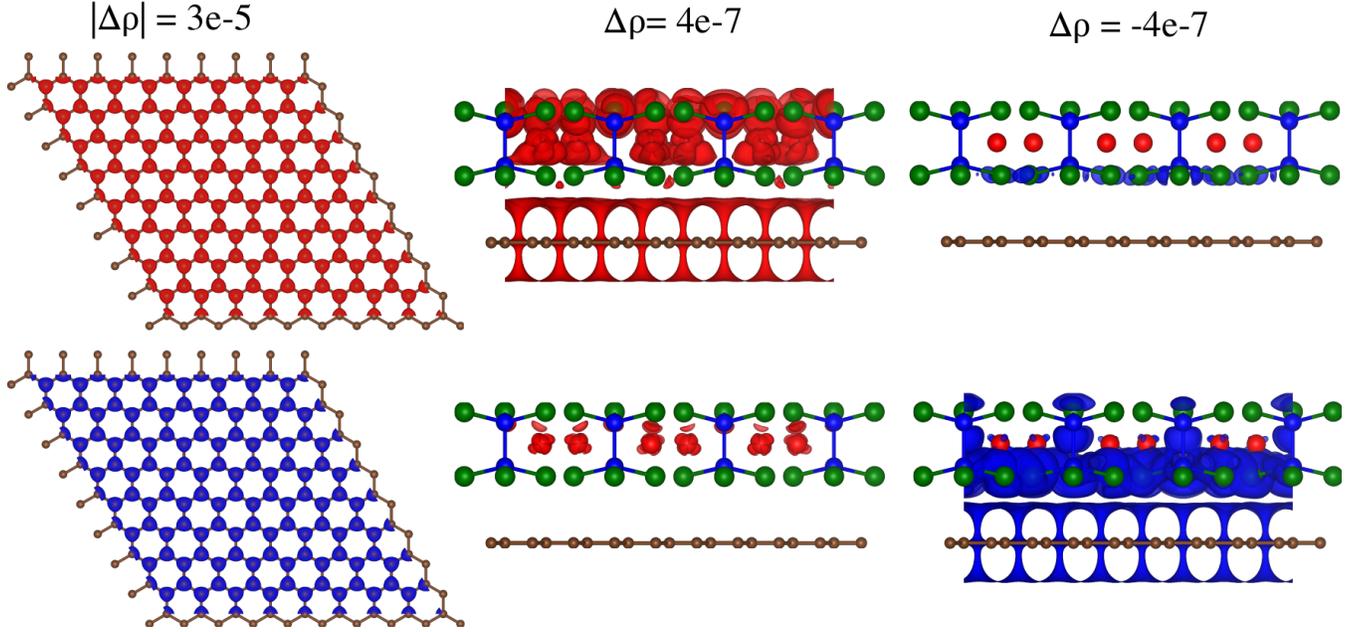

FIG. S13. Same as Fig. S12, but for a twist angle of $0°$.

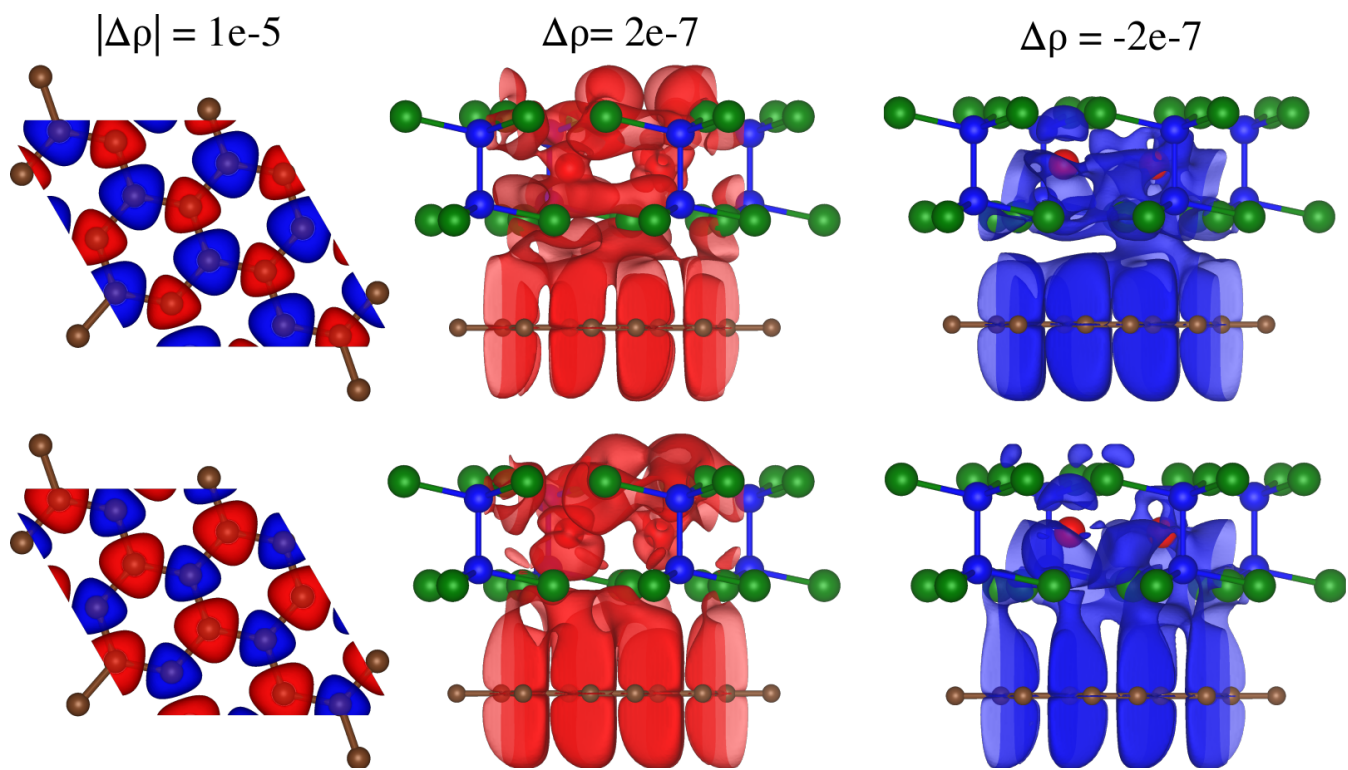

FIG. S14. Same as Fig. S12, but for a twist angle of 19.1°.

## S2. Backfolding of the Dirac point

Recent tight-binding considerations of twisted graphene/TMDC heterostructures [31, 32] show that the Dirac states couple to different $k$ points in the TMDC Brillouin zone for different twist angles, to ensure quasi momentum conservation. This strategy could be used to illustrate the the different contributions to the proximity exchange coupling we have found for our graphene/CGT heterostructures. In Fig. 5 in the main text we perform a similar analysis as in Refs. [31, 32]. Looking at the CGT monolayer band structure (Figs. S15, S16, and S17) and taking into account the backfolding of the graphene K point, as well as the band offsets for the different twist angles, we find that the Dirac states can couple to different bands of CGT.

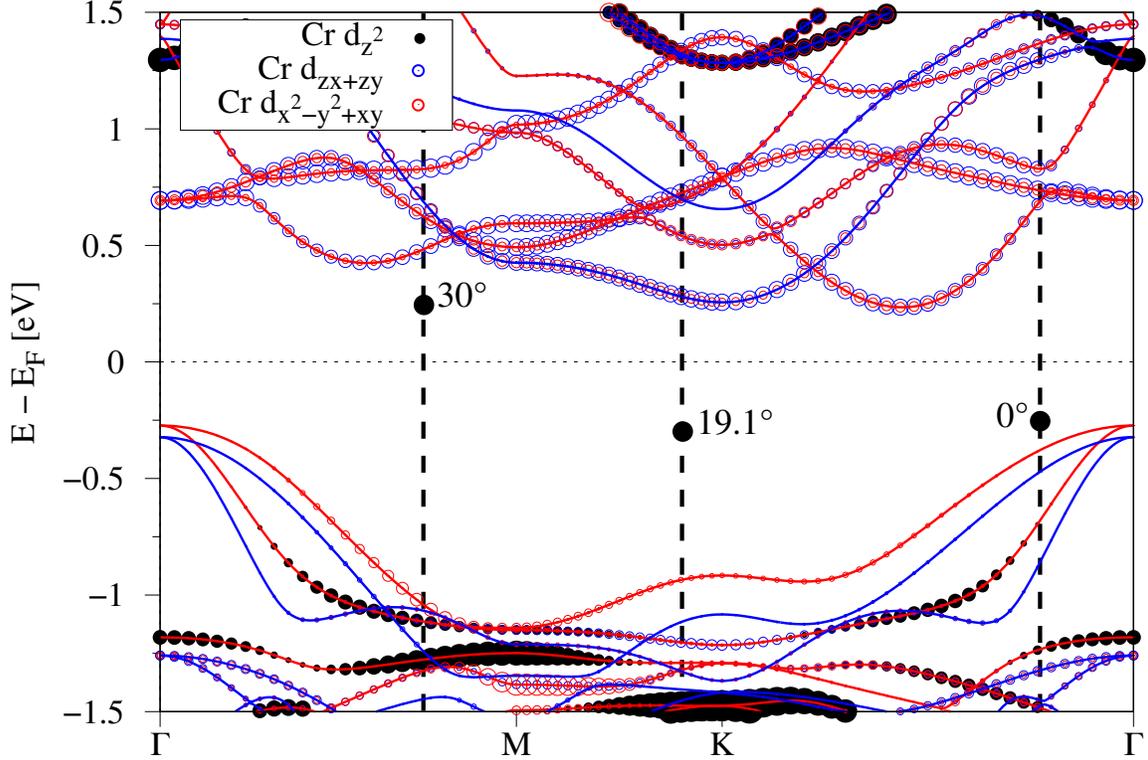

FIG. S15. DFT-calculated band structure of monolayer CGT, projected onto Cr $d$ orbitals. The size of the symbols represent the contribution of the orbitals to a specific band and at a given $k$ point. The color of the solid lines represent spin up (red) and spin down (blue). We also indicate the locations, to which the Dirac point backfolds, similar to Fig. 5.

From Fig. 5 it is evident that for 0°, the graphene Dirac states at the K point could predominantly couple to CGT valence band states near the Γ point. These CGT states are spin up and mainly formed by Te $p$ orbitals, see Fig. S17. For 30° twist angle, we find the Dirac states very close to CGT conduction band states near the M point. These CGT bands are formed mainly by Cr $d$ orbitals, see Fig. S15. For the 19.1° case, the Dirac states can couple to CGT bands near the K point. Here, the Dirac point is located nearly in the middle of CGT valence (formed by Te $p$ orbitals) and conduction (formed by Cr $d$ and Ge $p$ orbitals) bands, so that coupling to both seems reasonable. The competition between the couplings could effectively lead to antiferromagnetic proximity exchange. Our analysis can at least give indications about the different proximity exchange. What certainly also matters, as we have seen especially for the 19.1° twist angle, is the precise atomic registry of graphene above CGT.

In addition, the band offset of the Dirac bands with respect to the TMDC band edges was found to be important [31, 32]. Also here, the location of the Dirac bands with respect to the CGT bands plays a major role for the proximity exchange coupling, as we have found from our electric field results. From Table S6, we find that the closer the Dirac point is to the CGT conduction band edge, the larger in magnitude are the proximity exchange parameters (except for 12.2°).

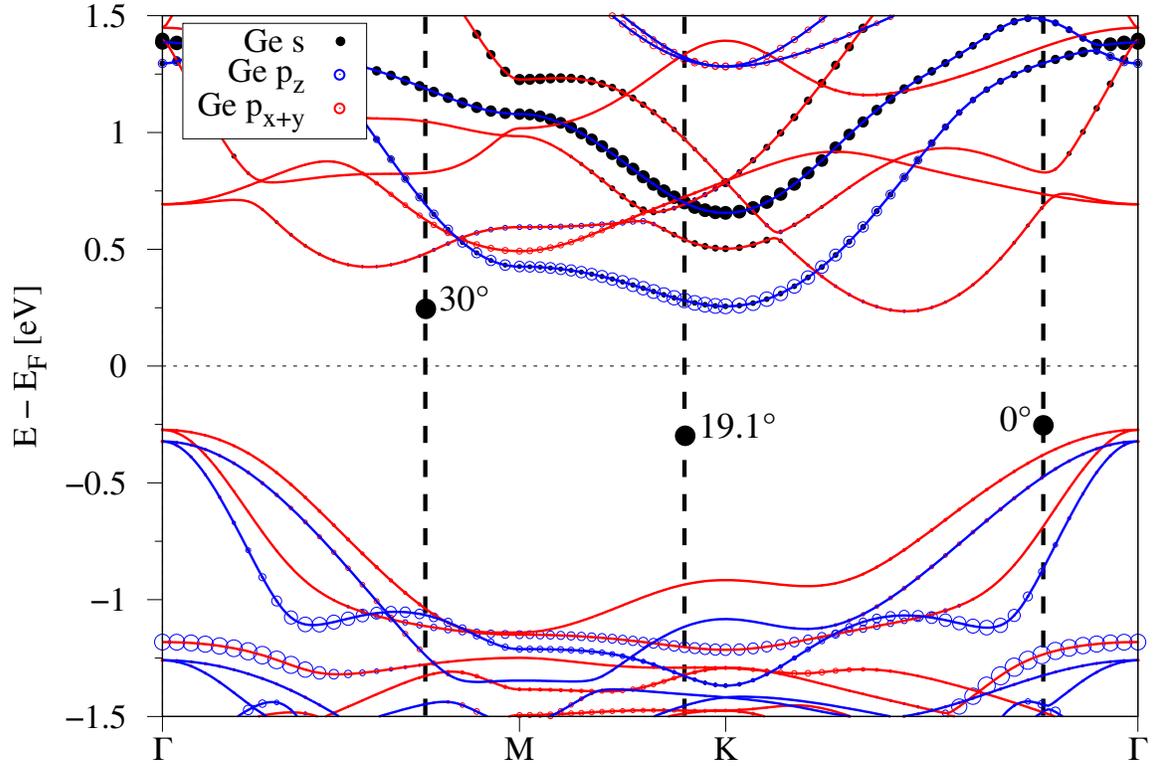

FIG. S16. Same as Fig. S15, but projected onto Ge $s$ and $p$ orbitals.

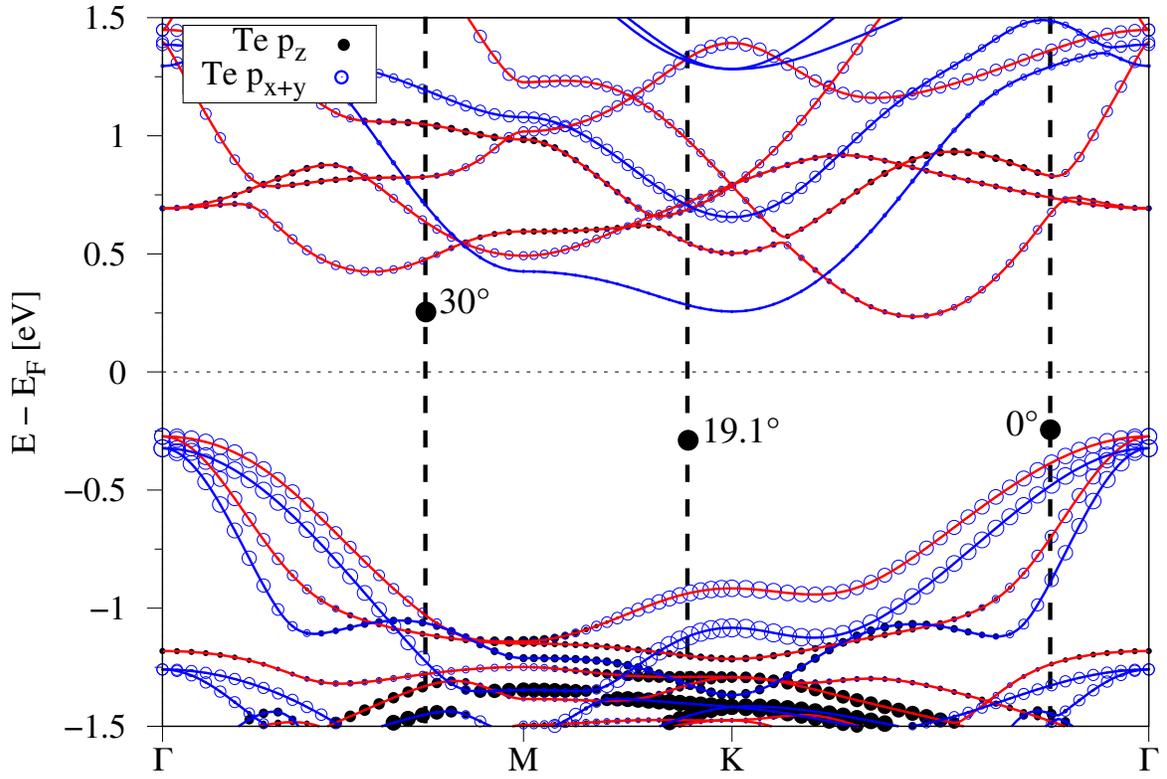

FIG. S17. Same as Fig. S15, but projected onto Te $p$ orbitals.

## S3. Anticrossings in projected band structure

Another helpful tool is to look for anticrossings between Dirac and CGT states in the projected band structure, see Fig. S18. We now believe, that this is most important to find out about the different proximity exchange couplings. For example, in the case of 30°, the Dirac states are very close to CGT spin up conduction bands, see Fig. S5(a). Also the backfolding picture would suggest a coupling to the energetically closest CGT spin up conduction bands, see Fig. 5. If such a coupling would be present, then spin up Dirac bands would be lowered in energy (considering simple perturbation theory), which is contrary to the low energy Dirac dispersion, see Fig. S5(b). From the projected band structure, we find that the coupling happens predominantly with CGT spin down conduction bands, as the pronounced anticrossings suggest [see the grey circles in Fig. S18]. The coupling to CGT spin down conduction bands lowers the spin down Dirac states in energy, consistent with the low energy dispersion for 30°. Such a band hybridization/anticrossing picture has been previously discussed in Ref. [33]. The coupling spin down states is also consistent with the local density of states and the spin polarization that we have found, see Fig. S11 and Fig. S12.

Let us do the same analysis for the other angles. In the case of 0°, the Dirac states are close to CGT spin up valence bands, see Fig. S7(a) and Fig. 5. This would lead to a raise of spin up Dirac bands in energy, which is again contrary to the calculated low energy dispersion, see Fig. S7(b). Looking at the projected band structure, the main anticrossing is between CGT spin up conduction bands with Dirac states. This coupling lowers the spin up Dirac bands in energy, consistent with the low energy dispersion for 0°. Again, this is also consistent with the local density of states and the spin polarization we have found, see Fig. S11 and Fig. S13, where a positive CGT spin polarization contributes to the conduction band Dirac states. In addition, a negative CGT spin polarization contributes to the valence band Dirac states, which leads to the same conclusion.

In the case of 19.1°, the first anticrossing in the conduction band is with CGT spin down bands, which lowers conduction band Dirac states in energy, consistent with the calculated low energy dispersion, see Fig. S6(b). For the valence band Dirac states, there are anticrossings with spin up and spin down CGT bands. In total, the interplay of all these couplings effectively leads to the observed antiferromagnetic proximity exchange in the low energy Dirac spectrum.

Another way to analyze the projected dispersion is to look at the percentage of graphene orbitals within all the bands of the heterostructure, see Fig. S19, to identify significant interlayer coupling. Looking at the CGT band manifold in the energy window from $-1$ to $1$ eV only at the K point, we find that for 30° spin up CGT bands contain nearly no graphene orbitals. In contrast, the spin down CGT bands contain much more graphene orbitals; hence a pronounced coupling is present. This coupling of graphene orbitals to CGT spin down conduction bands leads to a lowering of spin down Dirac bands for 30°. For 0°, the spin up CGT conduction bands contain more graphene orbitals than the spin down ones. The situation is reversed compared to 30° and the coupling leads to a lowering of spin up Dirac bands.

Finally, we would like to say a few words about the sensitivity of proximity exchange to the atomic registry in the case of 19.1°. In Fig. S20, we show the global band structures for two different lateral shifts, as listed in Table S7, where we also highlight band anticrossings. Without the shift ($x = 0, y = 0$) we get antiferromagnetic proximity exchange and barely anticrossings. For a different atomic registry ($x = 1/3, y = 2/3$), we get ferromagnetic proximity exchange and highly pronounced anticrossings. From that we can see that the atomic registry plays a highly important role for interlayer coupling and the proximity exchange. This is especially valid for small supercells.

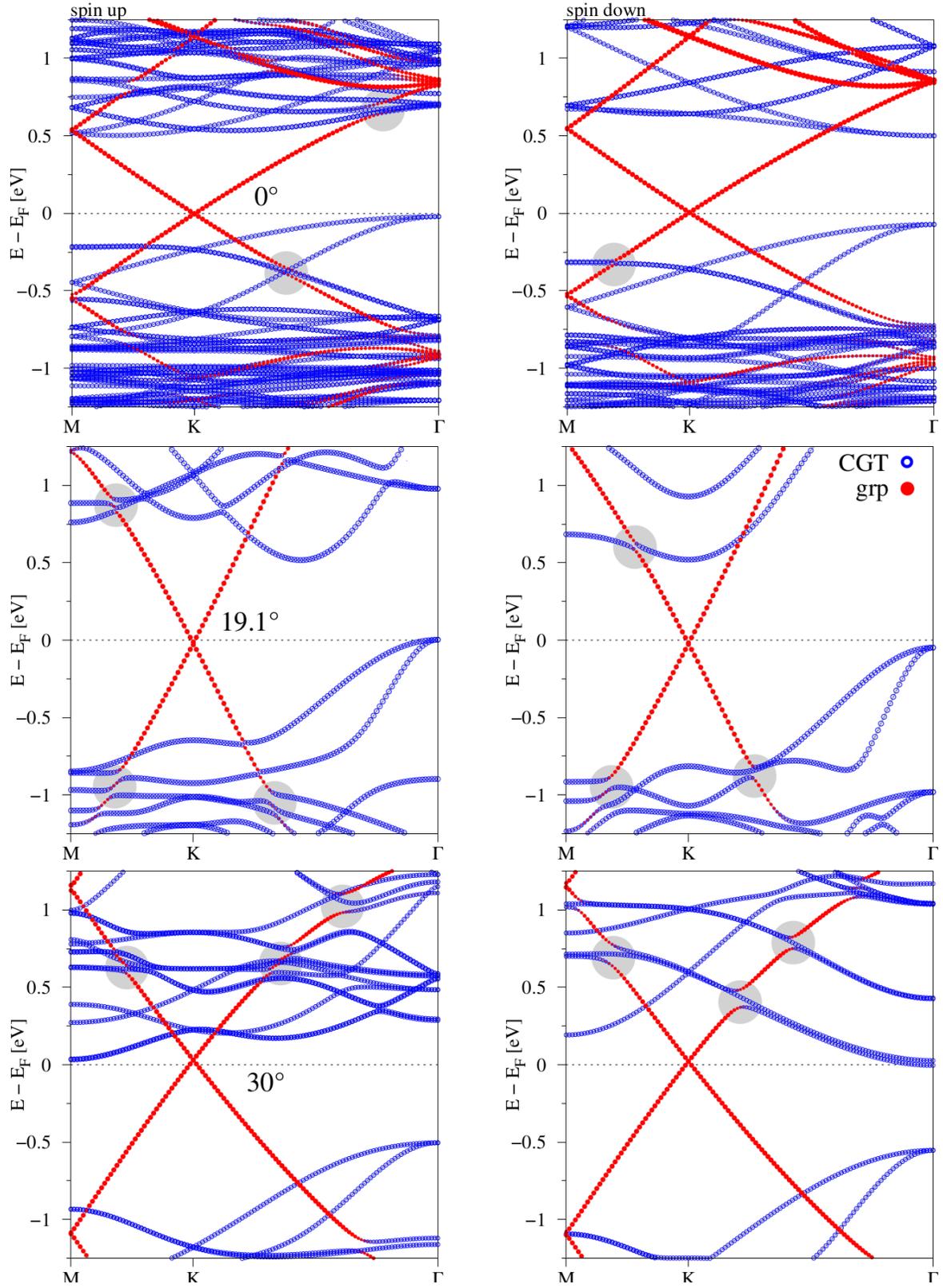

FIG. S18. DFT-calculated projected band structures of the graphene/CGT heterostructures for twist angles of 0°, 19.1°, and 30°. Left: Spin up bands. Right: Spin down bands. We project onto graphene (red) and CGT (blue) states. The grey circles are a guide for the eye to identify band anticrossings, corresponding to interlayer hybridization.

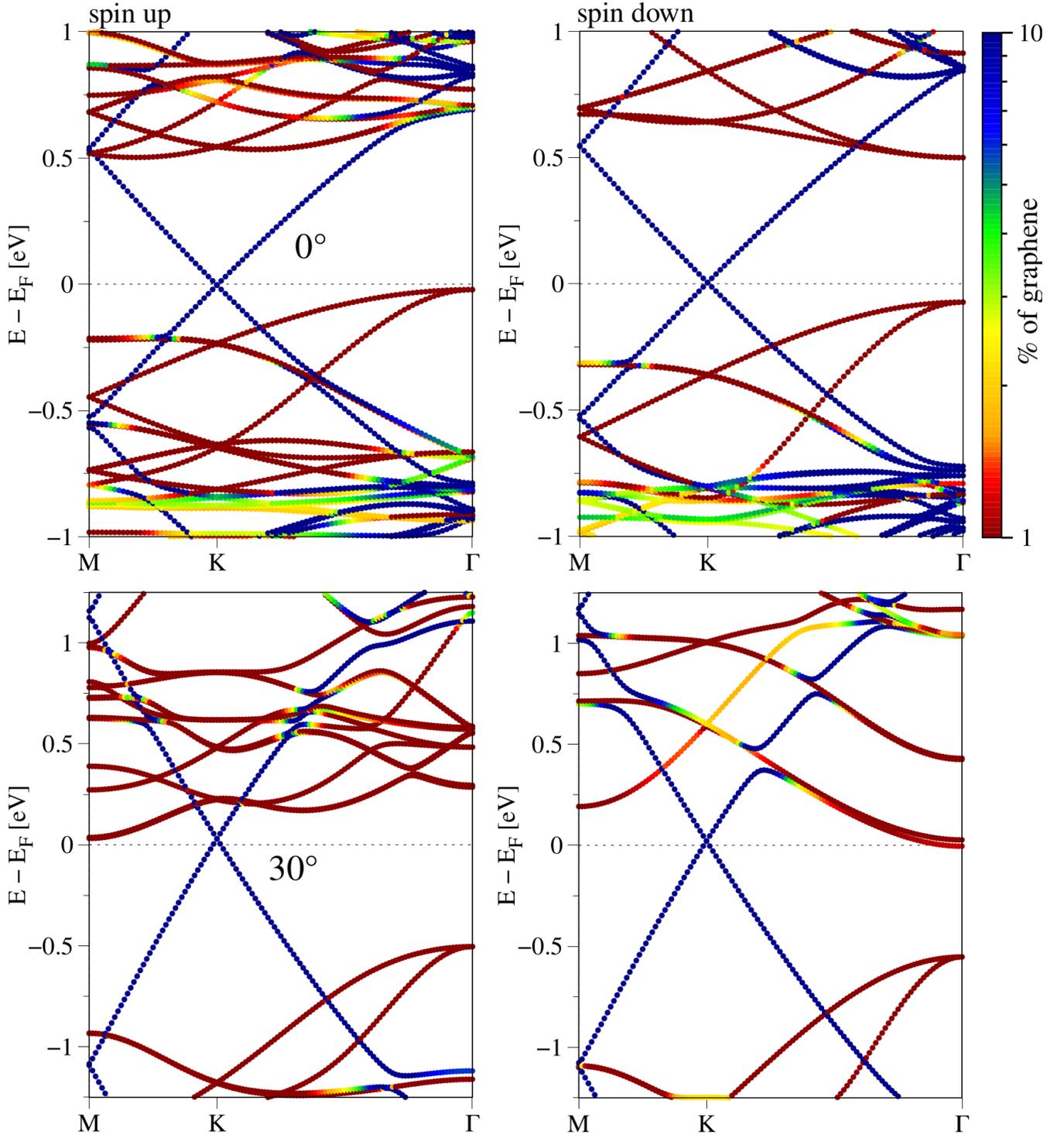

FIG. S19. DFT-calculated projected band structures of the graphene/CGT heterostructures for twist angles of 0° and 30°. Left: Spin up bands. Right: Spin down bands. The color code shows the contribution of graphene to the bands, i. e. the bands appear dark-reddish (blueish) when graphene orbitals are contained by less than 1% (at least 10%).

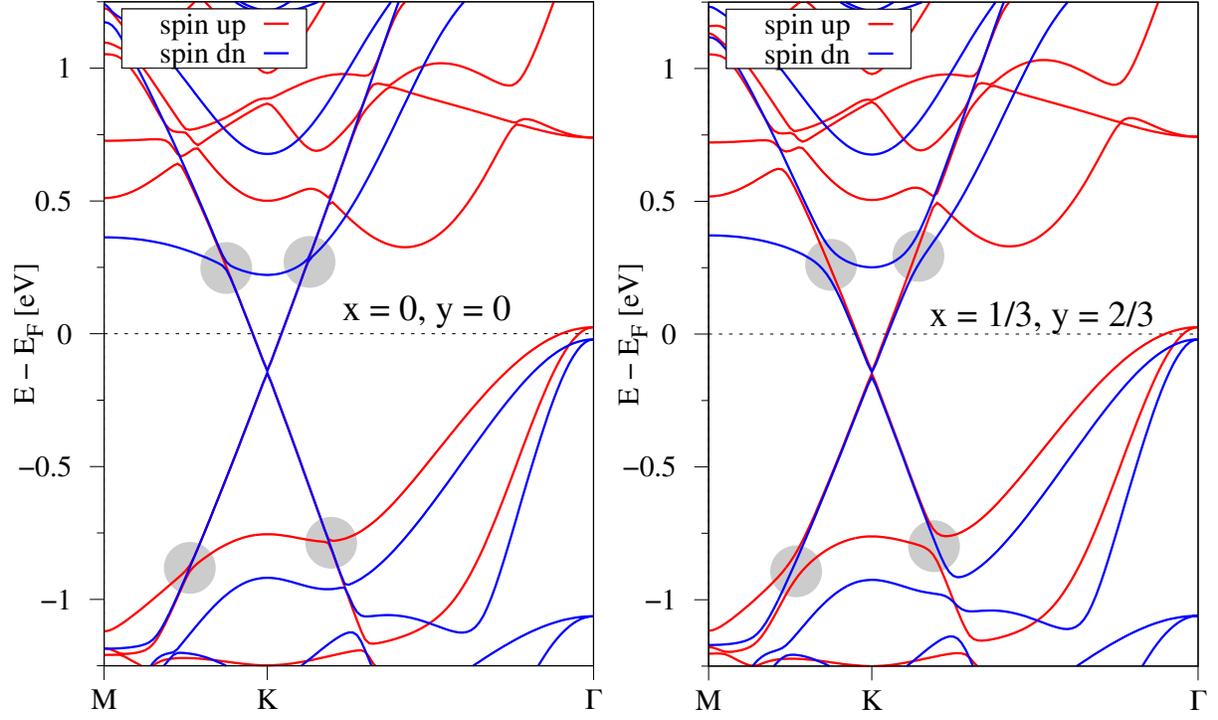

FIG. S20. DFT-calculated band structures of the graphene/CGT heterostructure for a twist angle of 19.1°, but different atomic registries as listed in Table S7. Left: No lateral shift leading to antiferromagnetic proximity exchange. Right: Lateral shift of $x = 1/3$ and $y = 2/3$ leading to ferromagnetic proximity exchange. The grey circles identify four anticrossings, that strongly depend on the atomic registry, and therefore different proximity exchange couplings are realized.

## S4. Predicting the twist-angle dependence of proximity exchange

One of the fundamental questions is, whether one can predict the twist-angle dependence of the proximity exchange coupling in graphene from the monolayer CGT dispersion. A reasonable approach is to consider $z$-extended orbitals from CGT that couple across the van der Waals gap to the $p_z$ orbitals of graphene and lead to a splitting of the Dirac states. Therefore, the coupling $C_{n,k}$, between a Dirac state $E_D$ and a CGT band $E_n$, at a given $k$ point could read $C_{n,k} = \sum_z P_{n,k}^z \cdot e^{-d_z}$. Here, $P_{n,k}^z$ is the projection onto a $z$-extended orbital from CGT, wich is additionally weighted by an exponential function, taking into account the interlayer distance between C atoms and different CGT atomic layers. The interlayer distances between graphene and the CGT atomic layers are roughly 3.55, 4.01, 5.24, 6.41, and 6.92 Å, with the atom order C - $\text{Te}_1$ - $\text{Ge}_1$ - Cr - $\text{Ge}_2$ - $\text{Te}_2$. The projections are calculated from the CGT monolayer dispersion, taking into account Te $p_z$, Ge $p_z$, and Cr $d_{z^2}+d_{xz+yz}$ as the relevant $z$-extended orbitals. With the couplings, one can apply second order perturbation theory to the unperturbed spin up and spin down Dirac states, $E_{D,k,\sigma}$, as $E'_{D,k,\sigma} = E_{D,k,\sigma} + \sum_n \frac{|C_{n,k,\sigma}|^2}{E_{D,k,\sigma}-E_n}$. The couplings are now spin dependent with $\sigma = \uparrow, \downarrow$ and the summation considers all CGT valence and conduction bands $E_n$. From that we obtain the perturbation on spin up and spin down Dirac states individually as function of the $k$ point. Due to the backfolding, one can associate different $k$ points with different twist angles. In Fig. S21, we summarize the results, showing the monolayer CGT band structure and the calculated perturbations, assuming $E_{D,k,\sigma} = E_F$ for the unperturbed Dirac states. For all $k$ points, and therefore all twist angles, the spin up perturbation is always smaller than the spin down one. In other words, our approach would predict the same sign of proximity exchange for all twist angles, but with variations in its magnitude. In conclusion, we cannot make qualitative predictions for the twist-angle dependence of proximity exchange by simply considering the magnetic substrate, and one has to apply the full DFT methodology for the heterostructures.

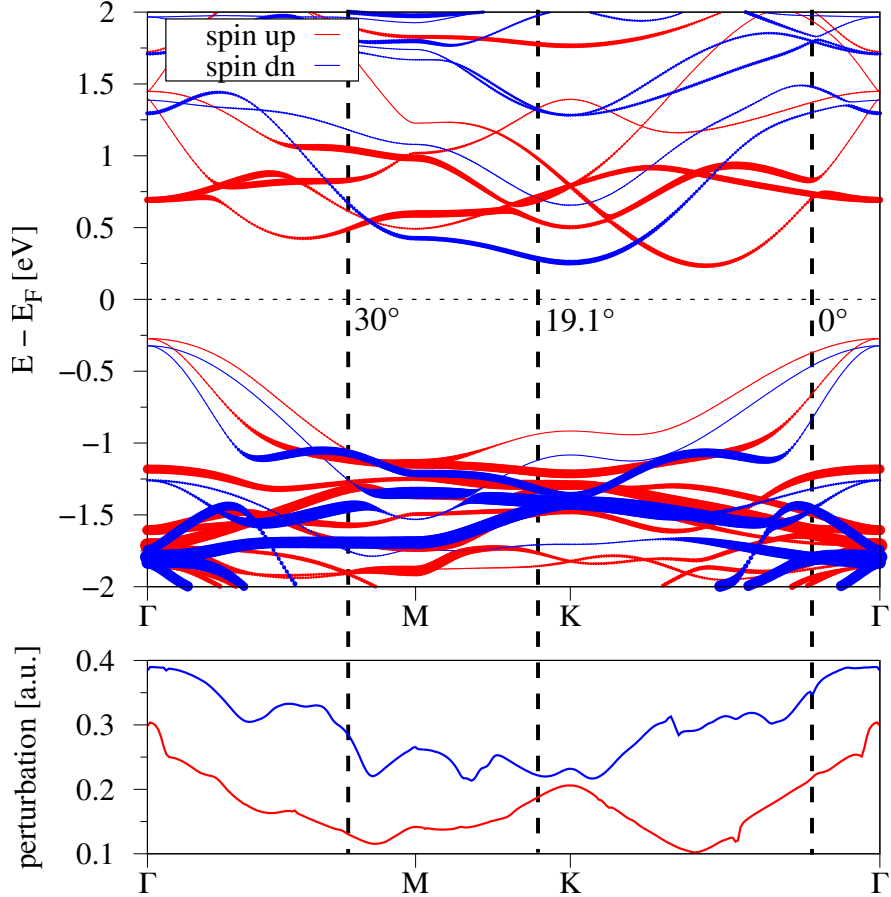

FIG. S21. Top: The DFT-calculated band structure of monolayer CGT, where the vertical dashed lines indicate the $k$-points, to which the Dirac states couple to, according to the backfolding. The line thicknesses are weighted by the couplings $C_{n,k}$. Bottom: The calculated perturbation on the spin up and spin down Dirac states, assuming $E_{D,k,\sigma} = E_F$ for the unperturbed Dirac states.

---

[1] S. R. Bahn and K. W. Jacobsen, "An object-oriented scripting interface to a legacy electronic structure code," Comput. Sci. Eng. **4**, 56 (2002).

[2] Predrag Lazic, "Cellmatch: Combining two unit cells into a common supercell with minimal strain," Computer Physics Communications **197**, 324 – 334 (2015).

[3] Daniel S Koda, Friedhelm Bechstedt, Marcelo Marques, and Lara K Teles, "Coincidence lattices of 2d crystals: heterostructure predictions and applications," The Journal of Physical Chemistry C **120**, 10895–10908 (2016).

[4] Stephen Carr, Shiang Fang, and Efthimios Kaxiras, "Electronic-structure methods for twisted moiré layers," Nature Reviews Materials **5**, 748–763 (2020).

[5] Y. Baskin and L. Meyer, "Lattice constants of graphite at low temperatures," Phys. Rev. **100**, 544 (1955).

[6] V Carteaux, D Brunet, G Ouvrard, and G Andre, "Crystallographic, magnetic and electronic structures of a new layered ferromagnetic compound Cr2Ge2Te6," J. Phys.: Condens. Mat. **7**, 69 (1995).

[7] P. Hohenberg and W. Kohn, "Inhomogeneous electron gas," Phys. Rev. **136**, B864 (1964).

[8] Paolo Giannozzi and et al., "Quantum espresso: a modular and open-source software project for quantum simulations of materials," J. Phys.: Cond. Mat. **21**, 395502 (2009).

[9] Cheng Gong, Lin Li, Zhenglu Li, Huiwen Ji, Alex Stern, Yang Xia, Ting Cao, Wei Bao, Chenzhe Wang, Yuan Wang, Z. Q. Qiu, R. J. Cava, Steven G. Louie, Jing Xia, and Xiang Zhang, "Discovery of intrinsic ferromagnetism in two-dimensional van der Waals crystals," Nature **546**, 265 (2017).

[10] G. Kresse and D. Joubert, "From ultrasoft pseudopotentials to the projector augmented-wave method," Phys. Rev. B **59**, 1758 (1999).

[11] John P. Perdew, Kieron Burke, and Matthias Ernzerhof, "Generalized gradient approximation made simple," Phys. Rev. Lett. **77**, 3865 (1996).

[12] Jiayong Zhang, Bao Zhao, Yugui Yao, and Zhongqin Yang, "Robust quantum anomalous hall effect in graphene-based van der waals heterostructures," Phys. Rev. B **92**, 165418 (2015).

[13] Klaus Zollner, Marko D. Petrović, Kapildeb Dolui, Petr Plecháč, Branislav K. Nikolić, and Jaroslav Fabian, "Scattering-induced and highly tunable by gate damping-like spin-orbit torque in graphene doubly proximitized by two-dimensional magnet cr2ge2te6 and monolayer ws2," Phys. Rev. Research **2**, 043057 (2020).

[14] Bogdan Karpiak, Aron W. Cummings, Klaus Zollner, Marc Vila, Dmitrii Khokhriakov, Anamul Md Hoque, André Dankert, Peter Svedlindh, Jaroslav Fabian, Stephan Roche, and Saroj P. Dash, "Magnetic proximity in a van der Waals heterostructure of magnetic insulator and graphene," 2D Mater. **7**, 015026 (2019).

[15] Stefan Grimme, "Semiempirical gga-type density functional constructed with a long-range dispersion correction," J. Comput. Chem. **27**, 1787 (2006).

[16] Stefan Grimme, Jens Antony, Stephan Ehrlich, and Helge Krieg, "A consistent and accurate ab initio parametrization of density functional dispersion correction (DFT-D) for the 94 elements H-Pu," J. Chem. Phys. **132**, 154104 (2010).

[17] Vincenzo Barone, Maurizio Casarin, Daniel Forrer, Michele Pavone, Mauro Sambi, and Andrea Vittadini, "Role and effective treatment of dispersive forces in materials: Polyethylene and graphite crystals as test cases," J. Comput. Chem. **30**, 934 (2009).

[18] Lennart Bengtsson, "Dipole correction for surface supercell calculations," Phys. Rev. B **59**, 12301 (1999).

[19] Xingxing Li and Jinlong Yang, "CrXTe 3 (X = Si, Ge) nanosheets: two dimensional intrinsic ferromagnetic semiconductors," J. Mater. Chem. C **2**, 7071 (2014).

[20] Xiaofang Chen, Jingshan Qi, and Daning Shi, "Strain-engineering of magnetic coupling in two-dimensional magnetic semiconductor crsite3: Competition of direct exchange interaction and superexchange interaction," Phys. Letters A **379**, 60 (2015).

[21] Lifei Liu, Xiaohui Hu, Yifeng Wang, Arkady V Krasheninnikov, Zhongfang Chen, and Litao Sun, "Tunable electronic properties and enhanced ferromagnetism in cr2ge2te6 monolayer by strain engineering," Nanotechnology **32**, 485408 (2021).

[22] Johannes Christian Leutenantsmeyer, Alexey A Kaverzin, Magdalena Wojtaszek, and Bart J van Wees, "Proximity induced room temperature ferromagnetism in graphene probed with spin currents," 2D Mater. **4**, 014001 (2016).

[23] Petra Högl, Tobias Frank, Klaus Zollner, Denis Kochan, Martin Gmitra, and Jaroslav Fabian, "Quantum anomalous hall effects in graphene from proximity-induced uniform and staggered spin-orbit and exchange coupling," Phys. Rev. Lett. **124**, 136403 (2020).

[24] Talieh S. Ghiasi, Alexey A. Kaverzin, Avalon H. Dismukes, Dennis K. de Wal, Xavier Roy, and Bart J. van Wees, "Electrical and thermal generation of spin currents by magnetic bilayer graphene," Nature Nanotechnology **16**, 788–794 (2021).

[25] Matthew R Rosenberger, Hsun-Jen Chuang, Madeleine Phillips, Vladimir P Oleshko, Kathleen M McCreary, Saujan V Sivaram, C Stephen Hellberg, and Berend T Jonker, "Twist angle-dependent atomic reconstruction and moiré patterns in transition metal dichalcogenide heterostructures," ACS nano **14**, 4550–4558 (2020).

[26] Astrid Weston, Yichao Zou, Vladimir Enaldiev, Alex Summerfield, Nicholas Clark, Viktor Zólyomi, Abigail Graham, Celal Yelgel, Samuel Magorrian, Mingwei Zhou, et al., "Atomic reconstruction in twisted bilayers of transition metal dichalcogenides," Nature Nanotechnology **15**, 592–597 (2020).

[27] Bálint Fülöp, Albin Márffy, Simon Zihlmann, Martin Gmitra, Endre Tóvári, Bálint Szentpéteri, Máté Kedves, Kenji Watanabe, Takashi Taniguchi, Jaroslav Fabian, and et al., "Boosting proximity spin–orbit coupling in graphene/wse2 heterostructures via hydrostatic pressure," npj 2D Materials and Applications **5**, 82 (2021).

[28] Bálint Fülöp, Albin Márffy, Endre Tóvári, Máté Kedves, Simon Zihlmann, David Indolese, Zoltán Kovács-Krausz, Kenji Watanabe, Takashi Taniguchi, Christian Schönenberger, and et al., "New method of transport measurements on van der waals heterostructures under pressure," Journal of Applied Physics **130**, 064303 (2021).

[29] Tribhuwan Pandey, Avinash P. Nayak, Jin Liu, Samuel T. Moran, Joon-Seok Kim, Lain-Jong Li, Jung-Fu Lin, Deji Akinwande, and Abhishek K. Singh, "Pressure-induced charge transfer doping of monolayer graphene/mos2 heterostructure," Small **12**, 4063–4069 (2016).

[30] Klaus Zollner and Jaroslav Fabian, "Bilayer graphene encapsulated within monolayers of ws$_2$ or cr$_2$ge$_2$te$_6$: Tunable proximity spin-orbit or exchange coupling," Phys. Rev. B **104**, 075126 (2021).

[31] Yang Li and Mikito Koshino, "Twist-angle dependence of the proximity spin-orbit coupling in graphene on transition-metal dichalcogenides," Phys. Rev. B **99**, 075438 (2019).

[32] Alessandro David, Péter Rakyta, Andor Kormányos, and Guido Burkard, "Induced spin-orbit coupling in twisted graphene–transition metal dichalcogenide heterobilayers: Twistronics meets spintronics," Phys. Rev. B **100**, 085412 (2019).

[33] Klaus Zollner, Martin Gmitra, Tobias Frank, and Jaroslav Fabian, "Theory of proximity-induced exchange coupling in graphene on hbn/(co, ni)," Phys. Rev. B **94**, 155441 (2016).

\clearpage

\end{document}